\RequirePackage{fix-cm}

\documentclass[smallcondensed]{svjour3}     
\smartqed 
\usepackage{graphicx}

\journalname{Machine Learning}

\makeatletter
\DeclareRobustCommand\onedot{\futurelet\@let@token\@onedot}
\def\@onedot{\ifx\@let@token.\else.\null\fi\xspace}

\def\ie{\emph{i.e.}}

\def\etal{\emph{et al.~}}
\makeatother

\usepackage{url}
\usepackage{algorithm}
\usepackage{algorithmic}
\usepackage{color}
\usepackage{soul}
\usepackage{hyperref}

\begin{document}

\title{Weakly Supervised Change Detection Using Guided Anisotropic Diffusion}

\author{Rodrigo Caye Daudt         \and
        Bertrand Le Saux         \and
        Alexandre Boulch         \and
        Yann Gousseau
}

\institute{Rodrigo Caye Daudt \at
            Institute of Geodesy and Photogrammetry \\
              ETH Zurich, 8092 Zurich, Switzerland \\
              \email{rodrigo.cayedaudt@geod.baug.ethz.ch}
}

\date{Received: date / Accepted: date}

\maketitle

\begin{abstract}
Large scale datasets created from crowdsourced labels or openly available data have become crucial to provide training data for large scale learning algorithms. While these datasets are easier to acquire, the data are frequently noisy and unreliable, which is motivating research on weakly supervised learning techniques. In this paper we propose original ideas that help us to leverage such datasets in the context of change detection. First, we propose the guided anisotropic diffusion (GAD) algorithm, which improves semantic segmentation results using the input images as guides to perform edge preserving filtering. We then show its potential in two weakly-supervised learning strategies tailored for change detection. The first strategy is an iterative learning method that  combines model optimisation and  data cleansing using GAD to extract the useful information from a large scale change detection dataset generated from open vector data. The second one incorporates GAD within a novel spatial attention layer that increases the accuracy of weakly supervised networks trained to perform pixel-level predictions from image-level labels. Improvements with respect to state-of-the-art are demonstrated on 4 different public datasets.

\keywords{Remote sensing \and Change detection \and Weak supervision \and Neural networks \and Anisotropic diffusion}

\end{abstract}

\section{Introduction}

Change detection (CD) is one of the oldest problems studied in the field of remote sensing image analysis~\cite{hussain2013change,singh1989review}. It consists of comparing a pair or sequence of coregistered images and identifying the regions where meaningful changes have taken place between the first and last acquisitions. However, the definition of meaningful change varies depending on the application. Changes of interest can be, for example, new buildings and roads, forest fires, and growth or shrinkage of water bodies for environmental monitoring. Although exceptions exist, such as object-based and change vector analysis methods, it is common for change detection algorithms to predict a change label for each pixel in the provided images by modelling the task mathematically as a segmentation or clustering problem. 

Many variations of convolutional neural networks (CNNs)~\cite{lecun1998gradient}, notably fully convolutional networks (FCNs)~\cite{long2015fully}, have recently achieved excellent performances in change detection tasks~\cite{daudt2018fully,daudt2018hrscd,chen2018mfcnet,guo2018learning}. These methods require large amounts of training data to perform supervised training of the proposed networks~\cite{lecun2015deep}. Unsupervised change detection methods with CNNs also exist, which circumvent the problem of scarce annotations~\cite{8608001,9120230,luppino2020codealigned,alvarez2020s2cgan}. Open labelled datasets for change detection are extremely scarce and are predominantly very small compared to labelled datasets in other computer vision areas. Benedek and Szirányi~\cite{benedek2009change} created the Air Change dataset which contain 13 small annotated images, divided into three regions. Daudt \etal created the OSCD~\cite{daudt2018urban} dataset from Sentinel-2 multispectral images, with a total of 24 fully annotated image pairs. While these datasets allow for simple models to be trained in a supervised manner, training more complex models with these data would lead to overfitting.

\begin{figure}[b]
     
    \begin{minipage}{0.25\linewidth}
        \centering
        \includegraphics[width=\linewidth]{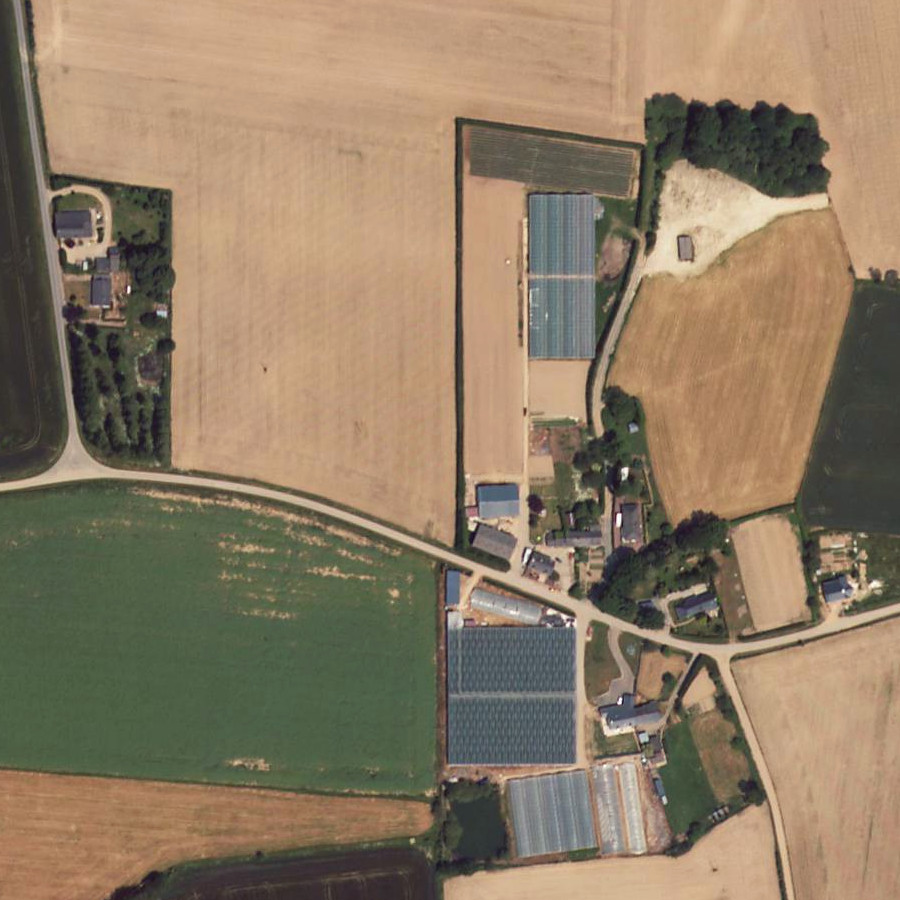}\\
        (a) Image 1 \newline
    \end{minipage}
    ~
    \begin{minipage}{0.25\linewidth}
        \centering
        \includegraphics[width=\linewidth]{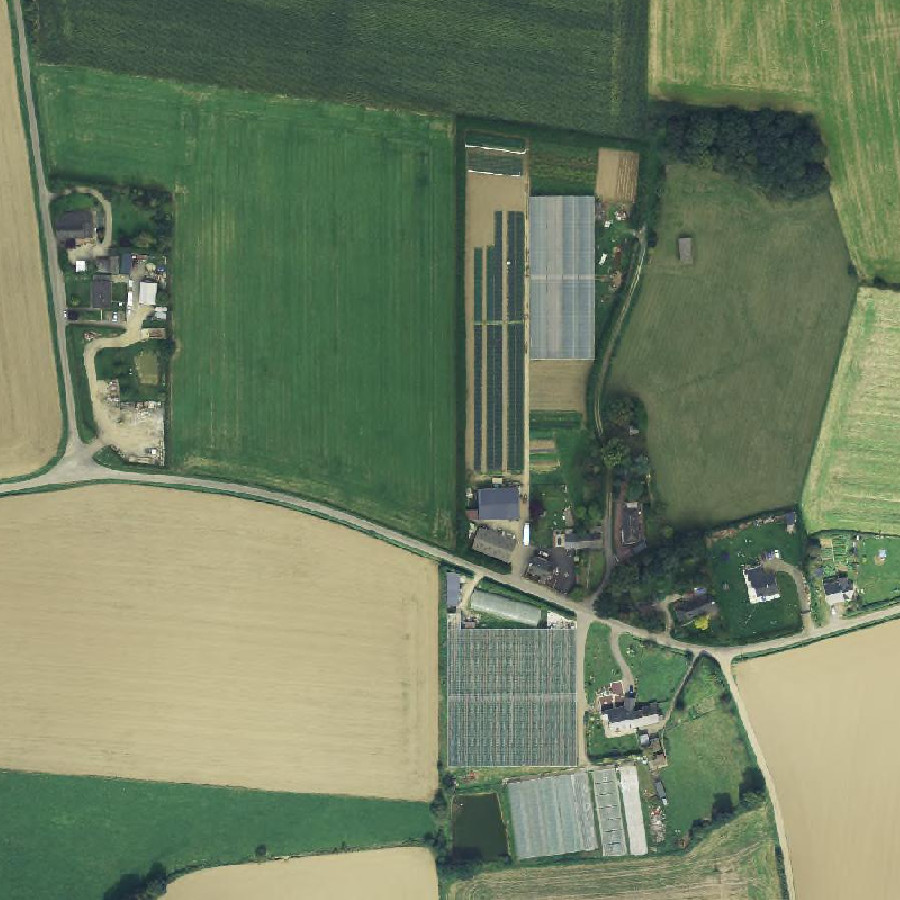}\\
        (b) Image 2 \newline
    \end{minipage}
    ~
    \begin{minipage}{0.25\linewidth}
        \centering
        \includegraphics[width=\linewidth]{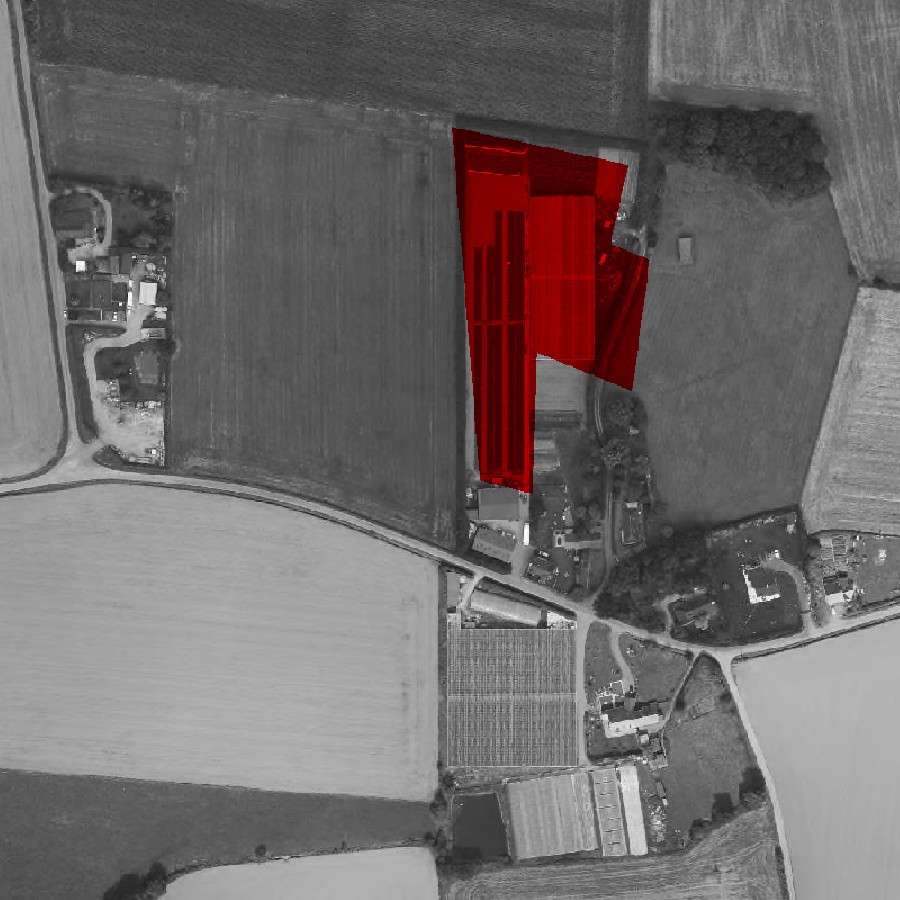}\\
        (c) Reference data
    \end{minipage}\break
    
    \begin{minipage}{0.25\linewidth}
        \centering
        \includegraphics[width=\linewidth]{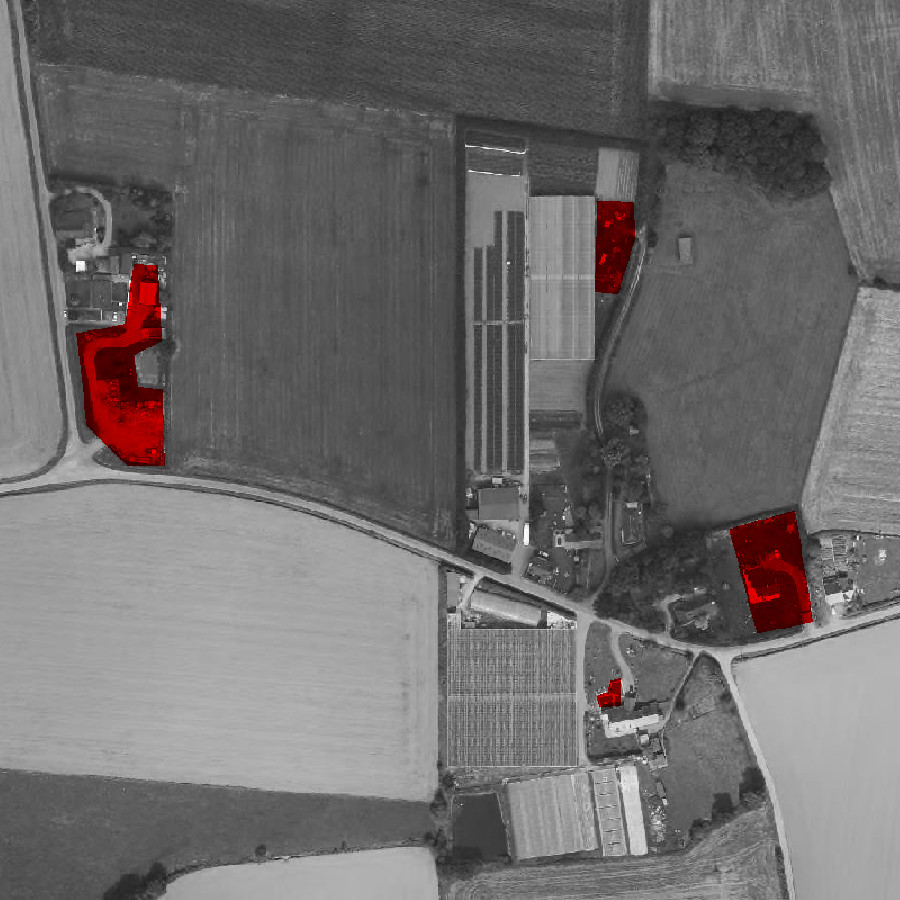}\\
        (d) Manual GT
    \end{minipage}
    ~
    \begin{minipage}{0.25\linewidth}
        \centering
        \includegraphics[width=\linewidth]{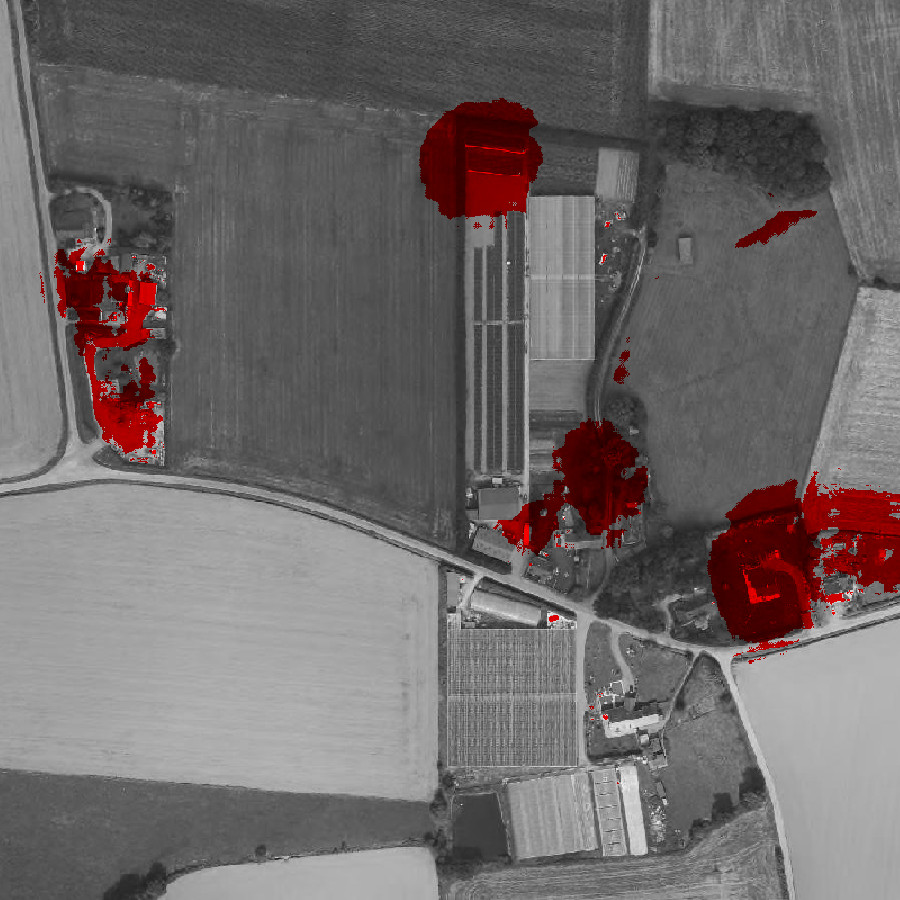}\\
        (e) Naive  \newline
    \end{minipage}
    ~
    \begin{minipage}{0.25\linewidth}
        \centering
        \includegraphics[width=\linewidth]{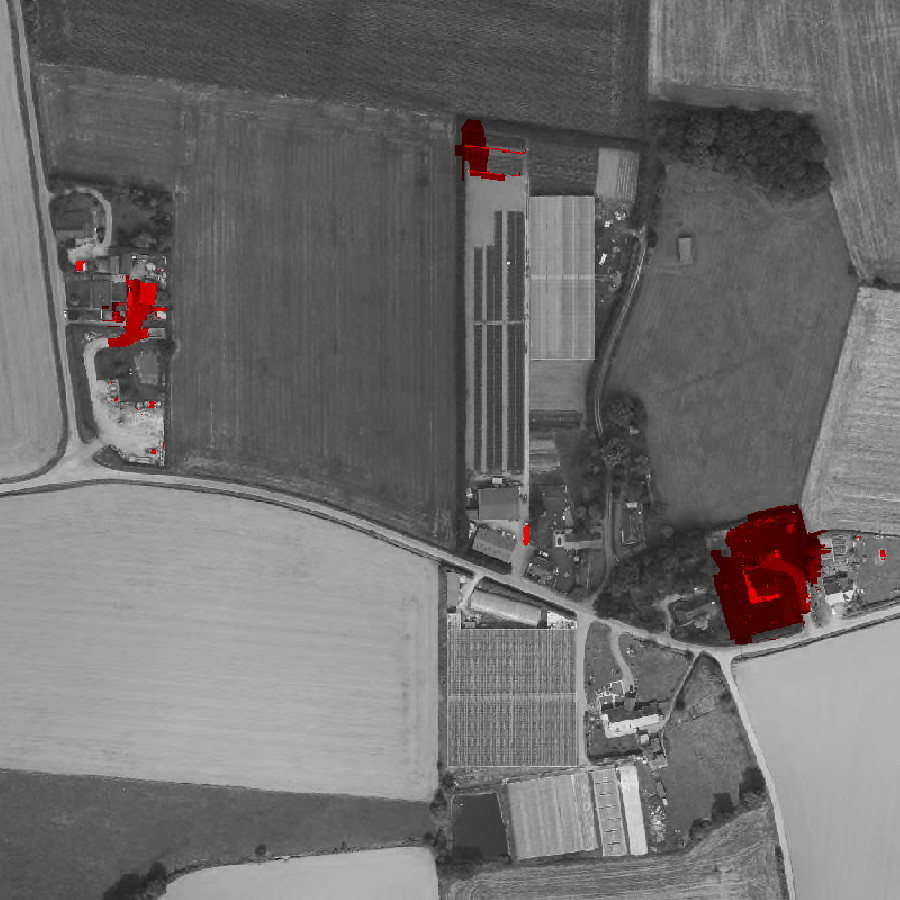}\\
        (f) Proposed \newline
    \end{minipage}\break
    
    \caption{(a)-(b) image pair, (c) change labels from the HRSCD dataset, (d) ground truth created by manually annotating changes, (e) result obtained by naive supervised training, (f) result obtained by our proposed method.}
    \label{fig:teaser}
\end{figure}

New datasets have recently appeared which change the scale of what is possible for machine learning approaches, but they also raise new issues which are illustrated by the two following examples. The High Resolution Semantic Change Detection (HRSCD) dataset~\cite{daudt2018hrscd} is a large scale change detection dataset that was generated by combining an aerial image database with open change and land cover data. Change maps and land cover maps were generated for 291 5000x5000 RGB image pairs, resulting in over 3000 times more annotated pixels than previous change detection datasets. This dataset, however, contains unreliable labels due to having been generated automatically. The effect of naively using these data for supervised learning of change detection networks is shown in Fig.~\ref{fig:teaser}. Inaccuracies in the reference data stem primarily from two causes: imperfections in the vector data at different semantic levels, and temporal misalignment between the annotations and the images. Naive supervision using such data leads to overestimation of the detected changes, as can be seen in Fig.~\ref{fig:teaser}(e). Nevertheless, there is much useful information in the available annotations that, if used adequately, can lead to better CD systems.
Due to the way the ground truth was generated, the labels in the dataset mark changes at a land parcel level with imprecise boundaries.
That is often the case when labels are extracted from parcel polygons.
While useful for global monitoring of changes in land cover, such labellings often do not delineate precise object-level changes.

Other change detection datasets rely on cross referencing data obtained by on-site surveys with available remote sensing imagery. Such is the case of the ABCD dataset~\cite{abcd}, which contains image pairs centered on buildings in a region that has been affected by a tsunami. Images before and after the event were taken with different sensors, and were registered and cropped around each known building in the area. Binary change labels for each image pair are available, but segmentation labels are not. This dataset contains over 8000 labelled image pairs, and is available in two versions: \textit{fixed scale}, where the spatial resolution of the images is kept constant, and \textit{resized}, where images are resized so that the length of the imaged building takes up roughly a third of the patch size.

We explore in this paper how to leverage and learn from imprecise or approximate labels for remote sensing image analysis. In particular, we propose the guided anisotropic diffusion (GAD) algorithm for label refinement. GAD is useful to improve the accuracy of prediction boundaries using the input images as guides, which is especially useful in a weakly supervised setting where accurate boundary predictions are often challenging to obtain. We validate its contribution in two weakly supervised learning settings that improve on previously proposed methods for semantic segmentation.  First, we use GAD in a training scheme that harnesses the useful information in the HRSCD dataset for parcel-wise change detection, attempting to refine the reference data while training a fully convolutional network. By acknowledging the presence of incorrect labels in the training dataset (with respect to our fine grained objective), we are able to focus on good data and ignore bad ones, improving the final results as seen in Fig.~\ref{fig:teaser}(f). A preliminary version of this idea has been proposed in \cite{Daudt_2019_CVPR_Workshops,daudt2018learning}. Second, we also assess GAD in a different task: performing pixel-level damage estimation in the ABCD dataset initially only designed for classification. GAD improves spatial attention weights, which are then used for classification and weakly supervised segmentation of changes. Finally, we evaluate the effectiveness of GAD as a postprocessing algorithm for enhancing the accuracy of region boundaries in semantic segmentation using fully convolutional networks.

\section{Related Work}

\textbf{Change detection} has a long history, being one of the early problems tackled in remote sensing image understanding~\cite{singh1989review}. It is done using coregistered image pairs or sequences, and consists of identifying areas in the images that have experienced significant modifications between the acquisitions. Many of the state-of-the-art ideas in pattern recognition have been used for change detection in the past, from pixel-level comparison of images, to superpixel segmentation, object-level image analysis, and image descriptors~\cite{hussain2013change}. In this paper we treat change detection as a two class semantic segmentation problem, in which a label is predicted for each pixel in the input images. With the rise of machine learning algorithms for semantic segmentation, notably convolutional neural networks, many algorithms have attempted to learn to perform change detection. Most algorithms circumvented the problem of the scarcity of training data through transfer learning by using pretrained networks to generate pixel descriptors~\cite{sakurada2015change,el2016convolutional,el2017zoom}. Fully convolutional networks trained end-to-end to perform change detection have recently been proposed by several authors independently, usually using Siamese architectures~\cite{zhan2017change,daudt2018fully,daudt2018hrscd,chen2018mfcnet,guo2018learning}. Unsupervised~\cite{8608001,9120230,luppino2020codealigned,alvarez2020s2cgan} and semi-supervised~\cite{9069898} alternatives have also been proposed to cope with the scarcity of accurately labelled data.

\textbf{Semantic segmentation} algorithms attempt to understand an input image and predict to which class among a known set of classes each pixel in an input image belongs. Change detection is modelled in this paper and many others as a semantic segmentation problem which takes as input two or more images. Long \etal proposed the first fully convolutional network for semantic segmentation, which achieved excellent performance and inference speed~\cite{long2015fully}. Since then, several improvements have been proposed for CNNs and FCNs. Ioffe and Szegedy have proposed batch normalization layers, which normalize activations and help avoid the vanishing/exploding gradient problem while training deep networks~\cite{ioffe2015batch}. Ronneberger \etal proposed the usage of skip connections that transfer details and boundary information from earlier to later layers in the network, which improves the accuracy around the edges between semantic regions~\cite{ronneberger2015u}. He \etal proposed the idea of residual connections, which have improved the performance of CNNs and FCNs and made it easier to train deep networks~\cite{he2016deep}.

\textbf{Noisy labels} for supervised learning is a topic that has already been widely explored~\cite{frenay2014comprehensive,frenay2014classification}. In many cases, label noise is random (by this we mean, following the literature terminology, independent of the data and not correlated)  and is modelled mathematically as such~\cite{natarajan2013learning,xiao2015learning,rolnick2017deep}. Rolnick \etal showed that supervised learning algorithms are robust to random label noise, and proposed strategies to further minimize the effect label noise has on training, such as increasing the training batch sizes~\cite{rolnick2017deep}. In the case presented in this paper, the assumption that the label noise is random does not hold. Incorrect change detection labels are usually around edges between regions or grouped together, which leads the network to learn to overestimate detected changes as seen in Fig.~\ref{fig:teaser}(e). Ignoring part of the training dataset, known as data cleansing (or cleaning), has already been proposed in different contexts~\cite{matic1992computer,john1995robust,guyon1996discovering,jeatrakul2010data}.

\textbf{Weakly supervised learning} is the name given to the group of machine learning algorithms that aim to perform different or more complex tasks than normally allowed by the training data at hand. Weakly supervised algorithms have recently gained popularity because they provide an alternative when data acquisition is too expensive. The problem of learning to perform semantic segmentation using only bounding box data or image level labels is closely related to the task discussed in this paper, since most methods propose the creation of an approximate semantic segmentation ground truth for training and dealing with its imperfections accordingly. Dai \etal proposed the BoxSup algorithm~\cite{dai2015boxsup} where region proposal algorithms are used to generate region candidates in each bounding box, before a semantic segmentation network is trained using these annotations and finally used to iteratively select better region proposal candidates. Khoreva \etal proposed improvements to the BoxSup algorithm that include using \textit{ad hoc} heuristics and an ignore class during training~\cite{khoreva2017simple}. They obtained best results using region proposal algorithms to create semantic segmentation training data directly from bounding boxes. Lu \etal modelled this problem as a simultaneous learning and denoising task through a  convex optimization problem~\cite{lu2017learning}. Ahn and Kwak proposed combining class activation maps, random walk and a learned network that predicts if pixels belong to the same region to perform semantic segmentation from image level labels~\cite{ahn2018learning}. Zhou \etal proposed the class activation mapping technique~\cite{Zhou_2016_CVPR}, which allows the networks to localize what regions in the image contribute to the prediction of each class, which can be harnessed for generating pixel-level predictions from image-level labels.

\textbf{Post-processing} methods that use information from guide images to filter other images, such as semantic segmentation results, have also been proposed~\cite{petschnigg2004digital,kopf2007joint,ferstl2013image}. A notable example is the Dense CRF algorithm proposed by Kr\"ahenb\"uhl and Koltun, in which an efficient solver is proposed for fully connected conditional random fields with Gaussian edge potentials~\cite{krahenbuhl2011efficient}. The idea of using a guide image for processing another is also the base of the Guided Image Filtering algorithm proposed by He \etal~\cite{he2013guided}, where a linear model that transforms a guide image into the best approximation of the filtered image is calculated, thus transferring details from the guide image to the filtered image. The use of joint filtering is popular in the field of computational photography, and has been used for several applications~\cite{petschnigg2004digital,kopf2007joint,ferstl2013image}. One of the building blocks of the filtering method we propose in this paper is the anisotropic diffusion, proposed by Perona and Malik~\cite{perona1990scale}, an edge preserving filtering algorithm in which the filtering of an image is modelled as a heat equation with a different diffusion coefficient at each edge between neighbouring pixels depending on the local geometry and contrast. However, to the best of our knowledge, this algorithm has not yet been used for guided filtering.

\section{Method}\label{sec:method}

The main contributions of this paper are: 1) the guided anisotropic diffusion algorithm, which uses information from the input images to filter and improve semantic segmentation results (section~\ref{sec:gad}), 2) an iterative training scheme that aims to efficiently learn from inaccurate and unreliable ground truth semantic segmentation data (section~\ref{sec:its}), and 3) a learned spatial attention layer that improves classification and weakly supervised semantic segmentation for datasets whose images have been cropped using the geographical coordinates of objects of interest. (section~\ref{sec:attn-layer}).  These contributions are described in detail below. While these ideas are presented in this paper in the context of change detection, the proposed methods' scope is broader and they could be used for other semantic segmentation problems, as we show in Sections \ref{sec:ee} and \ref{sec:exp-ee}.

\subsection{Guided Anisotropic Diffusion}
\label{sec:gad}

In their seminal paper, Perona and Malik proposed an anisotropic diffusion algorithm with the aim of performing scale space image analysis and edge preserving filtering~\cite{perona1990scale}. Their diffusion scheme has the ability to blur the inside of homogeneous regions while preserving or even enhancing edges. This is done by modelling the filtering as a diffusion equation with spatially variable coefficients. The corresponding equation is an extension of the linear heat equation, whose solution is mathematically equivalent to Gaussian filtering when diffusion coefficients are constant~\cite{koenderink1984structure}. Diffusion coefficients are set to be higher where the local contrast of the image is lower.  

More precisely, we consider the anisotropic diffusion equation
\begin{equation}
    \frac{\partial I}{\partial t} = div(c(x,y,t)\nabla I) = c(x,y,t)\Delta I + \nabla c \cdot \nabla I
\end{equation}
where $I$ is the input image, $c(x,y,t)$ is the coefficient diffusion at position $(x,y)$ and time $t$, $div$ represents the divergence, $\nabla$ represents the gradient, and $\Delta$ represents the Laplacian. In its original formulation, $c(x,y,t)$ is a function of the input image I. To perform edge preserving filtering, one approach is using the coefficient
\begin{equation}
    c(x,y,t) = \frac{1}{1 + \Big(\frac{||\nabla I(x,y,t)||}{K}\Big)^{2}},
\end{equation}
which approaches $1$ (strong diffusion) where the gradient is small, and approaches 0 (weak diffusion) for large gradient values. Other functions with these properties and bound in $[0,1]$ may also be used. The parameter $K$ controls the sensitivity to contrast in the image.

\begin{figure}[t]
     
    \begin{minipage}{0.25\linewidth}
        \centering
        \includegraphics[width=\linewidth]{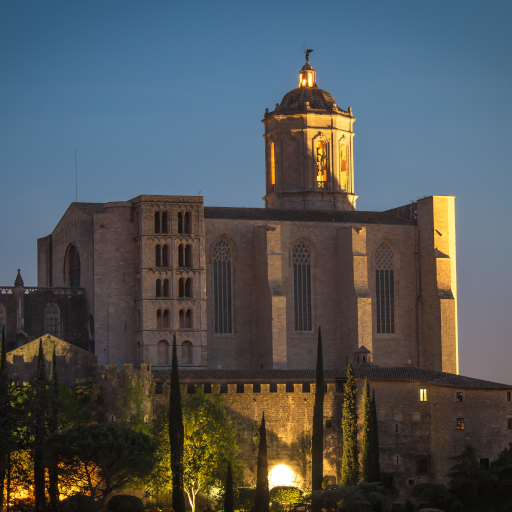}\\
        (a) Guide image
    \end{minipage}
    \hfill
    \begin{minipage}{0.25\linewidth}
        \centering
        \includegraphics[width=\linewidth]{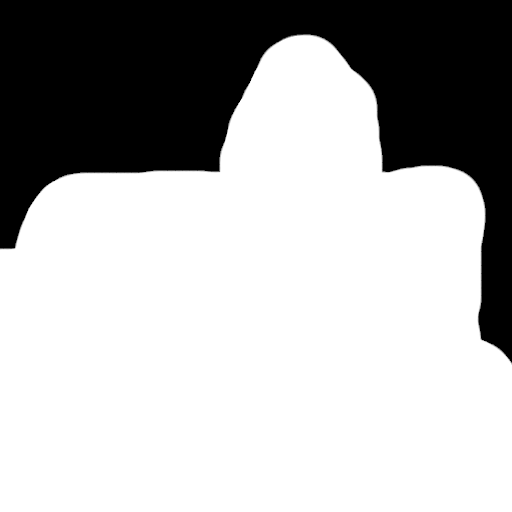}\\
        (b) Input image
    \end{minipage}
    \hfill
    \begin{minipage}{0.25\linewidth}
        \centering
        \includegraphics[width=\linewidth]{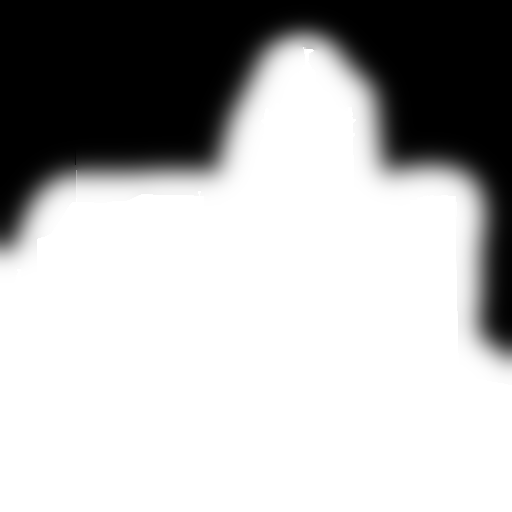}\\
        (c) 300 it. \newline
    \end{minipage}\break
    
    \begin{minipage}{0.25\linewidth}
        \centering
        \includegraphics[width=\linewidth]{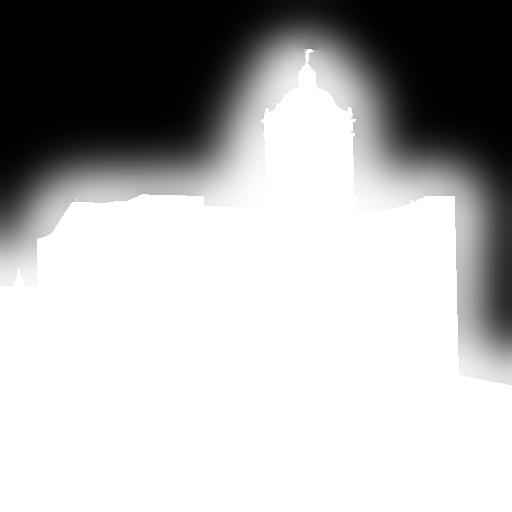}\\
        (d) 1000 it. \newline
    \end{minipage}
    \hfill
    \begin{minipage}{0.25\linewidth}
        \centering
        \includegraphics[width=\linewidth]{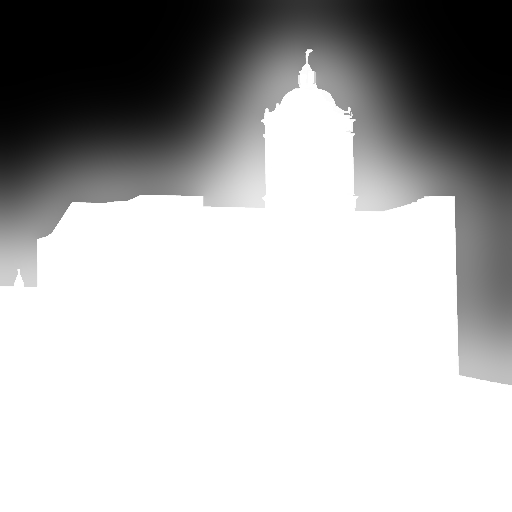}\\
        (e) 3000 it. \newline
    \end{minipage}
    \hfill
    \begin{minipage}{0.25\linewidth}
        \centering
        \includegraphics[width=\linewidth]{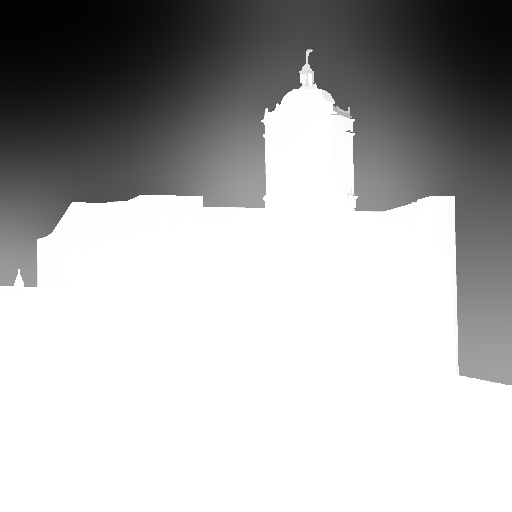}\\
        (f) 10000 it. \newline
    \end{minipage}\break
    
    \caption{Results of guided anisotropic diffusion. Edges in the guide image (a) are preserved in the filtered image (b). (c)-(f) show results using different numbers of iterations.}
    \label{fig:gad_triangle}
\end{figure}

In the \textit{guided} anisotropic diffusion (GAD) algorithm, the aim is to perform edge preserving filtering on an input image, but instead of preserving the edges in the filtered image we preserve edges coming from a separate guide image (or images). Doing so allows us to transfer properties from the guide image $I_g$ onto $I_{input}$, producing the filtered image $I_f$. An illustrative example is shown in Fig.~\ref{fig:gad_triangle}, where the image of a cathedral (a) is used as a guide to filter the image of a rough segmentation (b). The edges from the guide image $I_g$ are used to calculate $c(x,y,t)$, which in practice creates barriers in the diffusion of the filtered image $I_f$, effectively transferring details from $I_g$ to $I_f$. These edges effectively separate the image in two regions, inside and outside the region of interest, and the pixel values in each of these regions experience diffusion, but there is virtually no diffusion happening between them.

\begin{algorithm}[t]
    \caption{Guided Anisotropic Diffusion pseudocode.}
    \label{alg:gad}
    \begin{algorithmic}[1]
        \STATE \textbf{Input:}{$I_{g,1}$, $I_{g,2}$, $I_{input}$, $N$, $K$, $\lambda$}
        \STATE \textbf{Output:}{$I_f$}
        \STATE $I_f \gets I_{input}$
        \FOR {($i\gets1$; $i \leq N$; $i++$)}
            \FOR {($I_j = \{I_1, I_2\}$)}
                \STATE $\nabla I_j \gets$ Calculate gradient of $I_j$
                \STATE $c_{I_j} \gets$ Calculate diffusion coefficients using Eq.~\ref{eq:c}
                \STATE $I_j \gets I_j + \lambda \cdot \nabla I_j \cdot c_{I_j}$
            \ENDFOR
            \STATE $\nabla I_f \gets$ Calculate gradient of $I_f$
            \STATE $c_f \gets$ Calculate diffusion coefficients using Eq.~\ref{eq:multi_c}
            \STATE $I_f \gets I_f + \lambda \cdot \nabla I_f \cdot c_f$
        \ENDFOR
    \end{algorithmic}
\end{algorithm}

Our primary aim is to use this GAD algorithm to improve semantic segmentation results based on the input images. Weakly supervised learning methods are often used when there is an overestimation or underestimation of the target area: either the whole image is the starting point in classification to segmentation tasks, or the reference region is too large in parcel to region segmentation tasks, or a subset of pixels (points or squiggles) are used for supervision. GAD provides a way to improve these semantic segmentation results by making them more precisely fit the edges present in the input images. A few design choices were made to extend the anisotropic diffusion from gray level images to RGB image pairs. The extension to RGB images was done by taking the mean of the gradient norm at each location
\begin{equation}\label{eq:c}
    c_{I_g}(x,y,t) = \frac{1}{1 + \Big(\sum_{C \in \{R,G,B\}}\frac{||\nabla I_{g,C}(x,y,t)||}{3\cdot K}\Big)^{2}},
\end{equation}
so that edges in any of the color channels would prevent diffusion in the filtered image. To extend this further to be capable of taking multiple guide images simultaneously, which is necessary for the problem of change detection, the minimum diffusion coefficient at each position $(x,y,t)$ was used, once again to ensure that any edge present in any guide image would be transferred to the filtered image:
\begin{equation}\label{eq:multi_c}
    c_{I_{g,1},I_{g,2}}(x,y,t) = min_{i \in \{1,2\}} c_{I_i}(x,y,t).
\end{equation}

Guided anisotropic diffusion aims to improve semantic segmentation predictions by filtering the class probabilities yielded by a fully convolutional network. It is less adequate to correct for large classification mistakes, as opposed to non-local methods such as Dense CRF, but it leads to smoother predictions with more accurate edges. It can also be easily extended for any number of guide images by increasing the number of images considered in Eq.~\ref{eq:multi_c}. The pseudocode for the GAD algorithm can be found in Alg.~\ref{alg:gad}. As mentioned in the original anisotropic diffusion paper, the algorithm is unstable for $\lambda > 0.25$ when using 4-neighborhoods for the calculations. For more information the reader can refer to the mathematical derivations presented in \cite{perona1990scale,aubert2006mathematical}.

GAD parameters are tuned visually by performing anisotropic diffusion on guide images from the dataset. Each parameter offers different trade-offs:
\begin{itemize}
    \item $K$ allows us to choose the magnitude of gradients that should be considered as edges (and therefore diffusion should be restricted at that point).
    \item A larger number of iterations $N$ allows for the diffusion to mix pixel values at longer ranges (the "receptive field" radius is equal to the number if iterations).
    \item If $\lambda$ is set closer to 0 the algorithm approaches the continuous time solution of the equations, at the cost of diffusion speed. This has not been observed to improve results as long as $\lambda$ is kept below the threshold of stability at $\lambda=0.25$.
\end{itemize}

In the following sections, we show two ways to use GAD in approaches to learn to perform semantic segmentation with imperfect labels. First, we address in Section~\ref{sec:its} the inaccurate labelling problem with an iterative data cleansing scheme. Second, in Section~\ref{sec:attn-layer} we explore another use-case and use GAD to learn to segment changes from classification labels only.

\subsection{Iterative Training Scheme}
\label{sec:its}

The first use-case we investigate for deploying GAD tackles the type of  label noise present in parcel-based change detection datasets where pixel labels are generated from vector data . It is challenging due to its spatial structure and correlation between neighbors. In the taxonomy presented in \cite{frenay2014comprehensive,frenay2014classification}, this type of label noise would be classified as "noisy not at random" (NNAR). NNAR is the most complex among the label noise models in the taxonomy. This is the classification applied to noise when the samples that are mislabelled are not randomly dispersed, and the type of noise is also not random. For labels generated from parcel polygons, label noise will be concentrated around region boundaries and the type of noise will be defined by the classes of the imaged objects. In the case of HRSCD, most errors can be attributed to one of the following reasons: the available information is insufficient to perform labelling, errors on the part of the annotators, subjectiveness of the labelling task, and temporal misalignment between the databases used to create the HRSCD dataset.

\begin{figure}[t]
    \centering
    \includegraphics[width=0.5\linewidth]{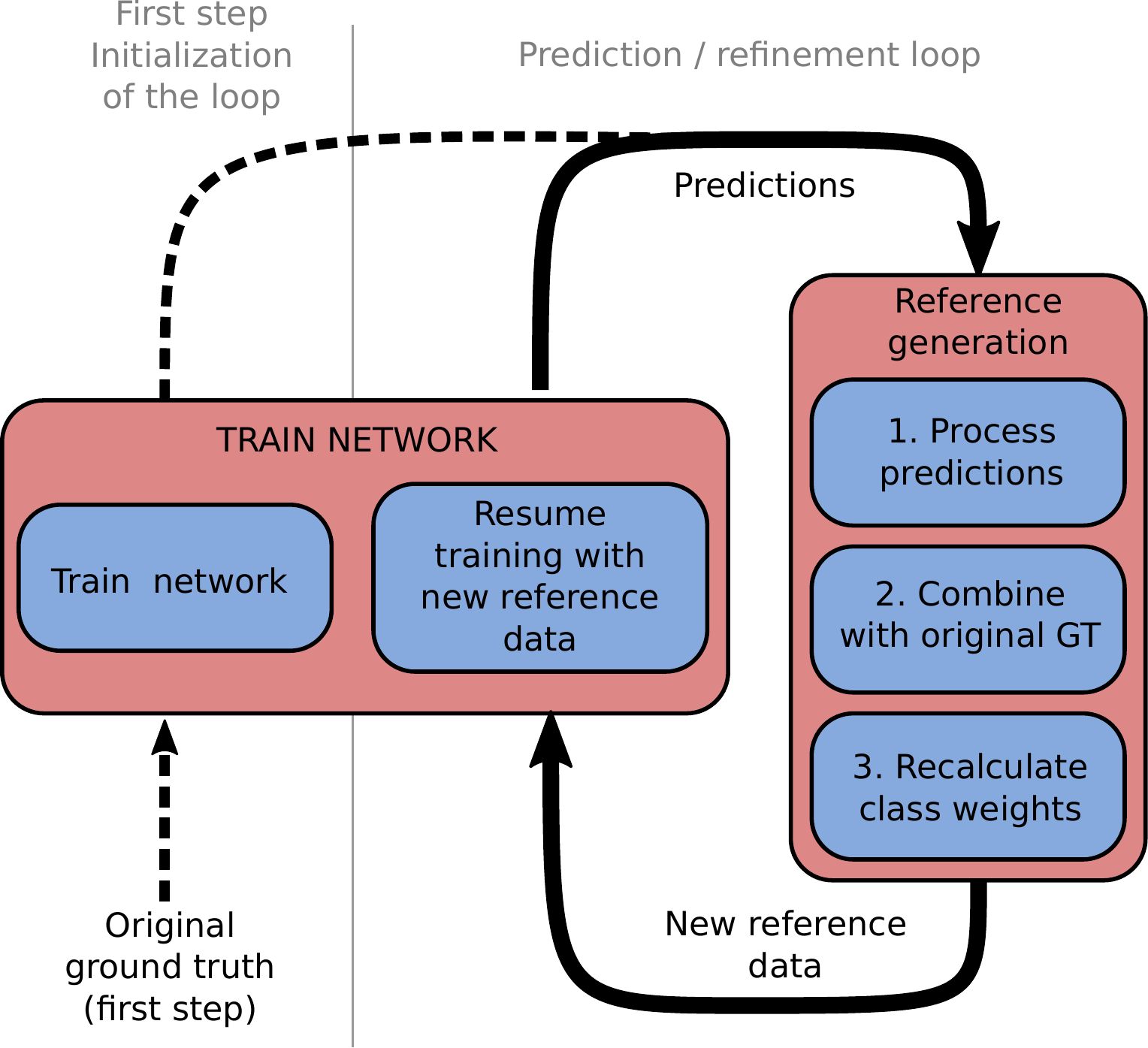}\\
    \caption{Iterative training method: alternating between training and data cleaning allows the network to simultaneously learn the desired task and to remove bad examples from the training dataset.}
    \label{fig:it_schematic}
\end{figure}

It is important to note that, as discussed by Fr{\'e}nay and Kab{\'a}n in \cite{frenay2014comprehensive}, label noise has an even more powerful damaging impact when a dataset is imbalanced since it alters the perceived, but not the real, class imbalance and therefore the methods used to mitigate class imbalance during training are less effective. In the case of change detection with the HRSCD dataset, the no change class outnumbers the change class 130 to 1, which means the label noise could significantly alter the calculated class weights used for training.

It has been noticed in \cite{daudt2018hrscd} and in the experiments presented in this paper that change detection networks trained directly on the HRSCD dataset had the capacity to detect changes in image pairs but tended to predict blobs around the detected change instances, as is depicted in Fig.~\ref{fig:gad_122}(c), likely in an attempt to minimize the loss for the training images where the surrounding pixels of true changes are also marked as having experienced changes. In many cases, it was observed that the network predictions were correct where the ground truth labels were not. Based on this observation, we propose a method for training the network that alternates between actual minimization of a loss function and using the network predictions to clean the reference data before continuing the training. A schematic that illustrates the main ideas of this method is shown in Fig.~\ref{fig:it_schematic}. For the remainder of this paper, the iteration cycles of training the network and cleaning of training data will be referred to as \textit{hyperepochs}.

\begin{algorithm}[t]
    \caption{Iterative training pseudocode.}
    \begin{algorithmic}
        \STATE \textbf{Input:} $\mathcal{I}$: Image pairs, $GT_o$: Original unreliable ground truths, $N$: Number of hyperepochs, $\Phi_r$: Initial random network weights.
        \STATE \textbf{Output:} $\Phi_N$: Trained network weights.
        \STATE $w_0 \gets$ calculate class weights inversely proportional to number of class examples
        \STATE $\Phi_0 \gets$ Train network with $\mathcal{I}$ and $GT_0$ until convergence or fixed number of epochs
        \FOR{($i\gets1$; $i \leq N$; $i++$)}
            \STATE $P_i \gets$ generate predictions for training dataset with current network
            \STATE $P_{i,pp} \gets$ Post-processing of predictions
            \STATE $GT_i \gets$ Combine $P_{i,pp}$ with $GT_0$ to generate cleaner ground truth data
            \STATE $\Phi_i \gets$ Continue training network from $\Phi_{i-1}$ using $\mathcal{I}$ and $GT_i$ until convergence
        \ENDFOR
    \end{algorithmic}
\end{algorithm}

Alternating between training a semantic segmentation network and using it to make changes to the training data has already been explored~\cite{dai2015boxsup,khoreva2017simple}. Such iterative methods are named "classification filtering"~\cite{frenay2014classification}. The main differences between the method proposed in this paper and previous ones are:
\begin{enumerate}
    \item \textbf{No bounding box information is available}: we work directly with pixel level annotations, which were generated form vector data;
    \item \textbf{Each annotated region may contain more than one instance}: the annotations often group several change instances together;
    \item \textbf{Annotations are not flawless}: the HRSCD dataset contains both false positives and false negatives in change annotations.
\end{enumerate}

It has also been shown by Khoreva \etal in \cite{khoreva2017simple} that simply using the outputs of the network as training data leads to degradation of the results, and that it is necessary to use priors and heuristics specific to the problem at hand to prevent a degradation in performance. In this paper we use two ways to avoid degradation of the results with iterative training. The first is using processing techniques that bring information from the input images into the predicted semantic segmentations, improving the results and providing a stronger correlation between inputs and predictions. The GAD algorithm presented in Section~\ref{sec:gad} serves this purpose, but other algorithms such as Dense CRF~\cite{krahenbuhl2011efficient} may also be used. The second way the degradation of results is avoided is by combining network predictions with the original reference data at each iteration, instead of simply using predictions as reference data.

\begin{figure}
    \centering
    \begin{minipage}{0.3\linewidth}
        \centering
        \hspace{20pt}Original GT
        
        \rotatebox[origin=c]{90}{Pred. \hspace{8pt}}
        \begin{tabular}{c|c|c}
              & \textbf{0} & \textbf{1} \\ \hline
            \textbf{0} & 0 & 0 \\ \hline
            \textbf{1} & 0 & 1 \\
        \end{tabular}
        
        (a)~Intersection
    \end{minipage}
    \begin{minipage}{0.3\linewidth}
        \centering
        \hspace{20pt}Original GT
        
        \rotatebox[origin=c]{90}{Pred. \hspace{8pt}}
        \begin{tabular}{c|c|c}
              & \textbf{0} & \textbf{1} \\ \hline
            \textbf{0} & 0 & 2 \\ \hline
            \textbf{1} & 0 & 1 \\
        \end{tabular}
        
        (b)~FN$\gets$ Ignore
    \end{minipage}
    \begin{minipage}{0.36\linewidth}
        \centering
        \hspace{20pt}Original GT
        
        \rotatebox[origin=c]{90}{Pred. \hspace{8pt}}
        \begin{tabular}{c|c|c}
              & \textbf{0} & \textbf{1} \\ \hline
            \textbf{0} & 0 & 2 \\ \hline
            \textbf{1} & 2 & 1 \\
        \end{tabular}
        
        (c)~FN$\cup$FP$\gets$~Ignore
    \end{minipage}
    \vspace{6pt}
    \caption{Proposed methods for merging original labels and network predictions. Classes: 0 is no change, 1 is change, 2 is ignore. (a) Intersection between original and detected changes. (b) Ignore false negatives from the perspective of original labels. (c) Ignore all pixels with label disagreements.}
    \label{ignore_strats}
\end{figure}

\begin{figure}[t]
    \hfill
    \begin{minipage}{0.25\linewidth}
        \centering
        \includegraphics[width=\linewidth]{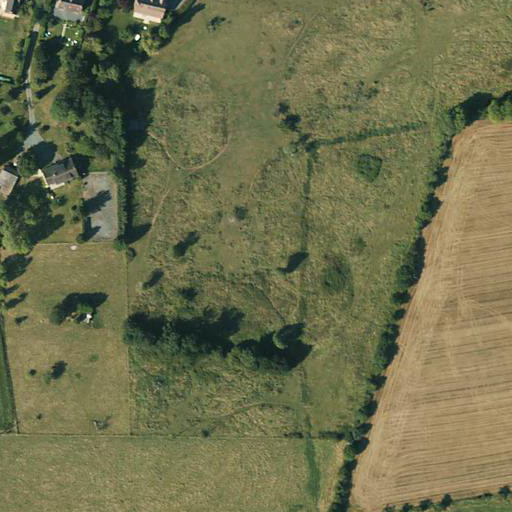}\\
        (a) Image 1
    \end{minipage}
    \hfill
    \begin{minipage}{0.25\linewidth}
        \centering
        \includegraphics[width=\linewidth]{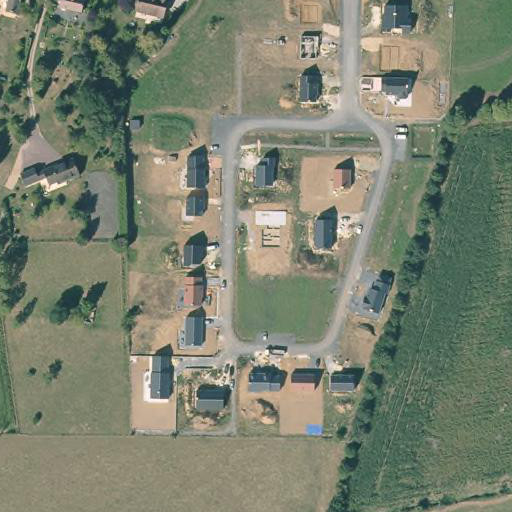}\\
        (b) Image 2
    \end{minipage}
    \hfill
    \begin{minipage}{0.25\linewidth}
        \centering
        \includegraphics[width=\linewidth]{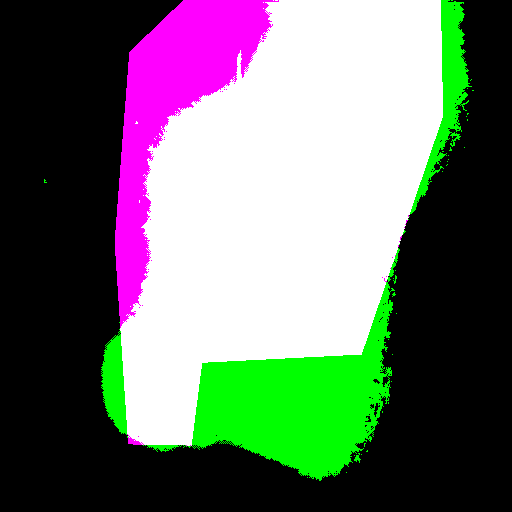}\\
        (c) GT and prediction
    \end{minipage}\hfill\break
    
    \hfill
    \begin{minipage}{0.25\linewidth}
        \centering
        \includegraphics[width=\linewidth]{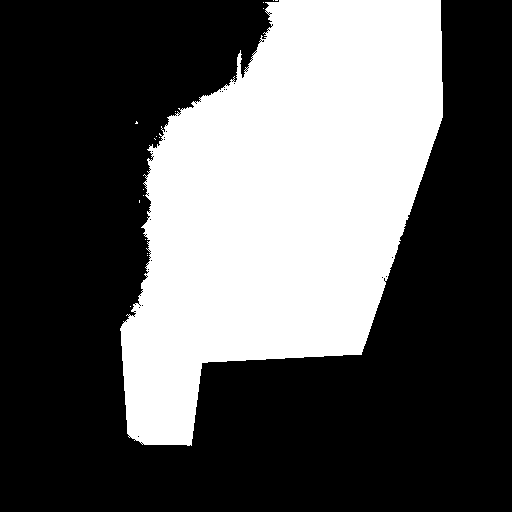}\\
        (d)~Intersection
    \end{minipage}
    \hfill
    \begin{minipage}{0.25\linewidth}
        \centering
        \includegraphics[width=\linewidth]{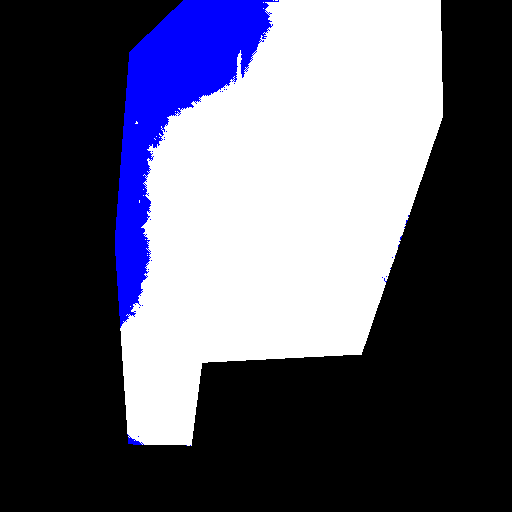}\\
        (e)~FN$\gets$ Ignore
    \end{minipage}
    \hfill
    \begin{minipage}{0.25\linewidth}
        \centering
        \includegraphics[width=\linewidth]{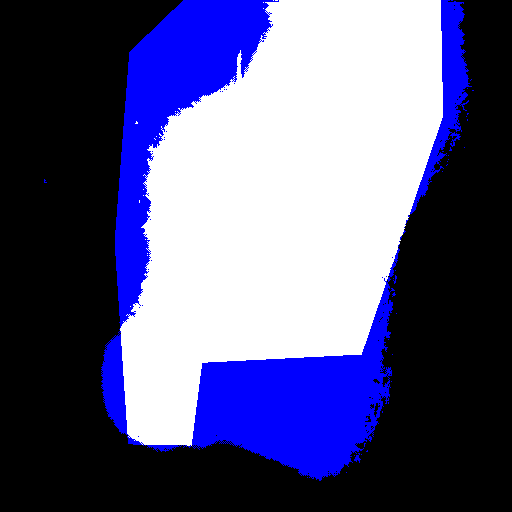}\\
        (f)~FN$\cup$FP$\gets$~Ign.
    \end{minipage}\hfill\break
    
    \caption{Example case of the three proposed merge strategies. In (c), black is true negative, white is true positive, magenta is false negative, and green is false positive. In (d)-(f) blue represents the ignore class.}
    \label{fig:merge_example}
\end{figure}

We propose three ways of merging the original labels with network predictions. When merging, each pixel will have a binary label from the original ground truth and a binary label from the network prediction. If these labels agree, there is no reason to believe the label for that pixel is wrong, and it is therefore kept unchanged. In case the labels disagree, the following options to decide the pixel's label are proposed:
\begin{enumerate}
    \item \textbf{The intersection of predicted and reference change labels is kept as change}: this strategy assumes all changes are marked in both the reference data and in the prediction. It also puts pixels with uncertain labels in the no change class, where they are more easily diluted during training due to the class imbalance.
    \item \textbf{Ignore false negatives}: using an ignore class for false negatives attempts to keep only good examples in the change class, improving the quality of the training data. It assumes all changes are marked in the original labels provided.
    \item \textbf{Ignore all disagreements}: marking all label disagreements to be ignored during training attempts to keep only clean labels for training at the cost of reducing the number of training examples. This approach is the only one that is class agnostic.
\end{enumerate}
In practice, the ignored pixels are marked as a different class that is given a class weight of 0 during the training. Tables for the three proposed methods can be found in Fig.~\ref{ignore_strats}, and an example can be found in Fig.~\ref{fig:merge_example}.

\subsection{Scene-Invariant Spatial Attention Layer}
\label{sec:attn-layer}

Our second change detection use-case for GAD-enhanced weak supervision addresses the more challenging task of inferring segmentation masks from image-level classification labels. Many datasets in remote sensing contain georeferenced data, such as patches cropped from large images using the coordinates of known objects for which a label is known. In such cases, objects to which the labels refer are located in the center of the images, while the characteristics of their surroundings are not directly related to the available labels. Pooling techniques such as max pooling and average pooling that are very often used in CNNs are invariant with respect to the image position. These operations fail to make use of the heuristic described above, and do not learn to prioritize some areas of the image over others when making classification predictions.

To increase the localization capability through the global average pooling operation, we propose here a learned spatial attention layer that can be used to allow the network to learn which positions of the images are more discriminative and should be prioritized over others when making inferences. We also propose to use the GAD algorithm to further focus the attention of the network on the most relevant features. Let's assume that a feature map $x$ of size $C \times M \times N$ is obtained after any number of convolution, pooling and other operations from an input image (or images in the case of change detection), where $C$ is the number of channels and $M$ and $N$ are spatial dimensions. We define a matrix $A$ of size $M \times N$ which will be learned by the network, and will serve as attention weights given to spatial locations. The attention operation $f(x)$ can then be defined as
\begin{equation}\label{eq:attention}
    f(x)_{c, i, j} = \alpha \cdot x_{c, i, j} \cdot \sigma(a_{i, j}) ,
\end{equation}
where $a_{i, j}$ denotes the element of $A$ in position $(i, j)$, $\sigma$ denotes the sigmoid function and $\alpha$ is a normalization term defined as
\begin{equation}\label{eq:attention_norm}
    \alpha = \sum_{i=1}^M \sum_{j=1}^N \sigma(a_{i, j}).
\end{equation}
The sigmoid function is used to ensure the attention weights given to each spatial location is in the range $(0,1)$. The matrix $A$ is initialized as a null matrix so that all spatial locations have equal attention values of $\sigma(a_{i, j}) = 0.5$. Random initialization of $A$ is neither necessary nor recommended.

The proposed attention layer is designed to be used after a softmax operation and before a global average pooling (GAP) layer. Doing so will force the network to produce per-pixel classification predictions, which are then put through a weighted average operation whose weights are learnable parameters which depend only on the spatial position of each feature. Global average pooling is preferable to max pooling at the end of a network when we want the network to be able to localize objects, as was discussed in \cite{Zhou_2016_CVPR}. Note that the number of learnable parameters introduced by this attention layer is only $M \cdot N$, which is extremely small in the context of deep neural networks. At inference time, the learned attention weights can be further adapted to the input images by using the GAD algorithm proposed in Section~\ref{sec:gad}. This helps the network to focus its attention at the building at the center of the image pair, further increasing classification performance.

\begin{figure}[t]
    \centering
    \includegraphics[width=0.9\linewidth]{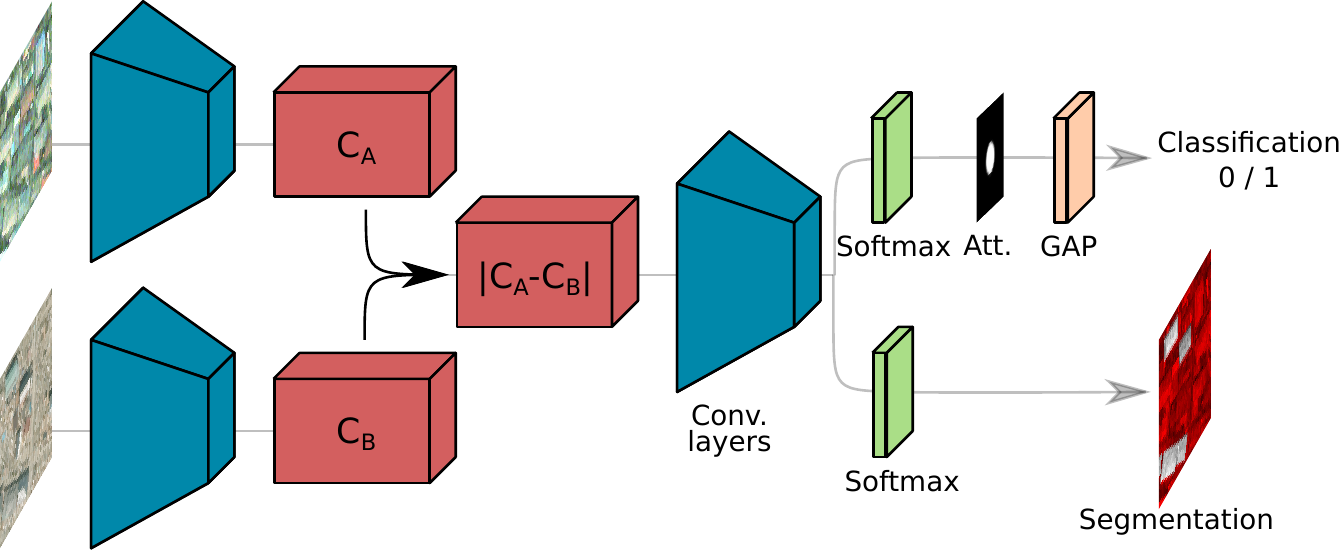}\\
    \caption{Basic schematic of the network used for weakly supervised change detection. Two paths can be taken: the classification path uses the proposed attention layer and global average pooling to produce a classification of the image, while the segmentation path avoids these steps to output pixel-level predictions. Supervision is only available on the classification path.}
    \label{fig:att_schematic}
\end{figure}

In this paper, we incorporate this attention layer into the classification branch of the architecture depicted in Fig.~\ref{fig:att_schematic}. The images are processed by two convolutions with stride $\frac{1}{2}$ and 5 residual blocks before their features are merged, and 4 residual blocks after. This architecture allows us to perform either classification or segmentation by choosing either of the paths at the end. This architecture is a straightforward Siamese extension of the ideas presented in \cite{Zhou_2016_CVPR}. Supervision is only available for the classification branch, but the structure of the network allows us to apply equivalent classification operations at each spatial locations by avoiding the attention and global average pooling layers, effectively performing semantic segmentation.

\subsection{Edge Enhancement Using Guided Anisotropic Diffusion for Segmentation Upsampling \label{sec:ee}}

\begin{figure}[t]
    \centering
    \includegraphics[width=0.9\linewidth]{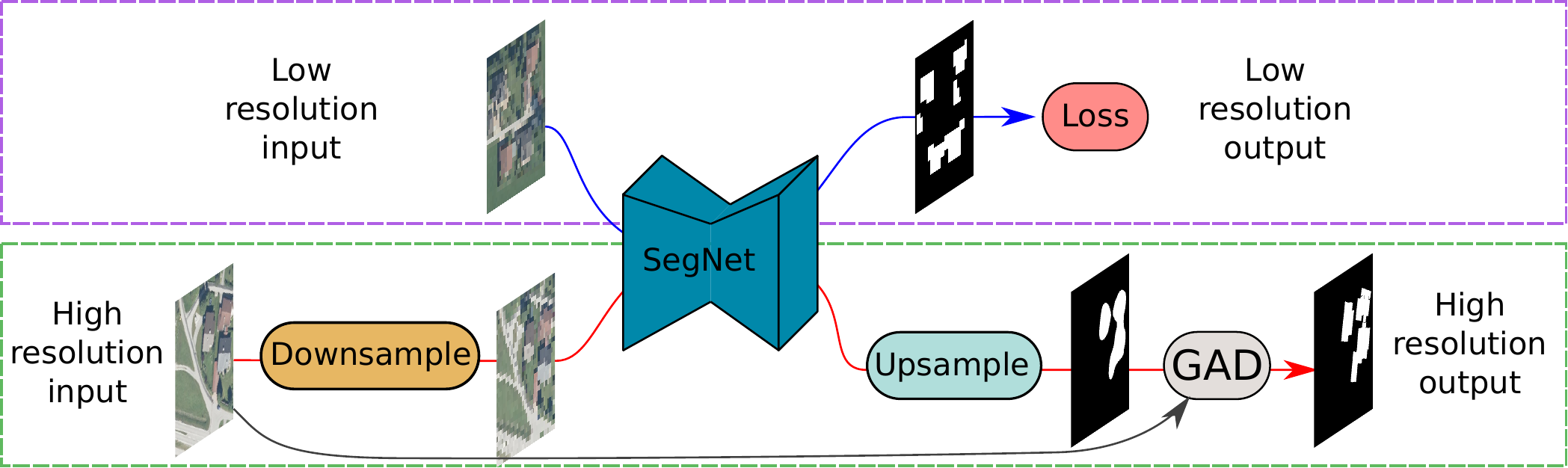}\\
    \caption{Schematic for the experiments performed to evaluate if GAD is able to compensate for supervision using images at a lower resolution and improve prediction accuracy around region boundaries.}
    \label{fig:ss_main_idea}
\end{figure}

We further propose an experiment to evaluate the efficacy of the proposed guided anisotropic diffusion algorithm in a more general semantic segmentation setting. In this case, we attempt to compensate for when the network is supervised with images with a lower ground sample distance (GSD) when it is applied to images with a higher GSD. In this case, a network that is supervised with images at a lower resolution struggles to predict region boundaries at higher resolutions.

To apply a CNN trained with images with a smaller GSD to images with higher GSDs, it is first necessary to downsample the higher resolution images. This is necessary because the network has learned to detect objects at a given scale, and changing the scale of the input images would pose additional problems due to the shift in dataset statistics. The downsampled image is then segmented using the fully convolutional network, which produces softmax activations, i.e. class probabilities, for each pixel. To bring these low resolution predictions back to the higher GSD, these class probabilities are upsampled back to the original resolution. This procedure leads to semantic region boundaries not being very accurate. In our experiments, we use the proposed guided anisotropic diffusion algorithm to recover boundary information using the high resolution as a guide for where the semantic boundaries should be located. A diagram that illustrates this procedure can be found in Fig.~\ref{fig:ss_main_idea}.

These experiments illustrate how GAD could be used to help networks adapt to new applications with different data. Resolution differences is very common in remote sensing due to images coming from different satellite or aerial sources, and thus coping with such variations is important.

\section{Experiments}\label{sec:experiments}

The experiments presented in this paper have been divided into three sections. The first one, in Section~\ref{sec:exp-its}, explores how the integration of GAD in the iterative training scheme of section~\ref{sec:its} allows us to refine approximate labels to obtain pixel-level change detection more accurately than through direct supervision. The second one, in Section~\ref{sec:exp-attn-layer}, shows the effectiveness of GAD combined with a spatial attention layer (Section~\ref{sec:attn-layer}) in performing weakly supervised change detection using only image-level labels to perform pixel-level predictions. Finally, we evaluate in Section~\ref{sec:exp-ee} the usage of GAD for adapting to multiple spatial resolutions for building and generic segmentation tasks, as defined in Section~\ref{sec:ee}.

\subsection{Label Refinement Through Iterative Learning \label{sec:exp-its}}

To validate the iterative training scheme proposed in Section~\ref{sec:its} we adopted the hybrid change detection and land cover mapping fully convolutional network presented in~\cite{daudt2018hrscd}, since it was already proven to work with the HRSCD dataset. We adopted \textit{strategy 4.2} described in~\cite{daudt2018hrscd}, in which the land cover mapping branches of the network are trained before the change detection one to avoid setting a balancing hyperparameter. The land cover mapping branches of the network were fixed to have the same parameter weights for all tests presented in this paper, and evaluating those results is not done here as the scope of this paper is restricted to the problem of change detection.

\begin{figure}[t]
     \hfill
    \begin{minipage}{0.155\linewidth}
        \centering
        \includegraphics[width=\linewidth]{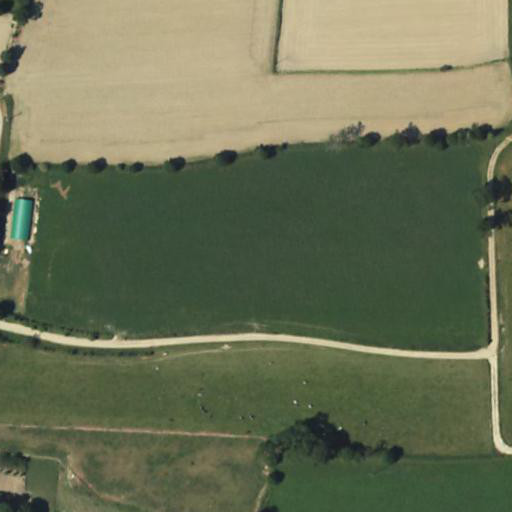}\\
    \end{minipage}
    \hfill
    \begin{minipage}{0.155\linewidth}
        \centering
        \includegraphics[width=\linewidth]{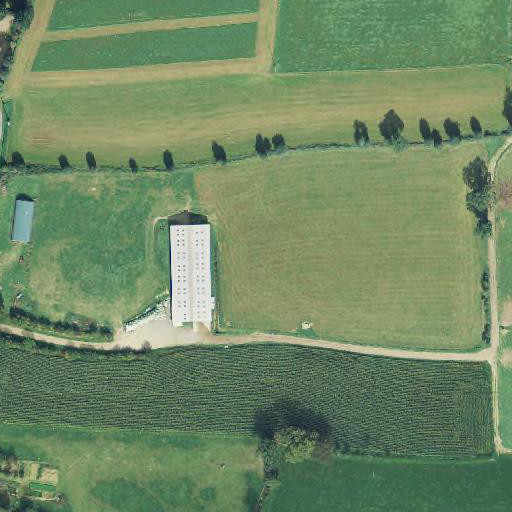}\\
    \end{minipage}
    \hfill
    \begin{minipage}{0.155\linewidth}
        \centering
        \includegraphics[width=\linewidth]{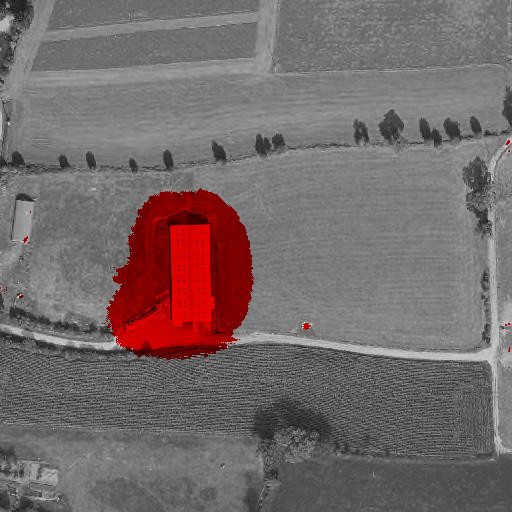}\\
    \end{minipage}
    \hfill
    \begin{minipage}{0.155\linewidth}
        \centering
        \includegraphics[width=\linewidth]{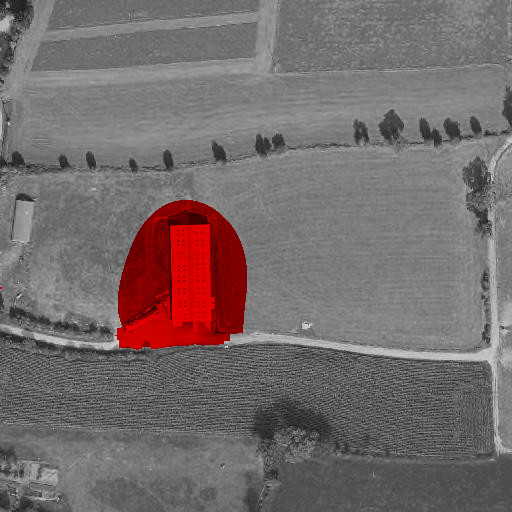}\\
    \end{minipage}
    \hfill
    \begin{minipage}{0.155\linewidth}
        \centering
        \includegraphics[width=\linewidth]{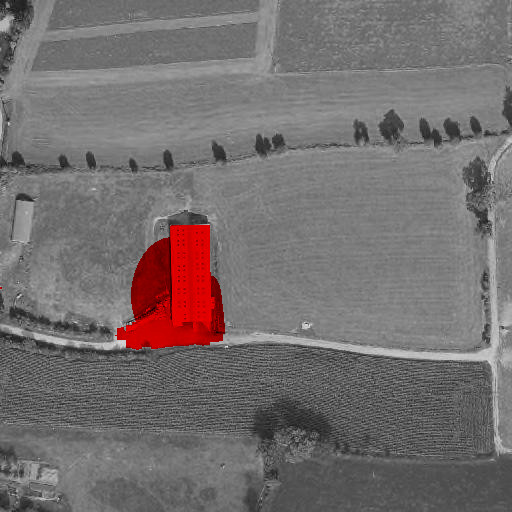}\\
    \end{minipage}
    \hfill
    \begin{minipage}{0.155\linewidth}
        \centering
        \includegraphics[width=\linewidth]{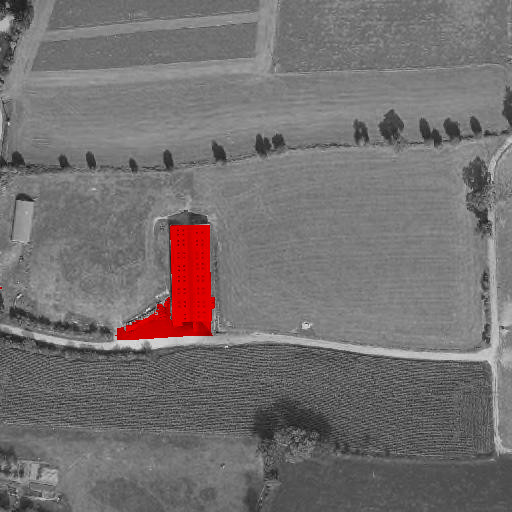}\\
    \end{minipage}\hfill\break
    
     \hfill
    \begin{minipage}{0.155\linewidth}
        \centering
        (a) Image 1 \newline
    \end{minipage}
    \hfill
    \begin{minipage}{0.155\linewidth}
        \centering
        (b) Image 2 \newline
    \end{minipage}
    \hfill
    \begin{minipage}{0.155\linewidth}
        \centering
        (c) Naive prediction
    \end{minipage}
    \hfill
    \begin{minipage}{0.155\linewidth}
        \centering
        (d) 2000 it. \newline
    \end{minipage}
    \hfill
    \begin{minipage}{0.155\linewidth}
        \centering
        (e) 5000 it. \newline
    \end{minipage}
    \hfill
    \begin{minipage}{0.155\linewidth}
        \centering
        (f) 20000 it. \newline
    \end{minipage}\hfill\break
    
    \caption{Guided anisotropic diffusion for filtering a real example of semantic segmentation. The diffusion allows edges from the guide images to be transferred to the target image, improving the results (change area in red).}
    \label{fig:gad_122}
\end{figure}

\begin{figure}[t]
     \hfill
    \begin{minipage}{0.155\linewidth}
        \centering
        \includegraphics[width=\linewidth]{ad_ex_122/01-I1.jpg}\\
    \end{minipage}
    \hfill
    \begin{minipage}{0.155\linewidth}
        \centering
        \includegraphics[width=\linewidth]{ad_ex_122/02-I2.jpg}\\
    \end{minipage}
    \hfill
    \begin{minipage}{0.155\linewidth}
        \centering
        \includegraphics[width=\linewidth]{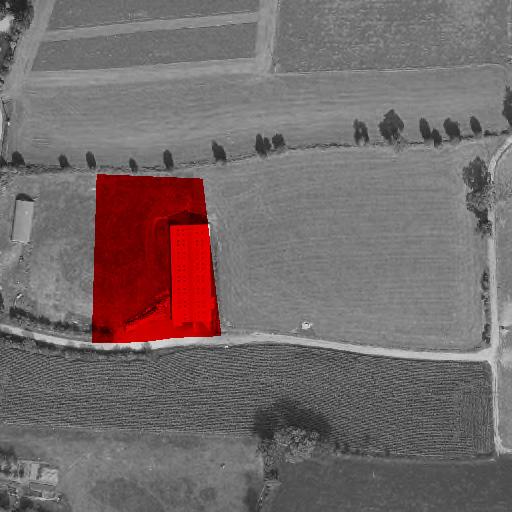}\\
    \end{minipage}
    \hfill
    \begin{minipage}{0.155\linewidth}
        \centering
        \includegraphics[width=\linewidth]{ad_ex_122/03-pred-00000.jpg}\\
    \end{minipage}
    \hfill
    \begin{minipage}{0.155\linewidth}
        \centering
        \includegraphics[width=\linewidth]{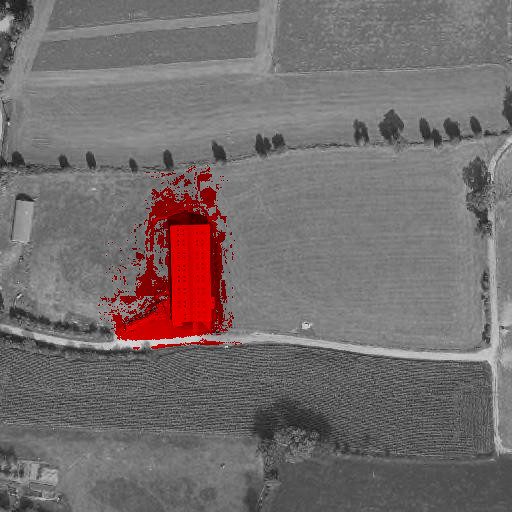}\\
    \end{minipage}
    \hfill
    \begin{minipage}{0.155\linewidth}
        \centering
        \includegraphics[width=\linewidth]{ad_ex_122/06-pred-20000.jpg}\\
    \end{minipage}
    \hfill
    \break
    
     \hfill
    \begin{minipage}{0.155\linewidth}
        \centering
        (a) Image 1 \newline
    \end{minipage}
    \hfill
    \begin{minipage}{0.155\linewidth}
        \centering
        (b) Image 2 \newline
    \end{minipage}
    \hfill
    \begin{minipage}{0.155\linewidth}
        \centering
        (c) Reference data
    \end{minipage}
    \hfill
    \begin{minipage}{0.155\linewidth}
        \centering
        (d) Naive prediction
    \end{minipage}
    \hfill
    \begin{minipage}{0.155\linewidth}
        \centering
        (e) Dense CRF
    \end{minipage}
    \hfill
    \begin{minipage}{0.155\linewidth}
        \centering
        (f) GAD \newline
    \end{minipage}
    \hfill
    \break
    
    \caption{Comparison between (c) original dataset ground truth, (e) prediction filtered by Dense CRF, and (f) prediction filtered with guided anisotropic diffusion for 20000 iterations. (changes in red).}
    \label{fig:gad_vs_dcrf}
\end{figure}

\begin{figure}[t]
    \centering
    \begin{minipage}{0.31\linewidth}
        \centering
        \includegraphics[width=\linewidth]{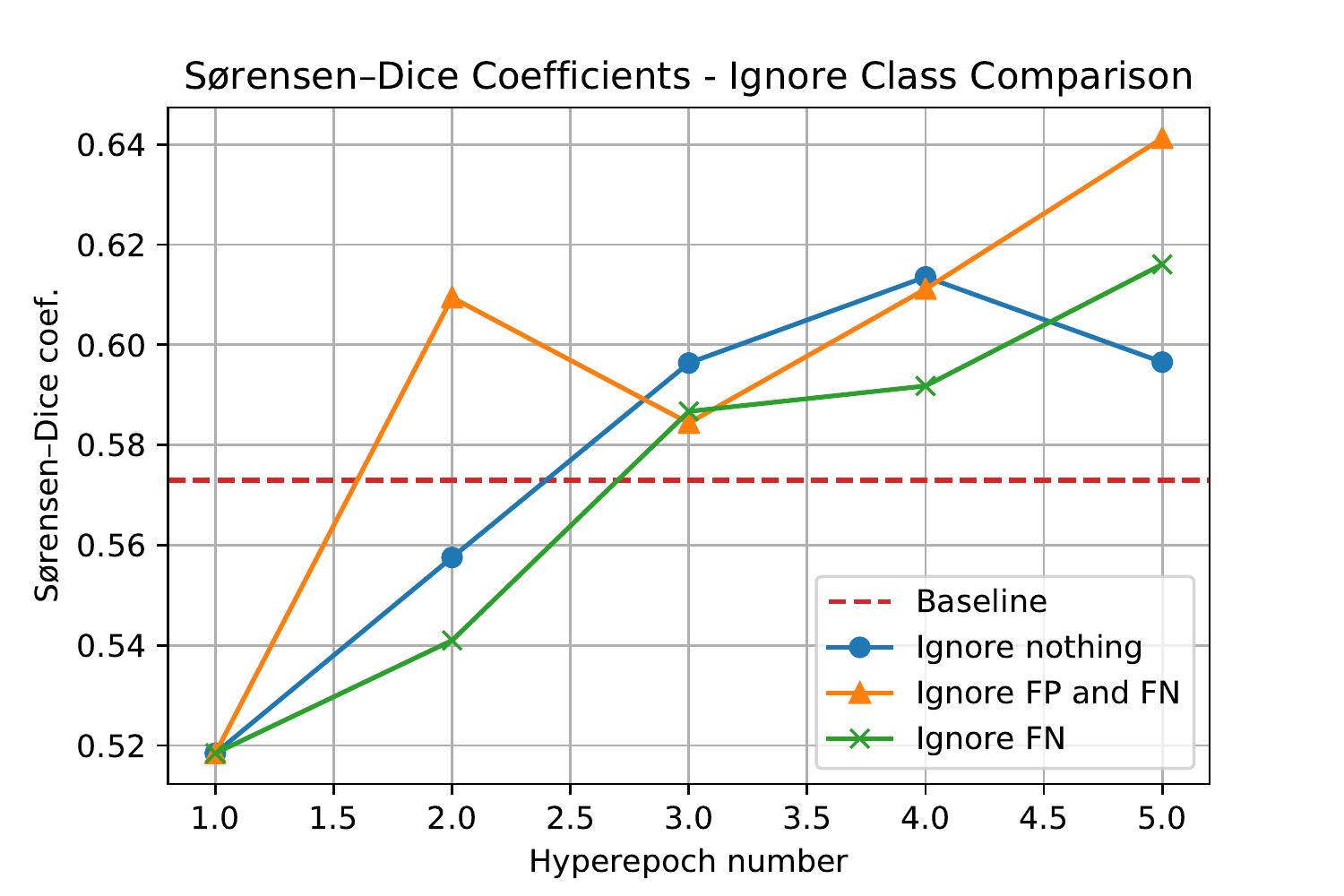}\\
    \end{minipage}
    ~
    \begin{minipage}{0.31\linewidth}
        \centering
        \includegraphics[width=\linewidth]{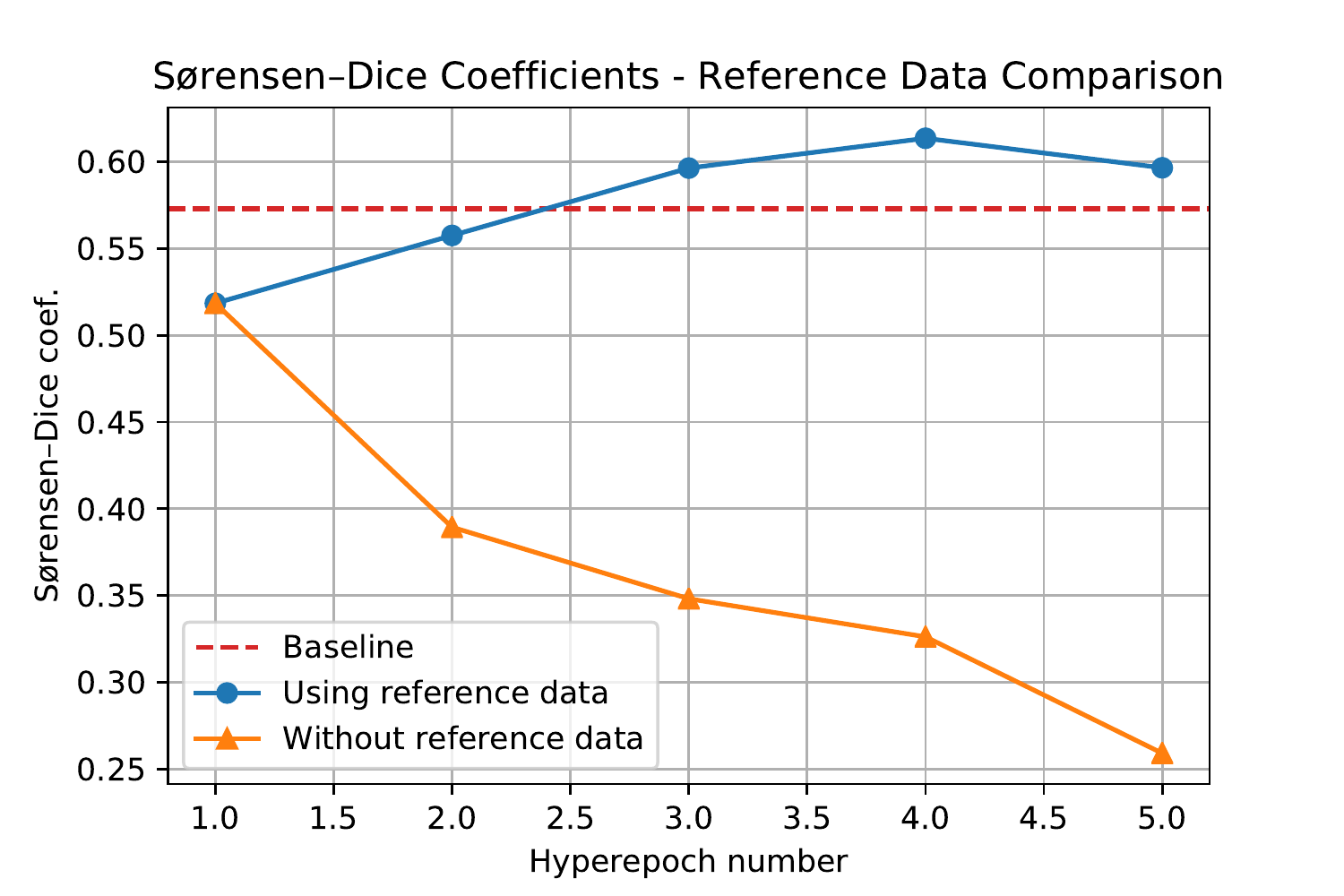}\\
    \end{minipage}
    ~
    \begin{minipage}{0.31\linewidth}
        \centering
        \includegraphics[width=\linewidth]{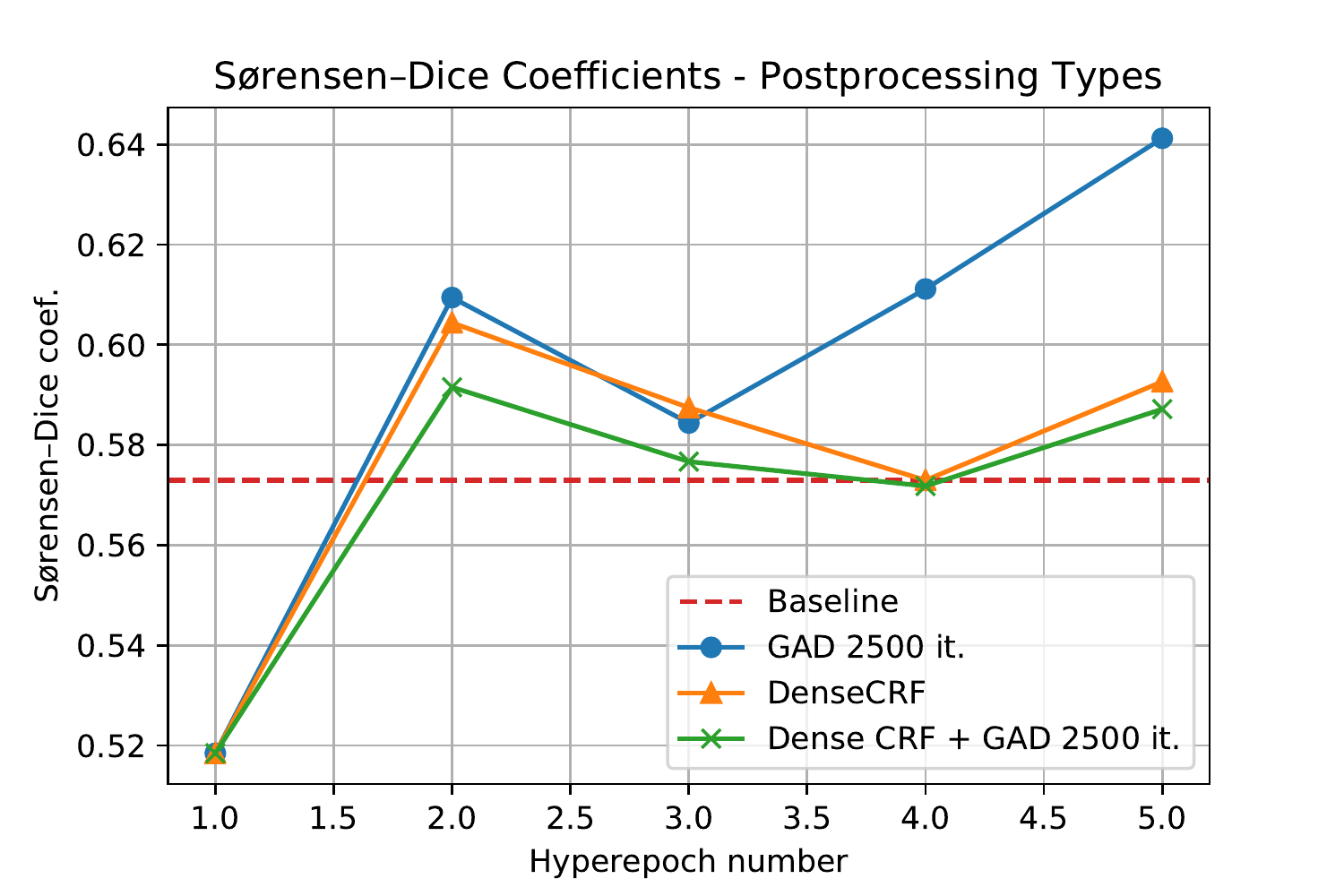}\\
    \end{minipage}
    
    \begin{minipage}{0.31\linewidth}
        \centering
        \includegraphics[width=\linewidth]{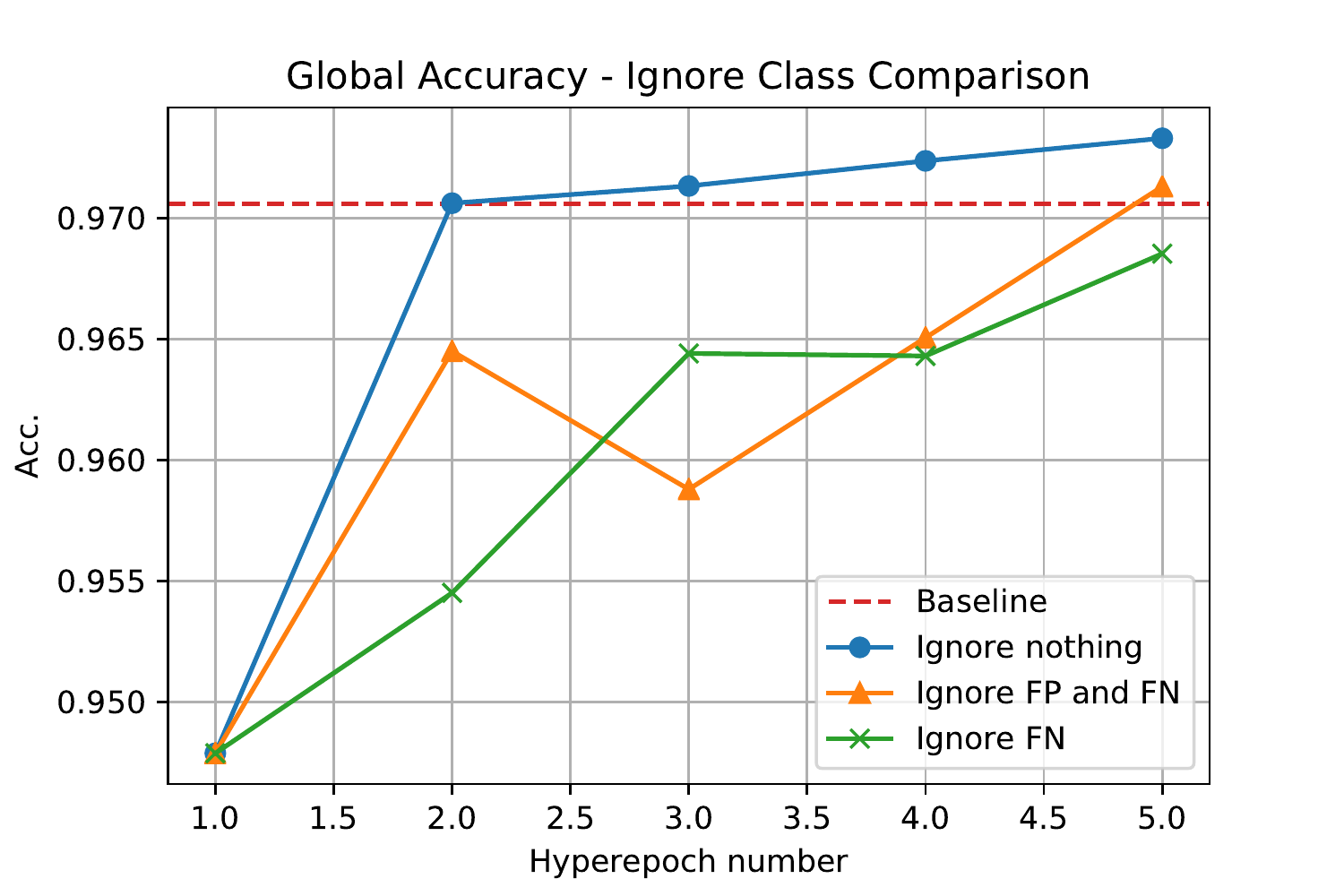}\\
        (a)
    \end{minipage}
    ~
    \begin{minipage}{0.31\linewidth}
        \centering
        \includegraphics[width=\linewidth]{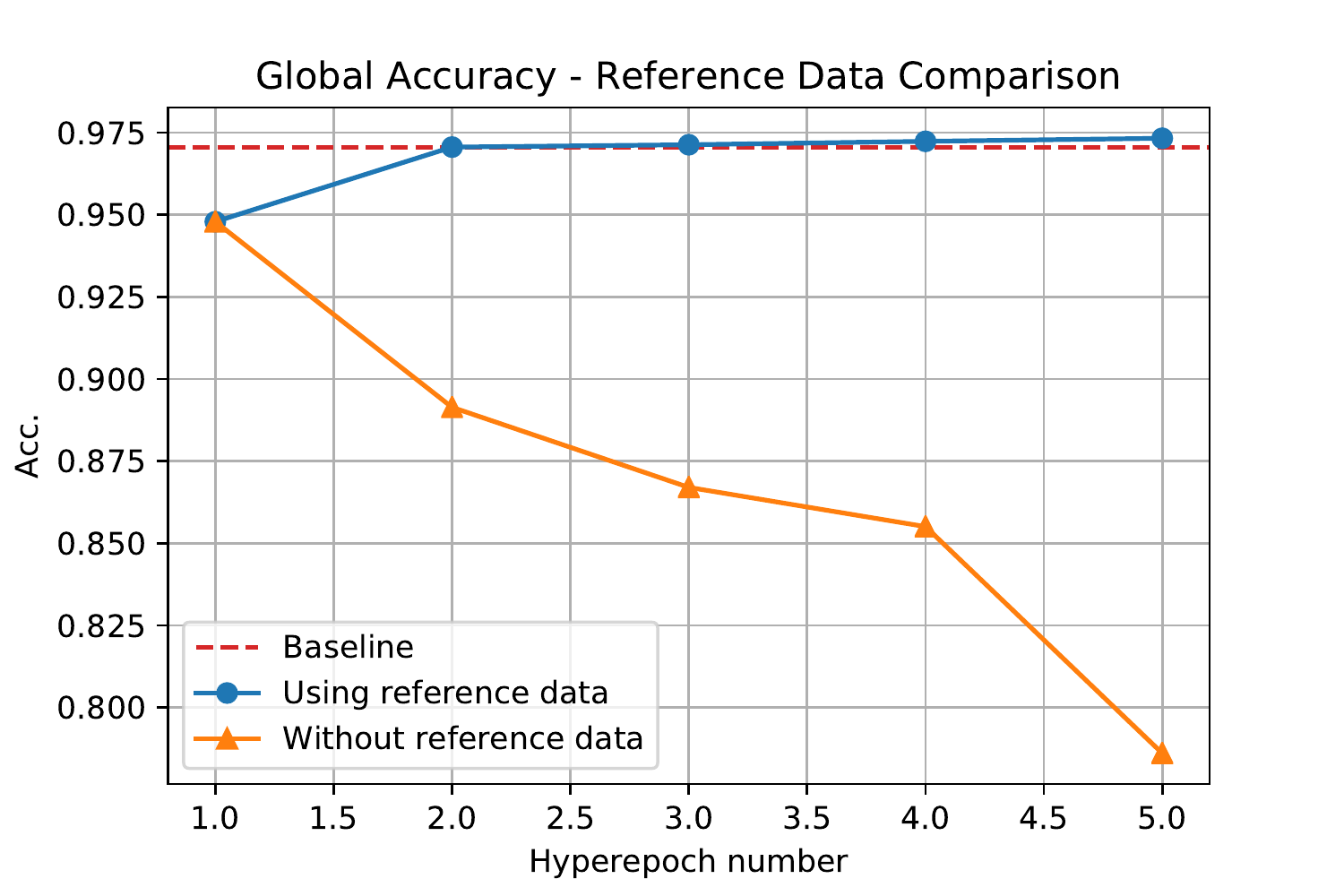}\\
        (b)
    \end{minipage}
    ~
    \begin{minipage}{0.31\linewidth}
        \centering
        \includegraphics[width=\linewidth]{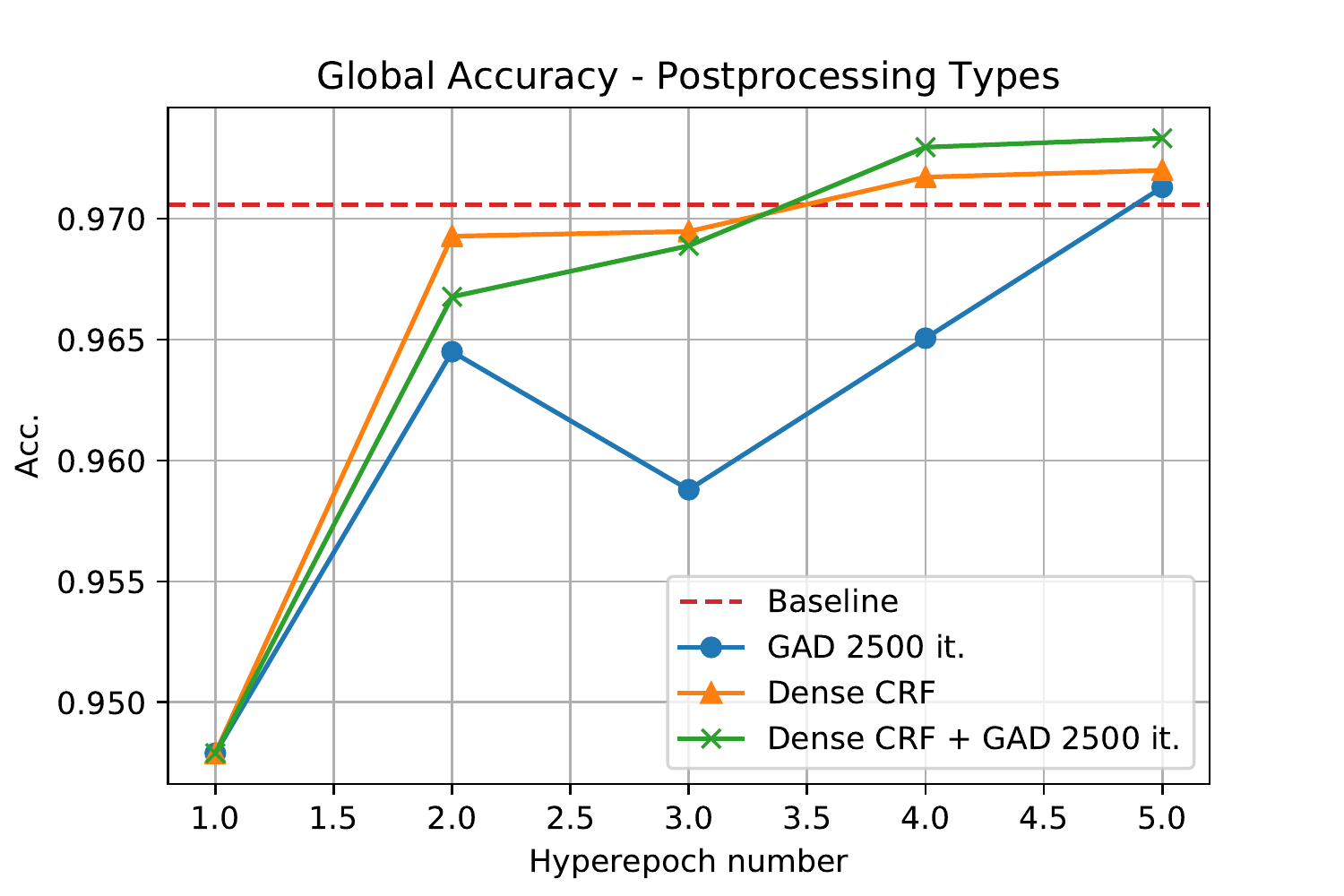}\\
        (c)
    \end{minipage}
    
    \caption{Ablation studies. (a) Comparison between strategies for merging network predictions and reference data. (b) Comparison between iterative training with and without the usage of original reference data. (c) Comparison between GAD and Dense CRF. Top row contains Dice scores, bottom row contains global accuracy curves.}
    \label{fig:ablation}
\end{figure}

We applied the GAD algorithm to the predictions from a network trained directly on the reference data from HRSCD to evaluate its performance. In Fig.~\ref{fig:gad_122} is displayed an example of the obtained results. As noted before, we can see in (c) that the change is detected but that unchanged pixels around it are also classified as changes by the network. In (d)-(f) it can be clearly seen how the GAD algorithm improves the results by diffusing the labels across similar pixels while preserving edges from the input images in the semantic segmentation results. As expected, more iterations of the algorithm lead to a stronger erosion of incorrect labels. For these results, GAD was applied with $K = 5$ and $\lambda = 0.24$. In Fig.~\ref{fig:gad_vs_dcrf} we can see a comparison between GAD and the Dense CRF\footnote{\url{https://github.com/lucasb-eyer/pydensecrf}} algorithm~\cite{krahenbuhl2011efficient}. While the non-local nature of fully connected CRFs is useful in some cases, we can see the results are less precise and significantly noisier than the ones obtained by using GAD.

To perform quantitative analysis of results, it would be meaningless to use the test data in the HRSCD dataset. Using a GeForce GTX 1060 GPU, applying GAD to a 512x512 image for 100 iterations took approximately 230~ms. Indeed, we are attempting to perform a task which is not the one for which ground truth data are available since \ie we are attempting to perform pixel-level precise change detection and not parcel-level change detection. For this reason we have manually annotated the changes as precisely as possible for two 10000x10000 image pairs in the dataset, for a total of 2$\cdot$10$^{8}$ test pixels, or 50~km$^2$. The image pairs were chosen before any tests were made to avoid biasing the results. Due to the class imbalance, total accuracy, \ie the percentage of correctly classified pixels, provides us with a skewed view of the results biased towards the performance on the class more strongly represented. Therefore, the S{\o}rensen-Dice coefficient (equivalent to the F1 score for binary problems) from the point of view of the change class was used~\cite{dice1945measures,sorensen1948method}. The S{\o}rensen-Dice coefficient score is defined as
\begin{equation}
    \mathit{Dice} = (2\cdot TP)/(2\cdot TP + FP + FN)
\end{equation}
where TP means true positive, FP means false positive, and FN means false negative. It serves as a balanced measurement of performance even for unbalanced data.

All tests presented here were done using PyTorch~\cite{paszke2017automatic}. At each hyperepoch, the network was trained for 100 epochs with an ADAM algorithm for stochastic optimization~\cite{kingma2014adam}, with learning rate of $10^{-3}$ for the first 75 epochs and $10^{-4}$ for the other 25 epochs. The tests show the performance of networks trained with the proposed method for 5 hyperepochs (iterations of training and cleaning the data), where the first one is done directly on the available data from the HRSCD dataset. For accurate comparison of methods and to minimize the randomness in the comparisons, the obtained network at the end of hyperepoch 1 is used as a starting point for all the methods. This ensures all networks have the same initialization at the point in the algorithm where they diverge. A baseline network was trained for the same amount of epochs and hyperepochs but with no changes done to the training data. This serves as a reference point as to the performance of the fully convolutional network with no weakly supervised training methods.

\begin{figure*}[t]
    \centering
    \begin{minipage}{0.15\linewidth}
        \centering
        \includegraphics[width=\linewidth]{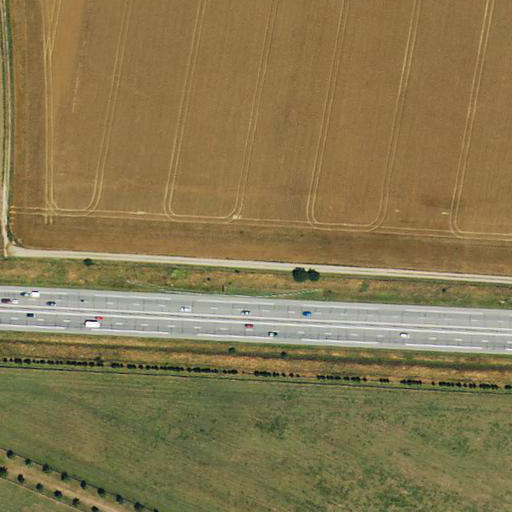}\\
    \end{minipage}
    \hfill
    \begin{minipage}{0.15\linewidth}
        \centering
        \includegraphics[width=\linewidth]{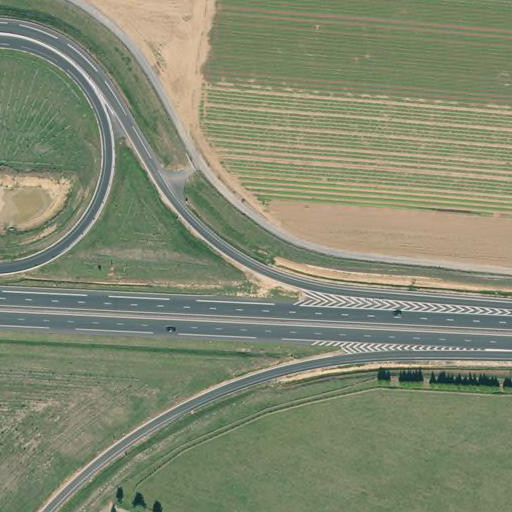}\\
    \end{minipage}
    \hfill
    \begin{minipage}{0.15\linewidth}
        \centering
        \includegraphics[width=\linewidth]{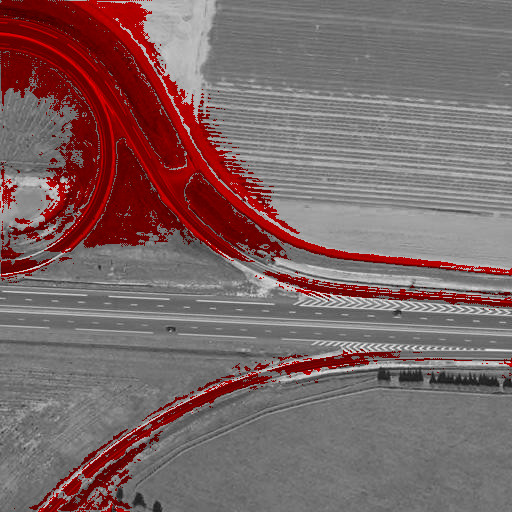}\\
    \end{minipage}
    \hfill
    \begin{minipage}{0.15\linewidth}
        \centering
        \includegraphics[width=\linewidth]{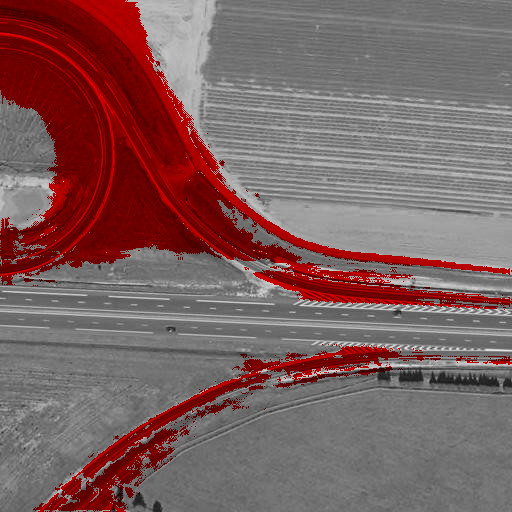}\\
    \end{minipage}
    \hfill
    \begin{minipage}{0.15\linewidth}
        \centering
        \includegraphics[width=\linewidth]{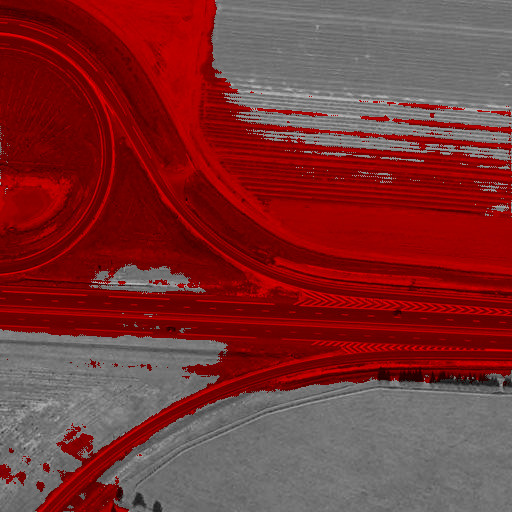}\\
    \end{minipage}
    \hfill
    \begin{minipage}{0.15\linewidth}
        \centering
        \includegraphics[width=\linewidth]{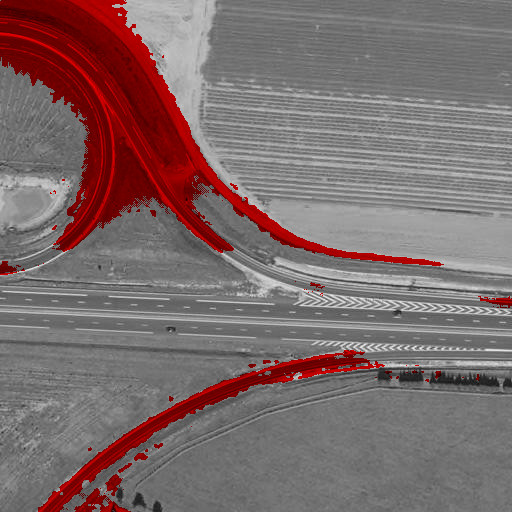}\\
    \end{minipage}
    \break
    
    \begin{minipage}{0.15\linewidth}
        \centering
        \includegraphics[width=\linewidth]{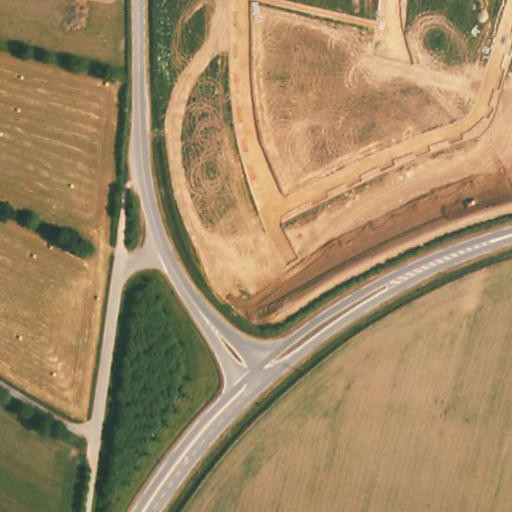}\\
    \end{minipage}
    \hfill
    \begin{minipage}{0.15\linewidth}
        \centering
        \includegraphics[width=\linewidth]{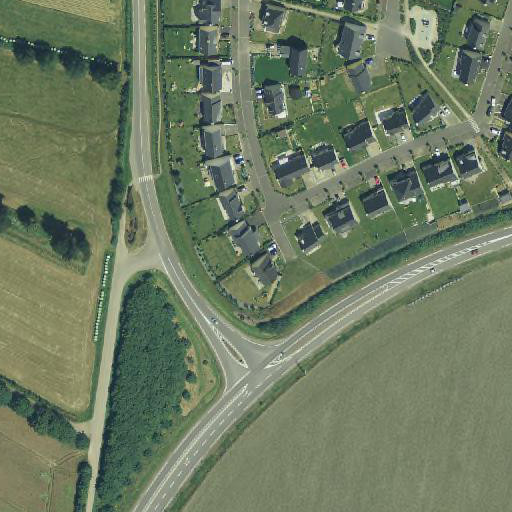}\\
    \end{minipage}
    \hfill
    \begin{minipage}{0.15\linewidth}
        \centering
        \includegraphics[width=\linewidth]{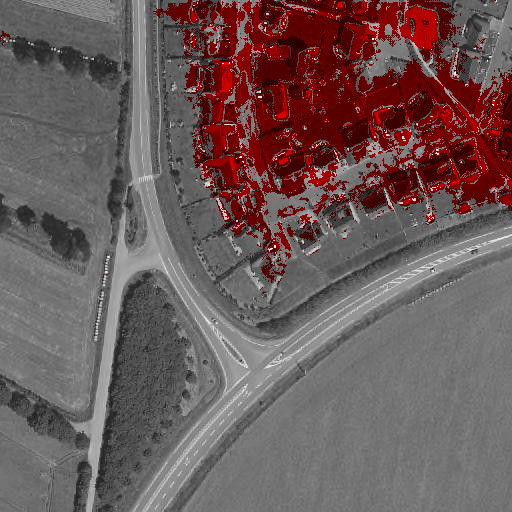}\\
    \end{minipage}
    \hfill
    \begin{minipage}{0.15\linewidth}
        \centering
        \includegraphics[width=\linewidth]{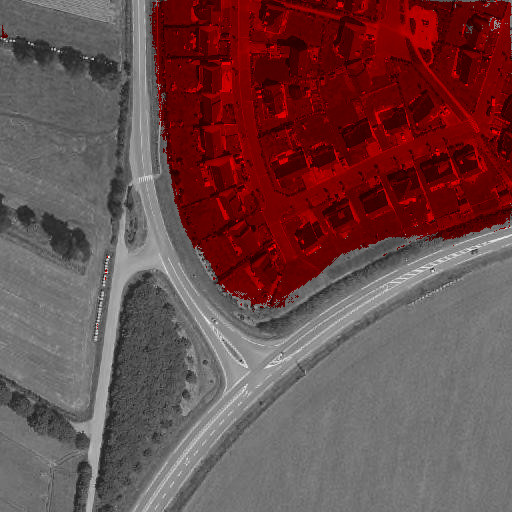}\\
    \end{minipage}
    \hfill
    \begin{minipage}{0.15\linewidth}
        \centering
        \includegraphics[width=\linewidth]{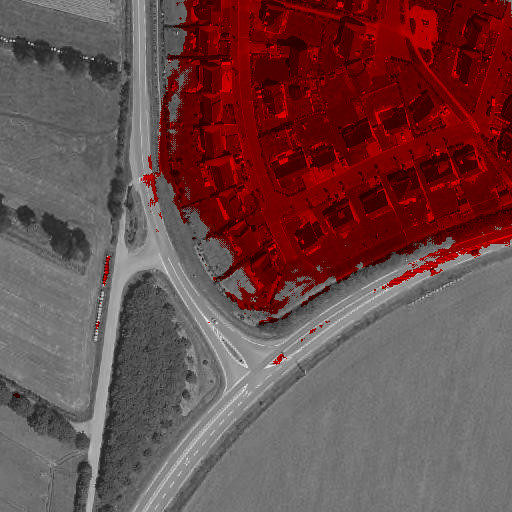}\\
    \end{minipage}
    \hfill
    \begin{minipage}{0.15\linewidth}
        \centering
        \includegraphics[width=\linewidth]{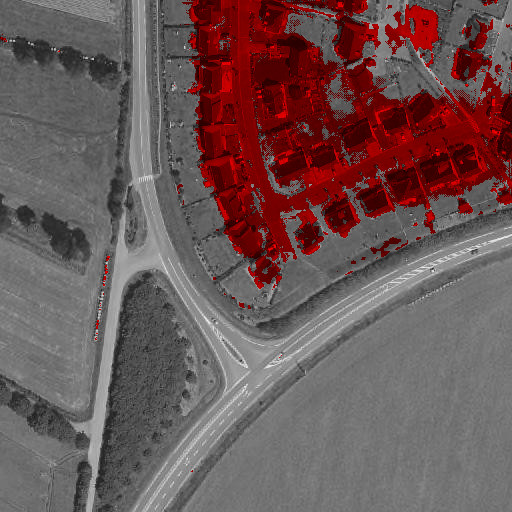}\\
    \end{minipage}
    \break
    
    \begin{minipage}{0.15\linewidth}
        \centering
        (a) Image 1 \newline
    \end{minipage}
    \hfill
    \begin{minipage}{0.15\linewidth}
        \centering
        (b) Image 2 \newline
    \end{minipage}
    \hfill
    \begin{minipage}{0.15\linewidth}
        \centering
        (c) Baseline \newline
    \end{minipage}
    \hfill
    \begin{minipage}{0.15\linewidth}
        \centering
        (d) GAD 2500 it.
    \end{minipage}
    \hfill
    \begin{minipage}{0.15\linewidth}
        \centering
        (e) No ref. constraint
    \end{minipage}
    \hfill
    \begin{minipage}{0.15\linewidth}
        \centering
        (f) Dense CRF
    \end{minipage}
    \break
    
    \caption{Change maps obtained by using different methods on two image pairs. Detected changes are marked in red color.}
    \label{fig:examples}
\end{figure*}

The first comparison, shown in Fig.~\ref{fig:ablation}(a), compares the three methods proposed in Section~\ref{sec:its} to combine the network predictions with the original ground truth from the HRSCD dataset. We notice that all three strategies surpass the baseline network using the proposed iterative training method, which validates the ideas presented earlier. In Fig.~\ref{fig:ablation}(b) we see a comparison between a training using the full training scheme proposed in this paper (without the usage of an ignore class) and the same method but without using the original reference data, \ie using only network predictions processed by GAD to continue training at each hyperepoch. Our results, which corroborate the ones in \cite{khoreva2017simple}, show that referring back to the original data at each hyperepoch is essential to avoid a degradation in performance.

\begin{figure}[t]
     \hfill
    \begin{minipage}{0.19\linewidth}
        \centering
        \includegraphics[width=\linewidth]{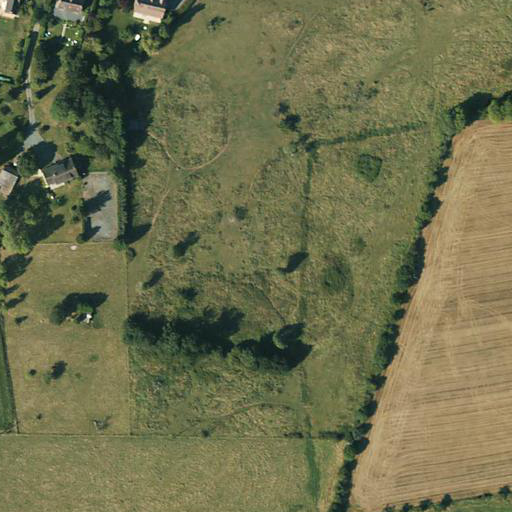}\\
    \end{minipage}
    \hfill
    \begin{minipage}{0.19\linewidth}
        \centering
        \includegraphics[width=\linewidth]{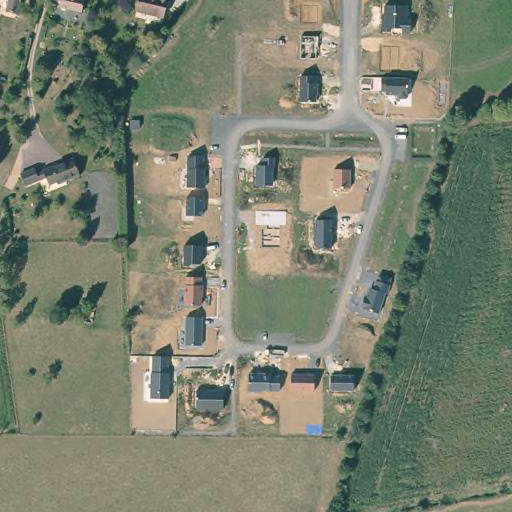}\\
    \end{minipage}
    \hfill
    \begin{minipage}{0.19\linewidth}
        \centering
        \includegraphics[width=\linewidth]{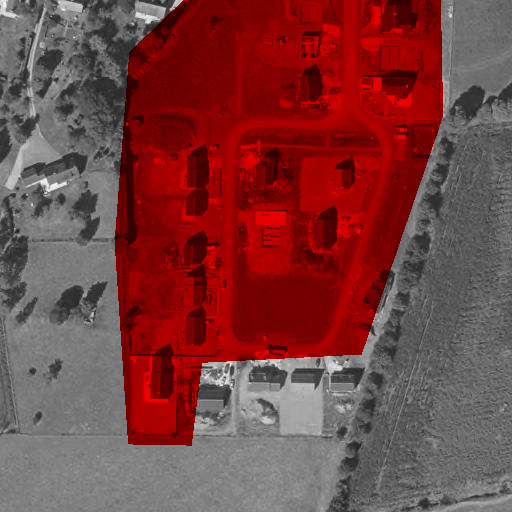}\\
    \end{minipage}
    \hfill
    \begin{minipage}{0.19\linewidth}
        \centering
        \includegraphics[width=\linewidth]{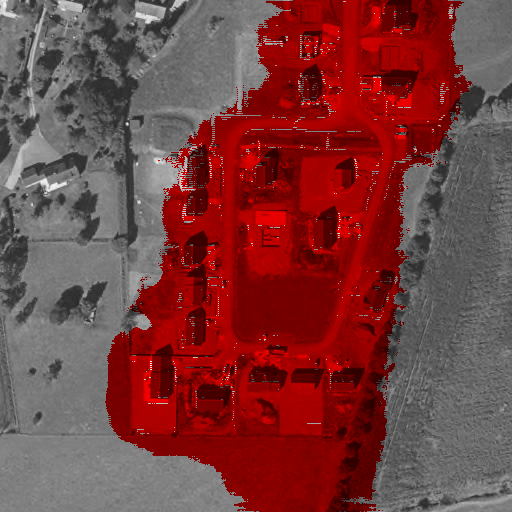}\\
    \end{minipage}
    \hfill
    \begin{minipage}{0.19\linewidth}
        \centering
        \includegraphics[width=\linewidth]{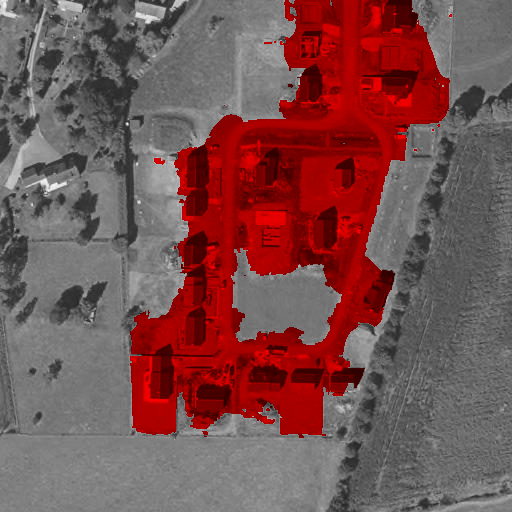}\\
    \end{minipage}
    \hfill
    \break

     \hfill
    \begin{minipage}{0.19\linewidth}
        \centering
        \includegraphics[width=\linewidth]{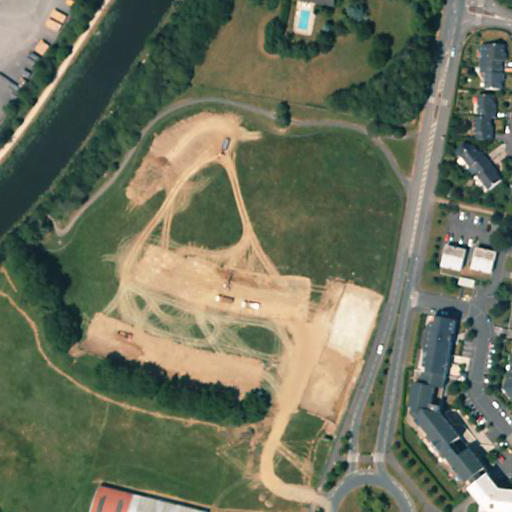}\\
    \end{minipage}
    \hfill
    \begin{minipage}{0.19\linewidth}
        \centering
        \includegraphics[width=\linewidth]{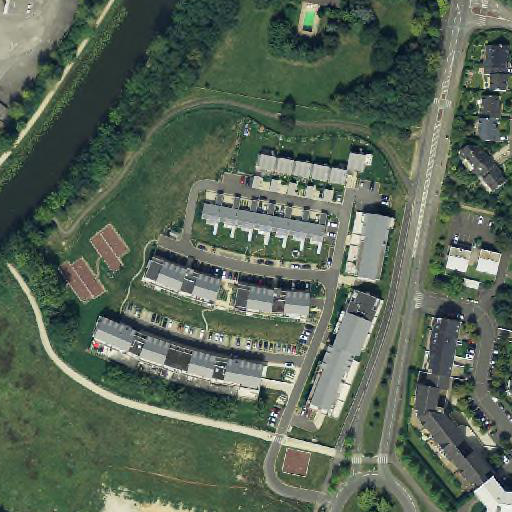}\\
    \end{minipage}
    \hfill
    \begin{minipage}{0.19\linewidth}
        \centering
        \includegraphics[width=\linewidth]{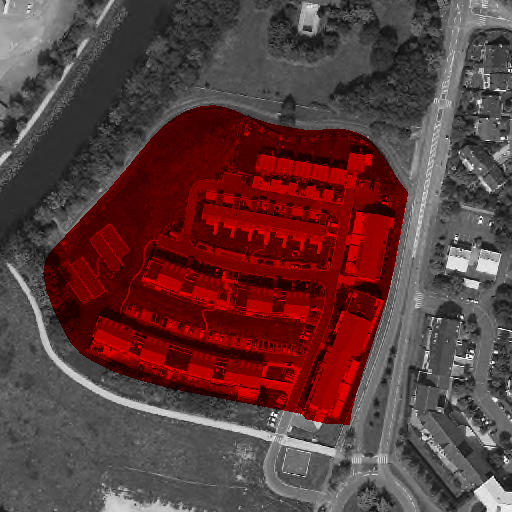}\\
    \end{minipage}
    \hfill
    \begin{minipage}{0.19\linewidth}
        \centering
        \includegraphics[width=\linewidth]{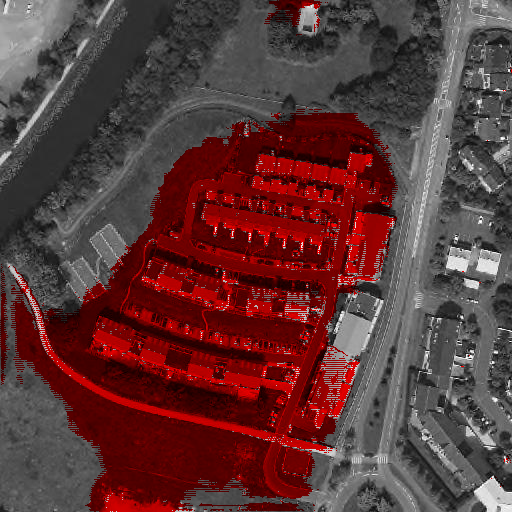}\\
    \end{minipage}
    \hfill
    \begin{minipage}{0.19\linewidth}
        \centering
        \includegraphics[width=\linewidth]{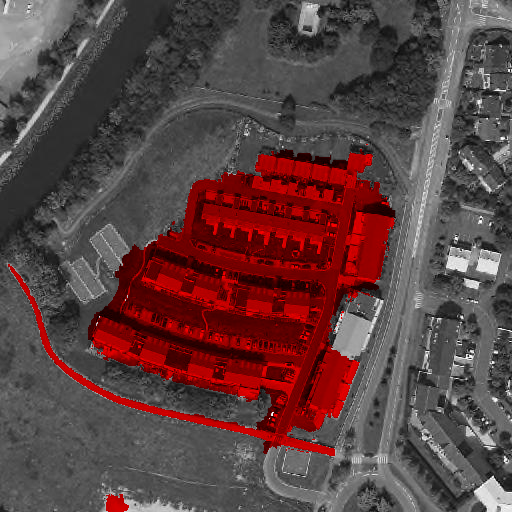}\\
    \end{minipage}
    \hfill
    \break

     \hfill
    \begin{minipage}{0.19\linewidth}
        \centering
        (a) Image 1 \newline
    \end{minipage}
    \hfill
    \begin{minipage}{0.19\linewidth}
        \centering
        (b) Image 2 \newline
    \end{minipage}
    \hfill
    \begin{minipage}{0.19\linewidth}
        \centering
        (c) Reference data
    \end{minipage}
    \hfill
    \begin{minipage}{0.19\linewidth}
        \centering
        (d) Naive prediction
    \end{minipage}
    \hfill
    \begin{minipage}{0.19\linewidth}
        \centering
        (e) GAD \newline
    \end{minipage}
    \hfill
    \break
    
    \caption{Results using the complete inference pipeline (changes in red). GAD is used to improve predictions during the iterative training process as well as for improving the final segmentations.}
    \label{fig:gad_exs}
\end{figure}

In Fig.~\ref{fig:ablation}(c) we show a comparison between using the proposed GAD algorithm versus the Dense CRF~\cite{krahenbuhl2011efficient} algorithm in the iterated training procedure, as well as using both together. We see that using the Dense CRF algorithm to process predictions leads to good performance in early hyperepochs, but is surpassed by GAD later on. This is likely explained by the non local nature of Dense CRF and its ability to deal with larger errors, but its inferior performance relative to GAD for finer prediction errors.

Figure~\ref{fig:examples} shows the predictions by networks trained by different methods on two example images. We see that the best results are obtained by using the full training scheme with GAD in (d)/(j), followed by Dense CRF, which also achieves good results shown in (f)/(l). The baseline results in (c)/(i), obtained by naively training the network in a supervised manner, and the ones without using the reference data as constraint in the iterative training scheme shown in (e)/(k) are significantly less accurate than those using GAD or Dense CRF. The final change maps that were produced by the proposed method for two test cases can be seen in Fig.~\ref{fig:gad_exs}.

\subsection{Scene-Invariant Spatial Attention Layer \label{sec:exp-attn-layer}}

We tested the proposed method using the ABCD dataset proposed by Fujita \etal \cite{abcd}. This dataset contains pairs of crops of images centered on buildings that have been surveyed to evaluate their destruction after a tsunami. We have followed the 5-fold cross validation that was defined by the dataset's creators. All networks were trained from scratch using only the ABCD dataset, using an initial learning rate of $0.005$ for 10 epochs, then with a linearly decaying learning rate for 90 epochs for a total of 100 epochs. The classification results for these tests are presented in Table~\ref{tab:abcd}. These results show that our network with the attention module performed very similarly to the ones presented in \cite{abcd}. It is also clear that the proposed attention module improved the classification accuracy of the networks significantly. The obtained results also show that filtering the attention weights using the GAD algorithm further increases the classification performance of the proposed network, improving the quality of the attention weights by using the input images as guides.

\begin{figure}[t]
    \begin{minipage}{0.19\linewidth}
        \centering
        \includegraphics[width=\linewidth]{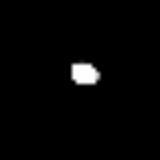}\\
    \end{minipage}
    \hfill
    \begin{minipage}{0.19\linewidth}
        \centering
        \includegraphics[width=\linewidth]{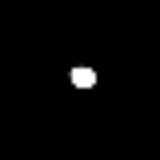}\\
    \end{minipage}
    \hfill
    \begin{minipage}{0.19\linewidth}
        \centering
        \includegraphics[width=\linewidth]{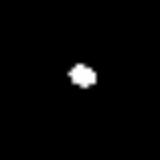}\\
    \end{minipage}
    \hfill
    \begin{minipage}{0.19\linewidth}
        \centering
        \includegraphics[width=\linewidth]{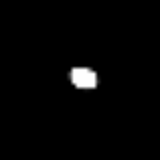}\\
    \end{minipage}
    \hfill
    \begin{minipage}{0.19\linewidth}
        \centering
        \includegraphics[width=\linewidth]{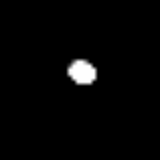}\\
    \end{minipage}\break

    \begin{minipage}{0.19\linewidth}
        \centering
        \includegraphics[width=\linewidth]{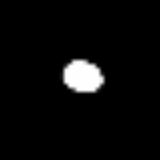}\\
    \end{minipage}
    \hfill
    \begin{minipage}{0.19\linewidth}
        \centering
        \includegraphics[width=\linewidth]{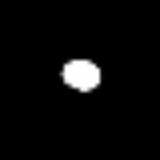}\\
    \end{minipage}
    \hfill
    \begin{minipage}{0.19\linewidth}
        \centering
        \includegraphics[width=\linewidth]{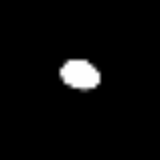}\\
    \end{minipage}
    \hfill
    \begin{minipage}{0.19\linewidth}
        \centering
        \includegraphics[width=\linewidth]{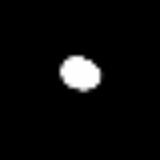}\\
    \end{minipage}
    \hfill
    \begin{minipage}{0.19\linewidth}
        \centering
        \includegraphics[width=\linewidth]{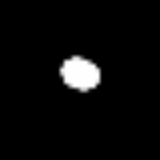}\\
    \end{minipage}\break

    \caption{Spatial attention weights that were learned in each of the cross-validation tests. Top row contains all 5 tests using fixed scale ABCD dataset, bottom row are the results using the rescaled version of the dataset. Note that the network was incredibly consistent in identifying the center of the images as most discriminative without any explicit knowledge. These attention matrices are of size $40 \times 40$.}
    \label{fig:attention_weights}
\end{figure}

Figure~\ref{fig:attention_weights} show the learned spatial attention weights learned in each of the performed tests. We can clearly see how consistent the network was in identifying that the most discriminative region of the images was located in the center. It is also apparent that the network identified that the scale of this discriminative region is larger in the \textit{resized} version of the dataset compared to the \textit{fixed-scale} version.

\begin{figure}[t]
    \begin{minipage}{0.155\linewidth}
        \centering
        \includegraphics[width=\linewidth]{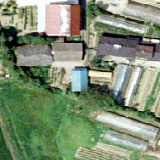}\\
    \end{minipage}
    \hfill
    \begin{minipage}{0.155\linewidth}
        \centering
        \includegraphics[width=\linewidth]{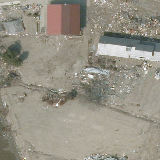}\\
    \end{minipage}
    \hfill
    \begin{minipage}{0.155\linewidth}
        \centering
        \includegraphics[width=\linewidth]{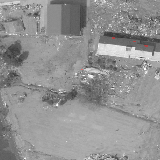}\\
    \end{minipage}
    \hfill
    \begin{minipage}{0.155\linewidth}
        \centering
        \includegraphics[width=\linewidth]{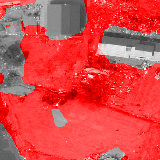}\\
    \end{minipage}
    \hfill
    \begin{minipage}{0.155\linewidth}
        \centering
        \includegraphics[width=\linewidth]{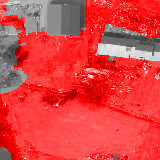}\\
    \end{minipage}
    \hfill
    \begin{minipage}{0.155\linewidth}
        \centering
        \includegraphics[width=\linewidth]{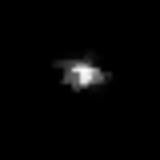}\\
    \end{minipage}\break

    \begin{minipage}{0.155\linewidth}
        \centering
        \includegraphics[width=\linewidth]{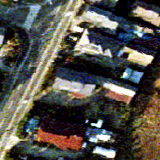}\\
    \end{minipage}
    \hfill
    \begin{minipage}{0.155\linewidth}
        \centering
        \includegraphics[width=\linewidth]{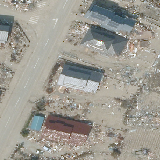}\\
    \end{minipage}
    \hfill
    \begin{minipage}{0.155\linewidth}
        \centering
        \includegraphics[width=\linewidth]{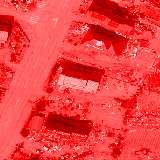}\\
    \end{minipage}
    \hfill
    \begin{minipage}{0.155\linewidth}
        \centering
        \includegraphics[width=\linewidth]{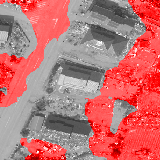}\\
    \end{minipage}
    \hfill
    \begin{minipage}{0.155\linewidth}
        \centering
        \includegraphics[width=\linewidth]{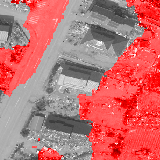}\\
    \end{minipage}
    \hfill
    \begin{minipage}{0.155\linewidth}
        \centering
        \includegraphics[width=\linewidth]{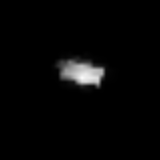}\\
    \end{minipage}\break

    \begin{minipage}{0.155\linewidth}
        \centering
        \includegraphics[width=\linewidth]{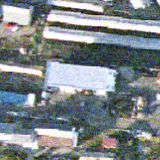}\\
    \end{minipage}
    \hfill
    \begin{minipage}{0.155\linewidth}
        \centering
        \includegraphics[width=\linewidth]{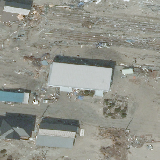}\\
    \end{minipage}
    \hfill
    \begin{minipage}{0.155\linewidth}
        \centering
        \includegraphics[width=\linewidth]{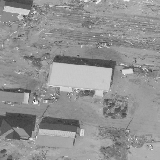}\\
    \end{minipage}
    \hfill
    \begin{minipage}{0.155\linewidth}
        \centering
        \includegraphics[width=\linewidth]{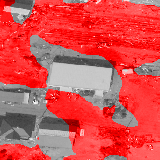}\\
    \end{minipage}
    \hfill
    \begin{minipage}{0.155\linewidth}
        \centering
        \includegraphics[width=\linewidth]{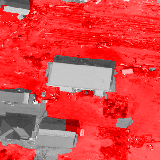}\\
    \end{minipage}
    \hfill
    \begin{minipage}{0.155\linewidth}
        \centering
        \includegraphics[width=\linewidth]{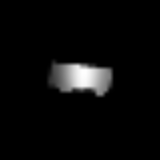}\\
    \end{minipage}\break

    \begin{minipage}{0.155\linewidth}
        \centering
        \includegraphics[width=\linewidth]{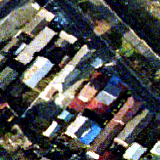}\\
    \end{minipage}
    \hfill
    \begin{minipage}{0.155\linewidth}
        \centering
        \includegraphics[width=\linewidth]{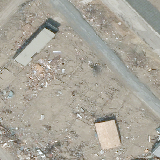}\\
    \end{minipage}
    \hfill
    \begin{minipage}{0.155\linewidth}
        \centering
        \includegraphics[width=\linewidth]{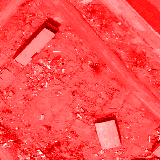}\\
    \end{minipage}
    \hfill
    \begin{minipage}{0.155\linewidth}
        \centering
        \includegraphics[width=\linewidth]{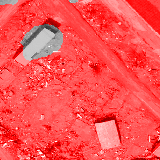}\\
    \end{minipage}
    \hfill
    \begin{minipage}{0.155\linewidth}
        \centering
        \includegraphics[width=\linewidth]{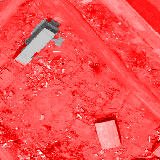}\\
    \end{minipage}
    \hfill
    \begin{minipage}{0.155\linewidth}
        \centering
        \includegraphics[width=\linewidth]{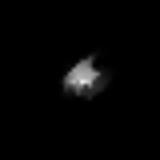}\\
    \end{minipage}\break

    \begin{minipage}{0.155\linewidth}
        \centering
        \includegraphics[width=\linewidth]{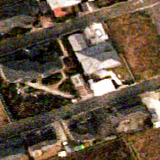}\\
        {\tiny Image 1}
    \end{minipage}
    \hfill
    \begin{minipage}{0.155\linewidth}
        \centering
        \includegraphics[width=\linewidth]{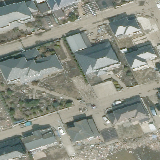}\\
        {\tiny Image 2}
    \end{minipage}
    \hfill
    \begin{minipage}{0.155\linewidth}
        \centering
        \includegraphics[width=\linewidth]{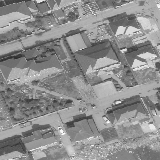}\\
        {\tiny No attention}
    \end{minipage}
    \hfill
    \begin{minipage}{0.155\linewidth}
        \centering
        \includegraphics[width=\linewidth]{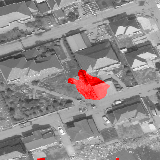}\\
        {\tiny Attention}
    \end{minipage}
    \hfill
    \begin{minipage}{0.155\linewidth}
        \centering
        \includegraphics[width=\linewidth]{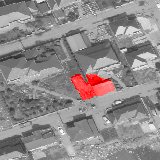}\\
        {\tiny Att. + GAD}
    \end{minipage}
    \hfill
    \begin{minipage}{0.155\linewidth}
        \centering
        \includegraphics[width=\linewidth]{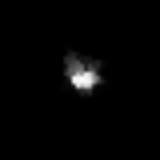}\\
        {\tiny Att. coef.}
    \end{minipage}\break

    \caption{Results obtained by using the proposed method. Note that when the attention layer is not used, the network does not learn to localize the features and tends to predict all pixels into the same class. The attention layer enables the network to localize features much more accurately, and the GAD post-processing further increases the spatial accuracy of such predictions. Changes are marked in red.}
    \label{fig:attention_results}
\end{figure}

Qualitative analysis of segmentation results show that the usage of the proposed spatial attention operation allowed the network to vastly increase its capacity to localize features in the input images, which led to much more accurate segmentation, as depicted in Fig.~\ref{fig:attention_results}. The results also show how using the GAD algorithm for post-processing further increased the spatial accuracy of the segmentation results. In these images, the application of our algorithm without an attention layer or GAD (column "No attention") can be seen as a simple Siamese extension of the CAM technique for handling two input images~\cite{Zhou_2016_CVPR}

\begin{table}
    \centering
    \caption{Accuracy and standard deviation for each test on ABCD dataset using 5-fold cross validation. Fixed scale and resized variations of the ABCD dataset were tested. Results from methods proposed by Fujita \etal are included for comparison.}
    \label{tab:abcd}       
    \begin{tabular}{l|l|l}
        \hline\noalign{\smallskip}
        Method & Fixed scale & Resized\\
        \noalign{\smallskip}\hline\noalign{\smallskip}
        6-ch~\cite{abcd} & $ 94.5 \pm 0.5$ & $ 94.7 \pm 0.3$ \\
        siam~\cite{abcd} & $ 94.8 \pm 0.3 $ & $ 94.9 \pm 0.4$ \\
        \noalign{\smallskip}\hline\noalign{\smallskip}
        No attention & $89.33 \pm 0.79$  & $90.96 \pm 0.65$\\
        Attention & $94.36 \pm 0.26$  & $94.88 \pm 0.18$\\
        Attention + GAD & $94.58 \pm 0.27$  & $94.90 \pm 0.22$\\
        \noalign{\smallskip}\hline
    \end{tabular}
\end{table}

These results suggest that there is a positive feedback loop that happens during the training process between the network's ability to localize discriminative features and the spatial attention operation. Once the network develops the ability to roughly localize discriminative features, this allows the training of the spatial attention layer, which leads the network to learn even more local features, and so on.

Two notable examples can be seen in Fig.~\ref{fig:attention_results}. The first one is the example in the fourth row, which shows that the network is not simply finding buildings in the second image and marking those as unchanged. In this example, a building is present in the second image but it is marked as a change nonetheless since it doesn't match the buildings in the first image. The second notable example is the one showed in the last row, where a very small change was detected in the center of the image, surrounded only by unchanged buildings. Since the position of this detected change coincided to the spatial attention position, the network was able to mark this image pair as a change, which is correct according to the ground truth label. The same was not accomplished by the network without the attention layer.

\subsection{Edge Enhancement for Segmentation Upsampling \label{sec:exp-ee}}

To further validate the effectiveness of GAD as a postprocessing tool in a more general setting, we perform the experiments described in Section~\ref{sec:ee}, where we study how GAD can be used for edge enhancement for upsampled softmax activations. To simulate a setting where data with different resolutions are available, we use subsampled images for training the network and study their predictions following the steps depicted in Fig.~\ref{fig:ss_main_idea}.

We performed these experiments using two datasets. The first one is the \textit{Inria Aerial Image Labeling Dataset}~\cite{maggiori2017dataset}, which contain RGB images of several urban areas in different countries and environments at a spatial resolution of $0.3~m/px$, along with binary pixel-level labels that indicate the presence of buildings. The train/validation split that was proposed by the dataset creators (i.e. keeping the first five images for each location for validation) was used, which results in 155 images for training and 25 images for validation, all of size $5000\times5000$ pixels. The second dataset used for testing this approach was the \textit{Vaihingen Dataset}\footnote{\url{https://www2.isprs.org/commissions/comm2/wg4/benchmark/2d-sem-label-vaihingen/}}, which contains false color images for the urban area of Vaihingen at a spatial resolution of $0.09~m/px$, as well as pixel-level semantic segmentation labels. We followed the train/validation split proposed by Audebert et al. in \cite{audebert_beyond_2017}. The code for this work\footnote{\url{https://github.com/nshaud/DeepNetsForEO}} was also used with the appropriate modification to perform the subsampling experiments. The standard SegNet architecture was used for all the experiments in this section~\cite{badrinarayanan2017segnet}. The parameters for the GAD algorithms were tuned visually using a few example images before being applied to the validation dataset. For the experiments presented in this section, the parameters that were used were: $\lambda = 0.24$, $N = 1000$, and $K = 0.002$ (for images normalized between 0 and 1).

\begin{figure}[ht]
    \begin{minipage}{0.15\linewidth}
        \centering
        \includegraphics[width=\linewidth]{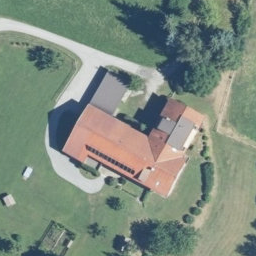}\\
        Input\\
        \vspace{3pt}
        \includegraphics[width=\linewidth]{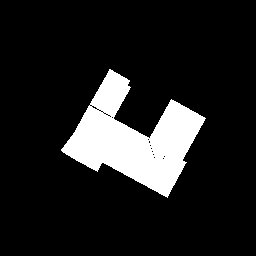}\\
        GT
    \end{minipage}
    ~
    \begin{minipage}{0.15\linewidth}
        \centering
        \includegraphics[width=\linewidth]{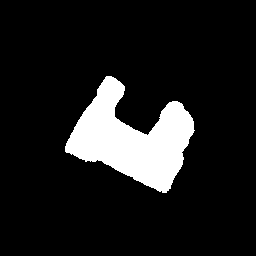}\\
        ss=1\\
        \vspace{3pt}
        \includegraphics[width=\linewidth]{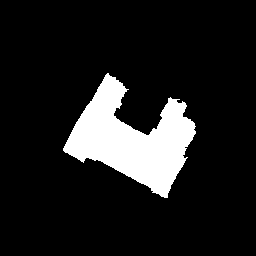}\\
        ss=1+GAD
    \end{minipage}
    ~
    \begin{minipage}{0.15\linewidth}
        \centering
        \includegraphics[width=\linewidth]{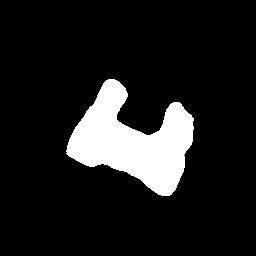}\\
        ss=4\\
        \vspace{3pt}
        \includegraphics[width=\linewidth]{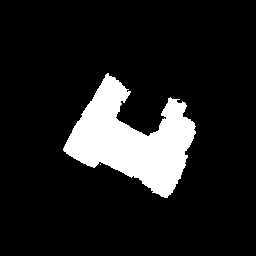}\\
        ss=4+GAD
    \end{minipage}
    ~
    \begin{minipage}{0.15\linewidth}
        \centering
        \includegraphics[width=\linewidth]{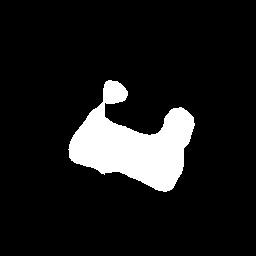}\\
        ss=8\\
        \vspace{3pt}
        \includegraphics[width=\linewidth]{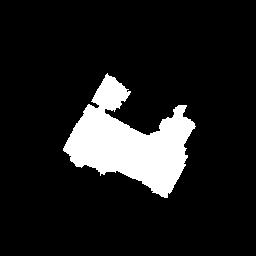}\\
        ss=8+GAD
    \end{minipage}
    ~
    \begin{minipage}{0.15\linewidth}
        \centering
        \includegraphics[width=\linewidth]{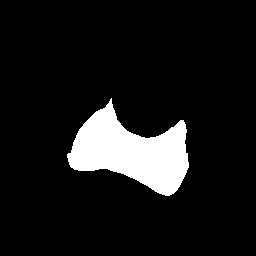}\\
        ss=12\\
        \vspace{3pt}
        \includegraphics[width=\linewidth]{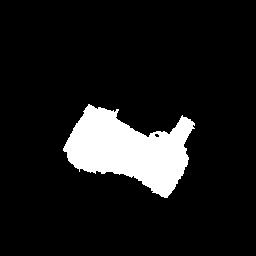}\\
        ss=12+GAD
    \end{minipage}
    ~
    \begin{minipage}{0.15\linewidth}
        \centering
        \includegraphics[width=\linewidth]{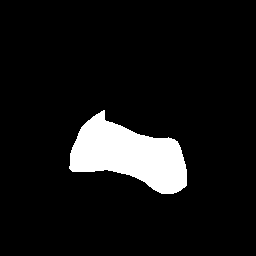}\\
        ss=16\\
        \vspace{3pt}
        \includegraphics[width=\linewidth]{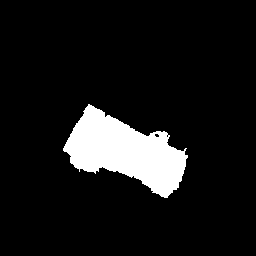}\\
        ss=16+GAD
    \end{minipage}

    \caption{Results obtained from experiments on the Inria Aerial Image Labeling Dataset. GAD successfully mitigated the accuracy loss in region boundaries at higher subsampling rates.}
    \label{fig:inria_results}
\end{figure}

\begin{figure}[ht]

    \begin{minipage}{0.46\linewidth}
        \centering
        \includegraphics[width=\linewidth]{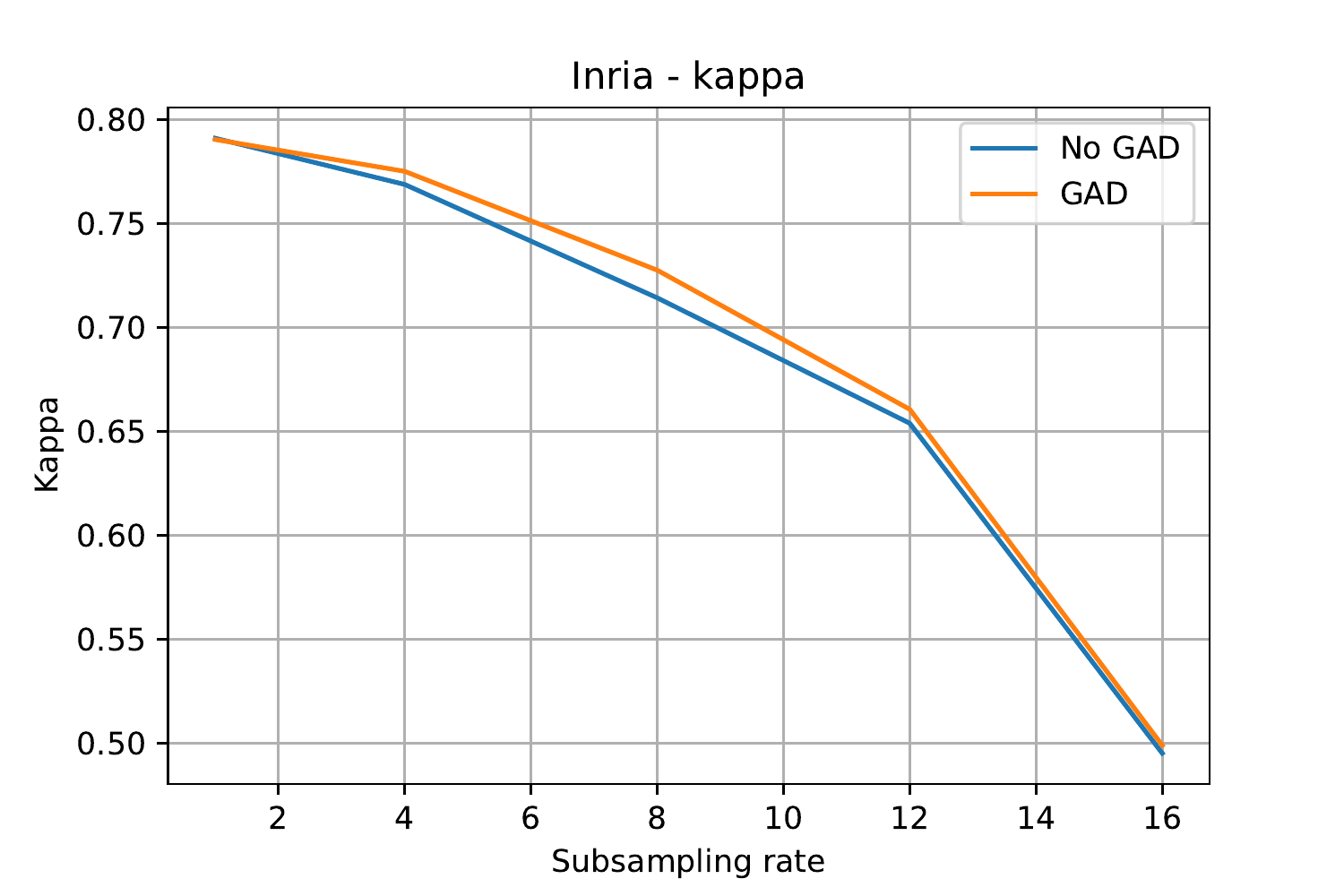}\\
    \end{minipage}
    \hfill
    \begin{minipage}{0.46\linewidth}
        \centering
        \includegraphics[width=\linewidth]{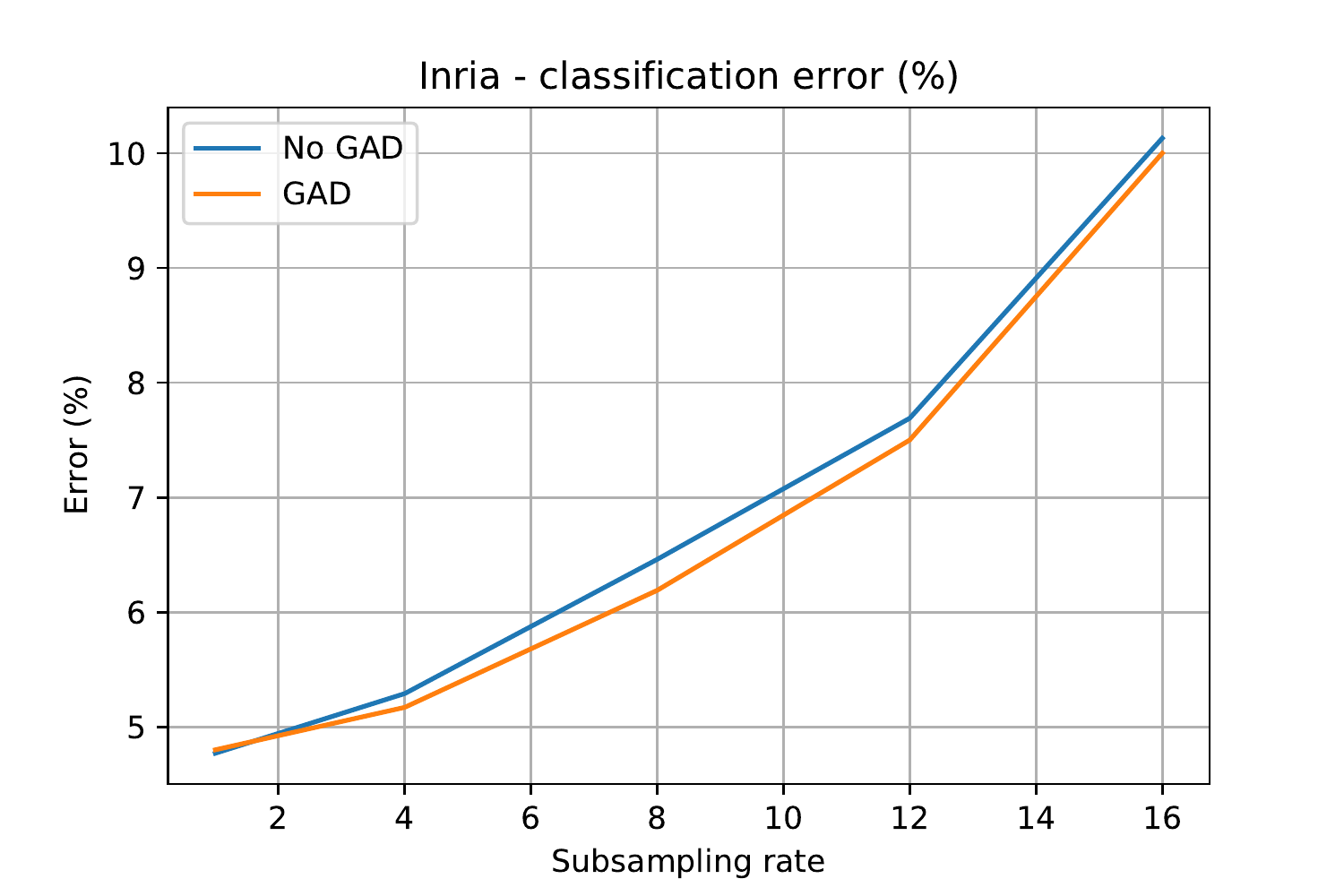}\\
    \end{minipage}\break

    \begin{minipage}{0.46\linewidth}
        \centering
        \includegraphics[width=\linewidth]{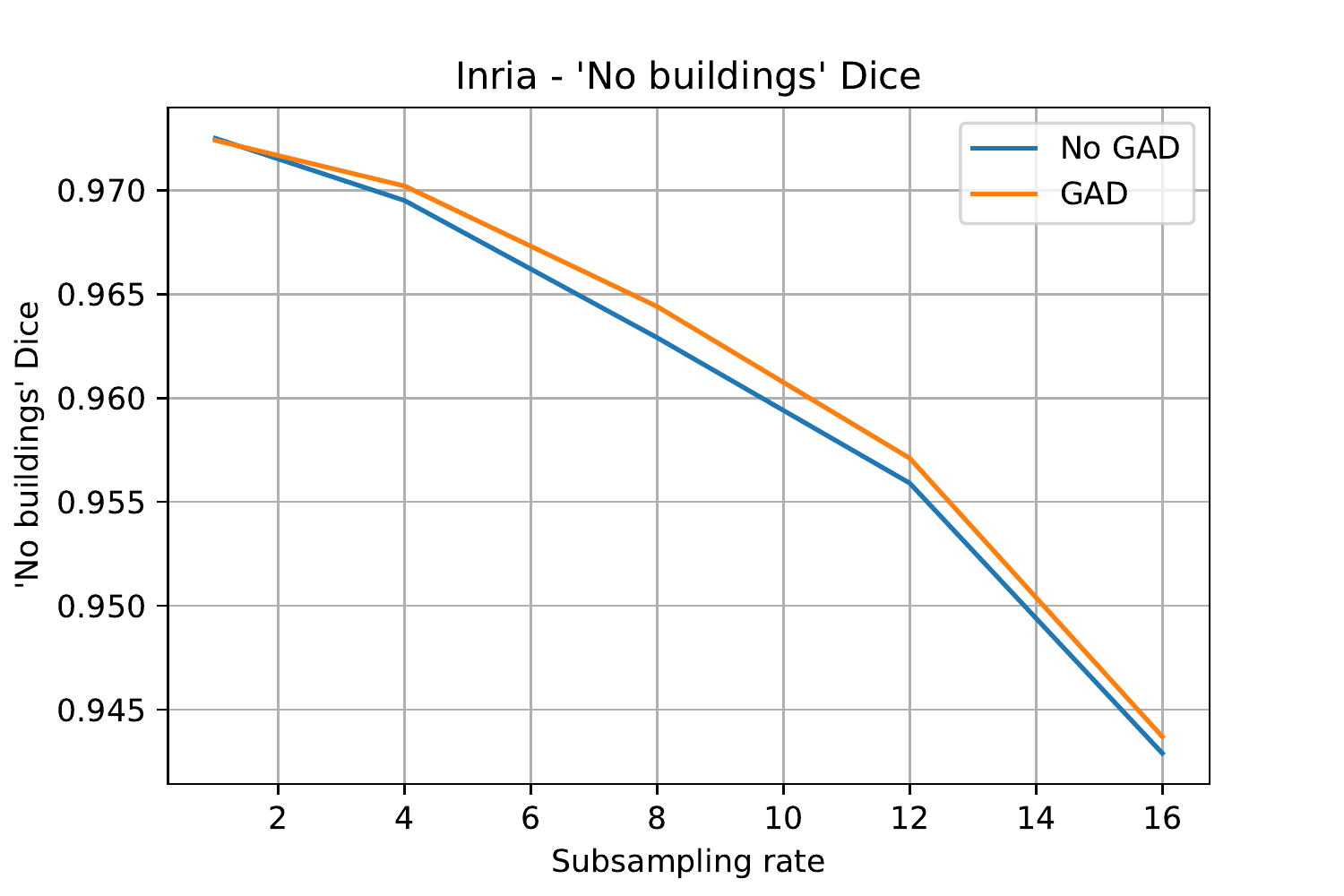}\\
    \end{minipage}
    \hfill
    \begin{minipage}{0.46\linewidth}
        \centering
        \includegraphics[width=\linewidth]{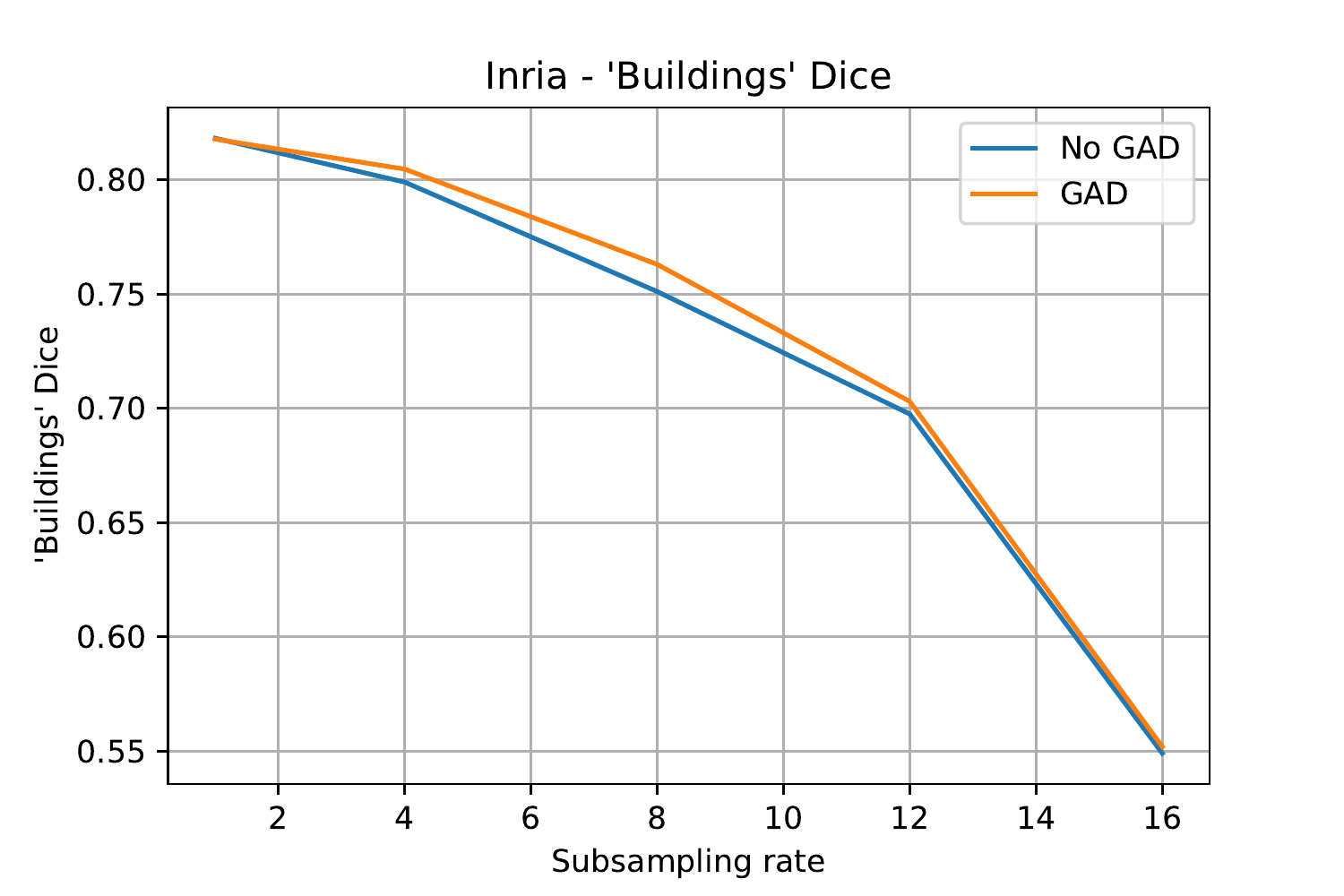}\\
    \end{minipage}\break

    \caption{Results for improving segmentation boundaries on the Inria Aerial Image Labeling Dataset. Consistent improvements across all metrics have been observed for all subsampling rates bigger than 1.}
    \label{fig:inria}
\end{figure}

Qualitative and quantitative results for the experiments performed on the Inria dataset can be seen in Figs.~\ref{fig:inria_results} and \ref{fig:inria}, respectively. Figure~\ref{fig:inria_results} clearly shows the efficacy of GAD to transfer edges from the guide image onto the predictions. It also shows that, as a postprocessing algorithm, the quality of the output is strongly linked to the quality of the input, and GAD is not able to accurately fix large errors in the predictions if the inputs miss large parts of the objects. These images illustrate that higher subsampling factors led to worse results, as is expected. Nevertheless, GAD was able to improve the precision of predicted region boundaries in all cases.

The quantitative analysis presented in Fig.~\ref{fig:inria} shows a small but consistent improvement due to the GAD algorithm. The small scale of these quantitative improvements can be explained by the fact that GAD only affects region boundaries, which is itself only a small fraction of the total number of pixels. But the consistency that is observed in the improvement of these results show that GAD is clearly improving the predictions. It is also important to note that the Dice scores are improved for both classes. According to these results, the subsampling rate at which GAD leads to the biggest gain is $ss=8$.

\begin{figure}[t]
    \begin{minipage}{0.15\linewidth}
        \centering
        \includegraphics[width=\linewidth]{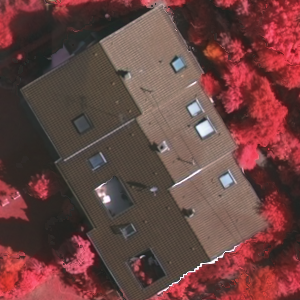}\\
        Input\\
        \vspace{3pt}
        \includegraphics[width=\linewidth]{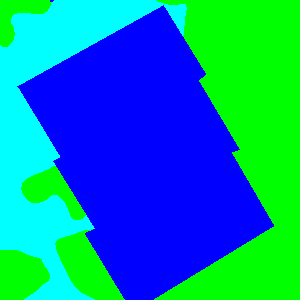}\\
        GT
    \end{minipage}
    ~
    \begin{minipage}{0.15\linewidth}
        \centering
        \includegraphics[width=\linewidth]{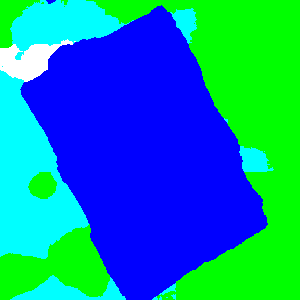}\\
        ss=1\\
        \vspace{3pt}
        \includegraphics[width=\linewidth]{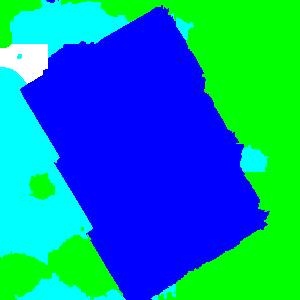}\\
        ss=1+GAD
    \end{minipage}
    ~
    \begin{minipage}{0.15\linewidth}
        \centering
        \includegraphics[width=\linewidth]{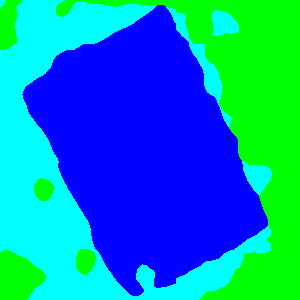}\\
        ss=4\\
        \vspace{3pt}
        \includegraphics[width=\linewidth]{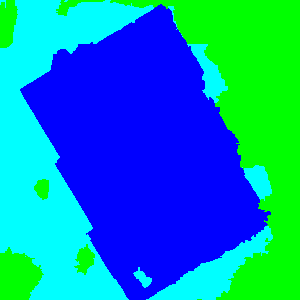}\\
        ss=4+GAD
    \end{minipage}
    ~
    \begin{minipage}{0.15\linewidth}
        \centering
        \includegraphics[width=\linewidth]{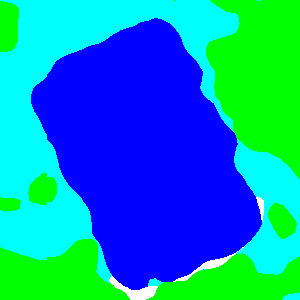}\\
        ss=8\\
        \vspace{3pt}
        \includegraphics[width=\linewidth]{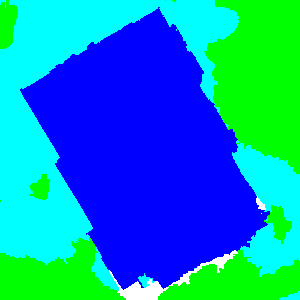}\\
        ss=8+GAD
    \end{minipage}
    ~
    \begin{minipage}{0.15\linewidth}
        \centering
        \includegraphics[width=\linewidth]{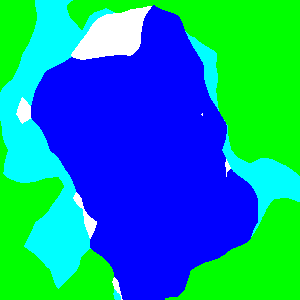}\\
        ss=12\\
        \vspace{3pt}
        \includegraphics[width=\linewidth]{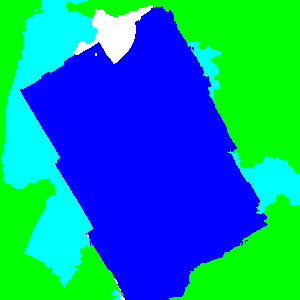}\\
        ss=12+GAD
    \end{minipage}
    ~
    \begin{minipage}{0.15\linewidth}
        \centering
        \includegraphics[width=\linewidth]{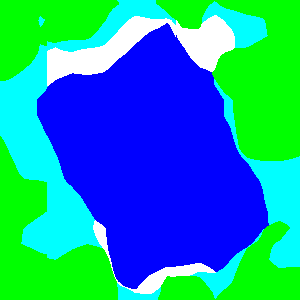}\\
        ss=16\\
        \vspace{3pt}
        \includegraphics[width=\linewidth]{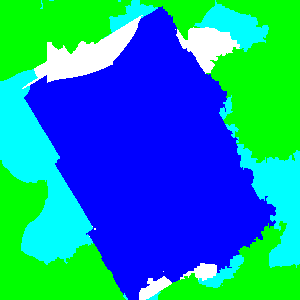}\\
        ss=16+GAD
    \end{minipage}

    \caption{Results obtained from experiments on Vaihingen dataset. GAD performs well in larger objects with visible boundaries in the guide image. Classes with weak color differences (e.g. trees next to low vegetation) don't provide the clear gradient maps that GAD needs to perform well.}
    \label{fig:vaihingen_results}
\end{figure}

\begin{figure}[ht]

    \begin{minipage}{0.32\linewidth}
        \centering
        \includegraphics[width=\linewidth]{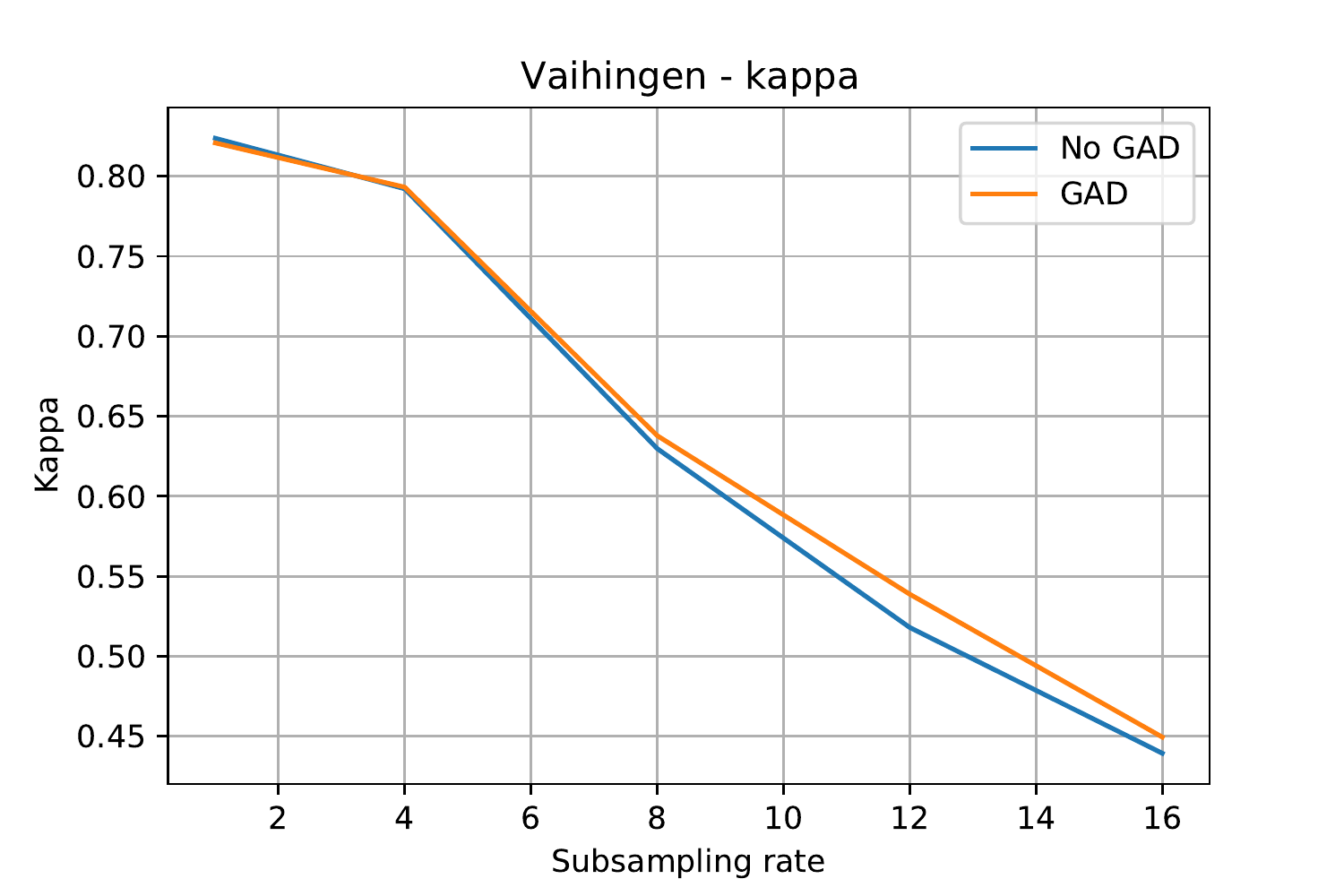}\\
    \end{minipage}
    \hfill
    \begin{minipage}{0.32\linewidth}
        \centering
        \includegraphics[width=\linewidth]{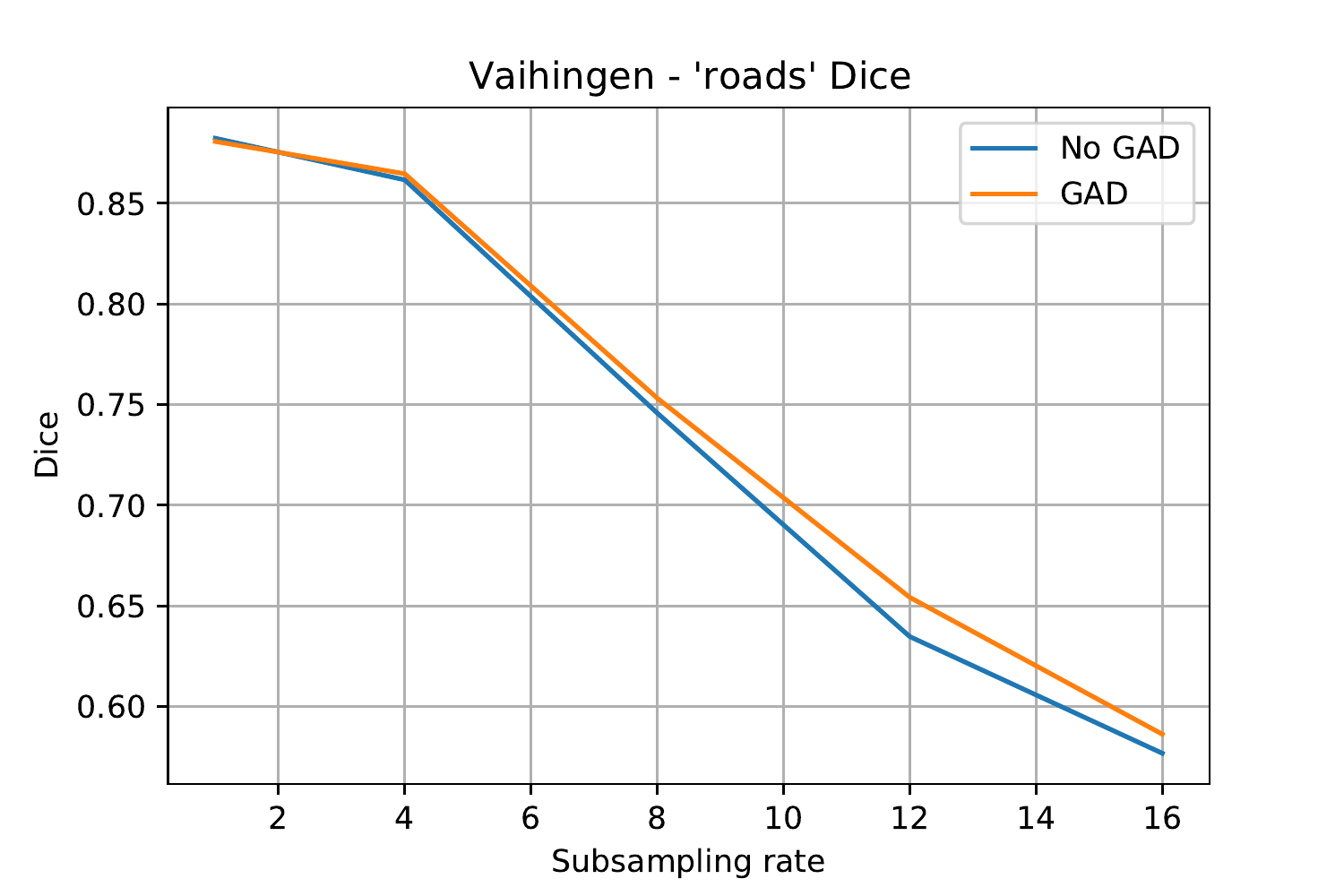}\\
    \end{minipage}
    \hfill
    \begin{minipage}{0.32\linewidth}
        \centering
        \includegraphics[width=\linewidth]{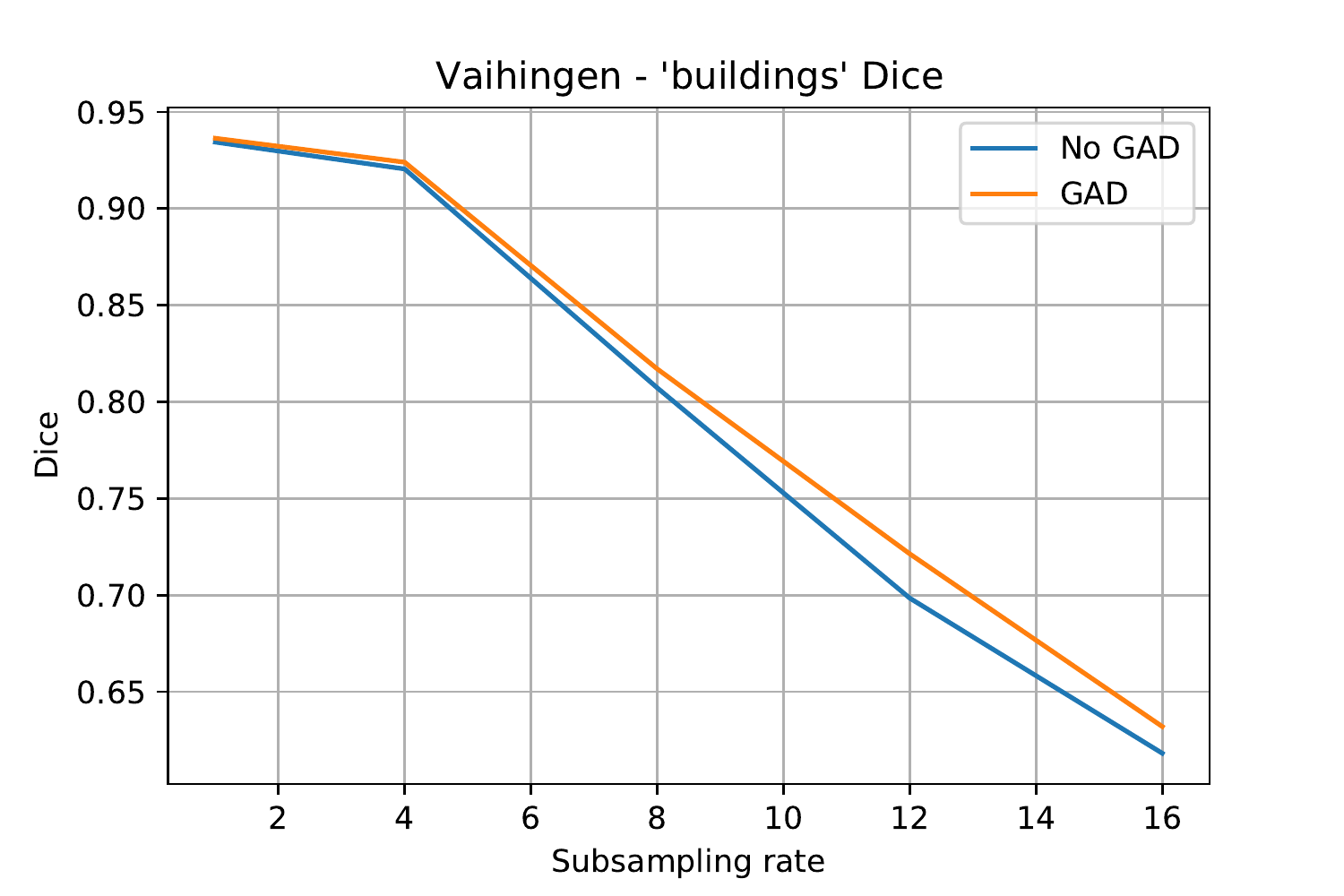}\\
    \end{minipage}\break
    
    \begin{minipage}{0.32\linewidth}
        \centering
        \includegraphics[width=\linewidth]{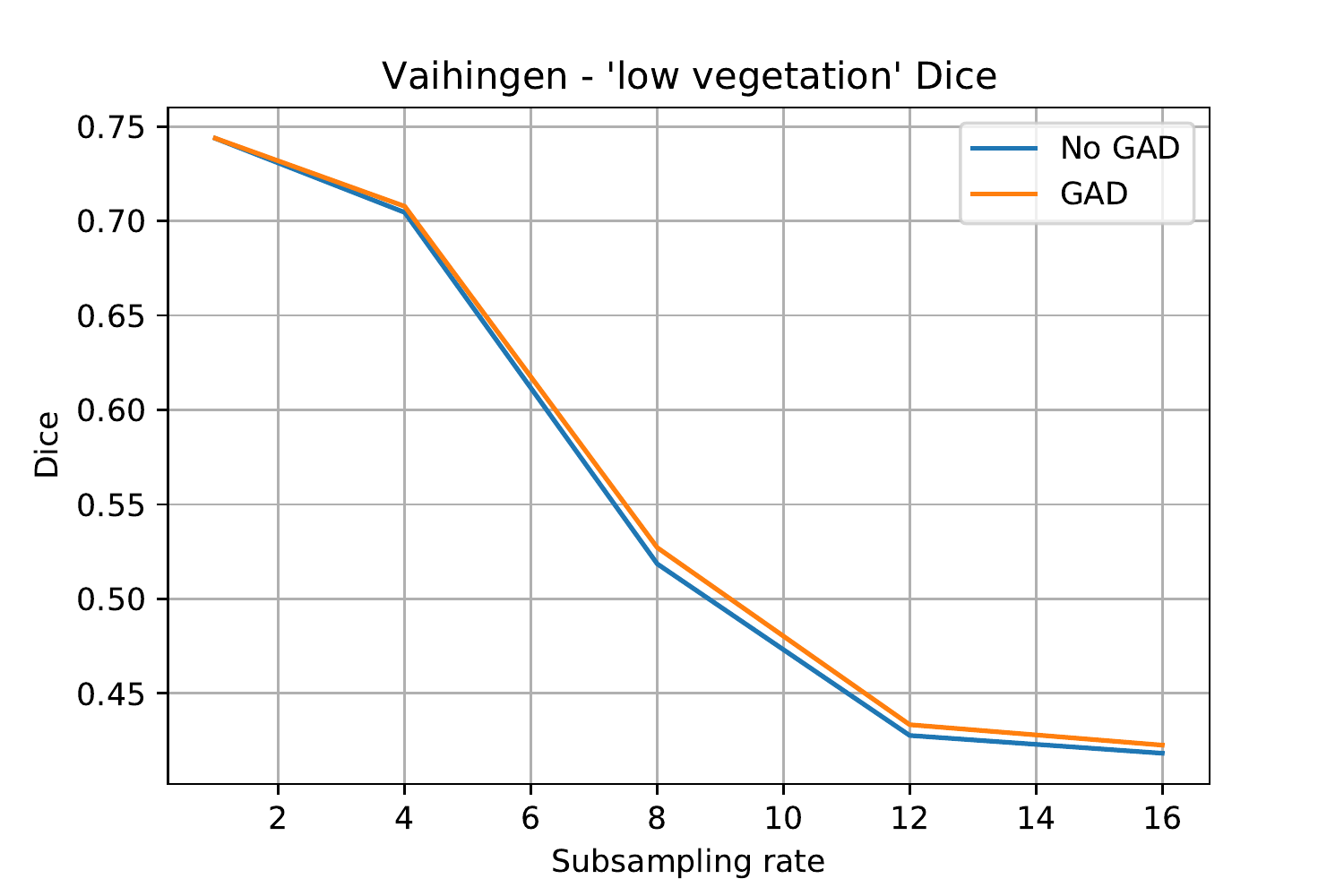}\\
    \end{minipage}
    \hfill
    \begin{minipage}{0.32\linewidth}
        \centering
        \includegraphics[width=\linewidth]{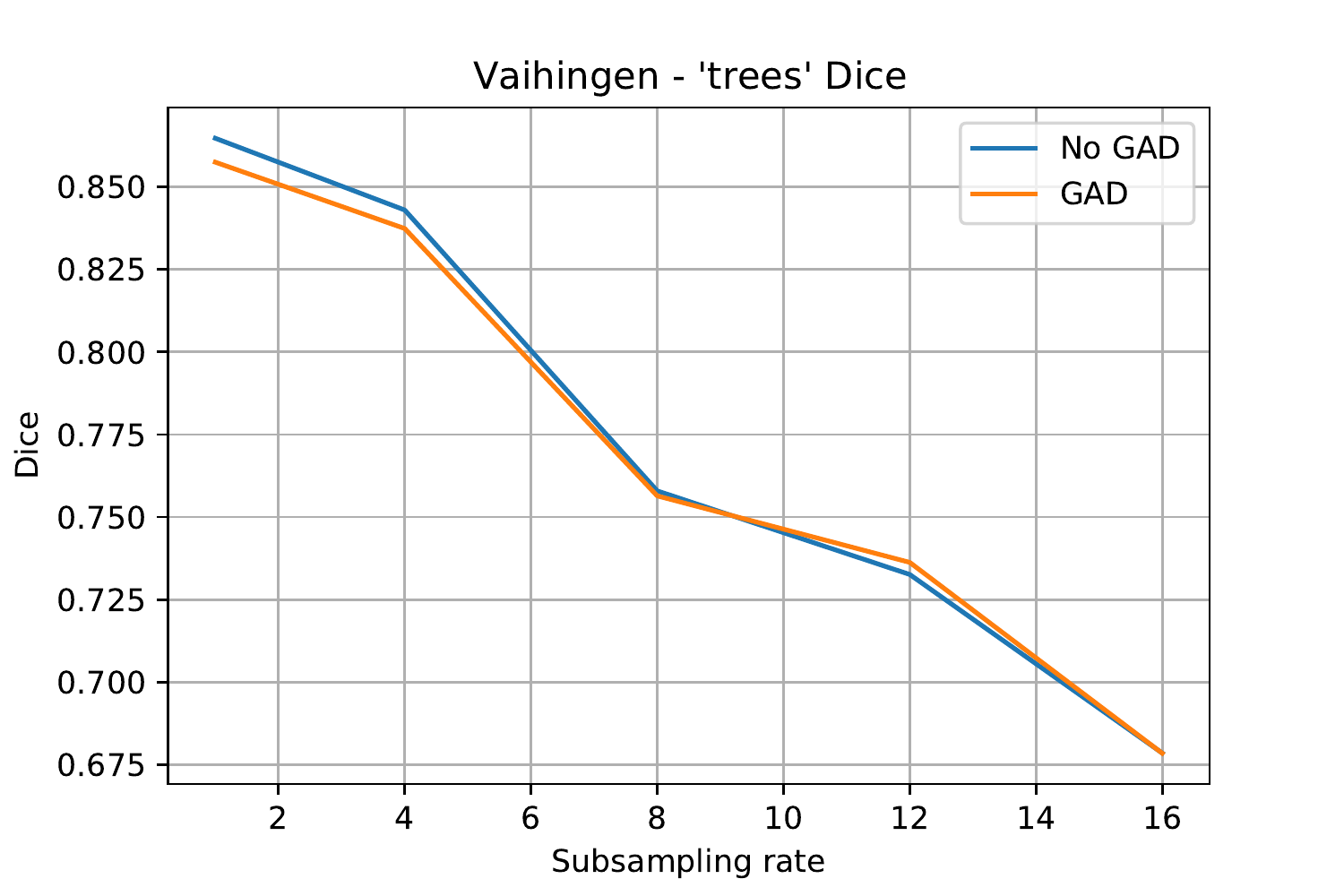}\\
    \end{minipage}
    \hfill
    \begin{minipage}{0.32\linewidth}
        \centering
        \includegraphics[width=\linewidth]{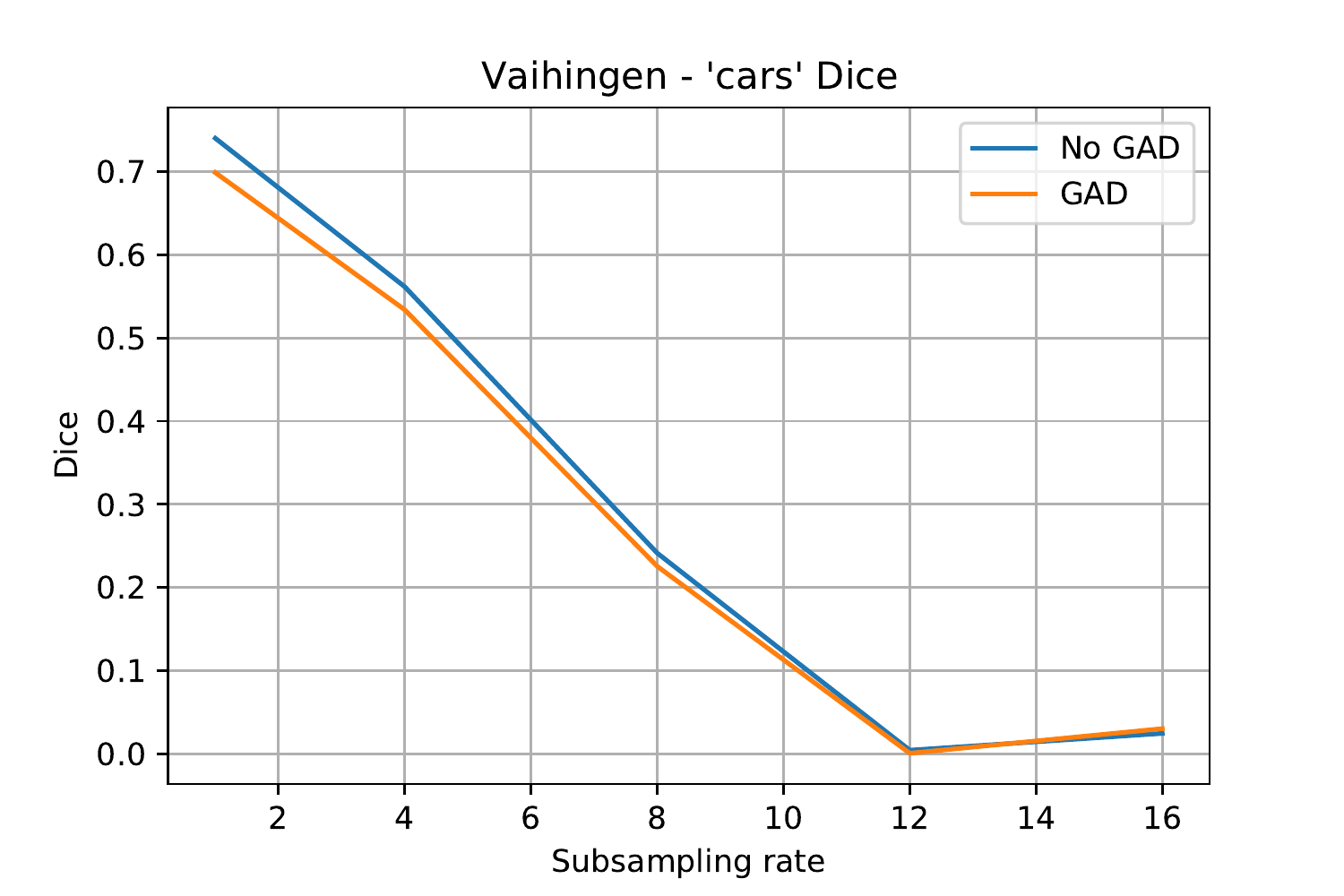}\\
    \end{minipage}\break
    
    \caption{Results for improving semantic segmentation on the Vaihingen dataset. Using GAD to postprocess the outputs consistently improves results for classes with larger objects and sharp visible edges.}
    \label{fig:vaihingen}
\end{figure}

\begin{figure}[ht]

    \begin{minipage}{0.32\linewidth}
        \centering
        \includegraphics[width=\linewidth]{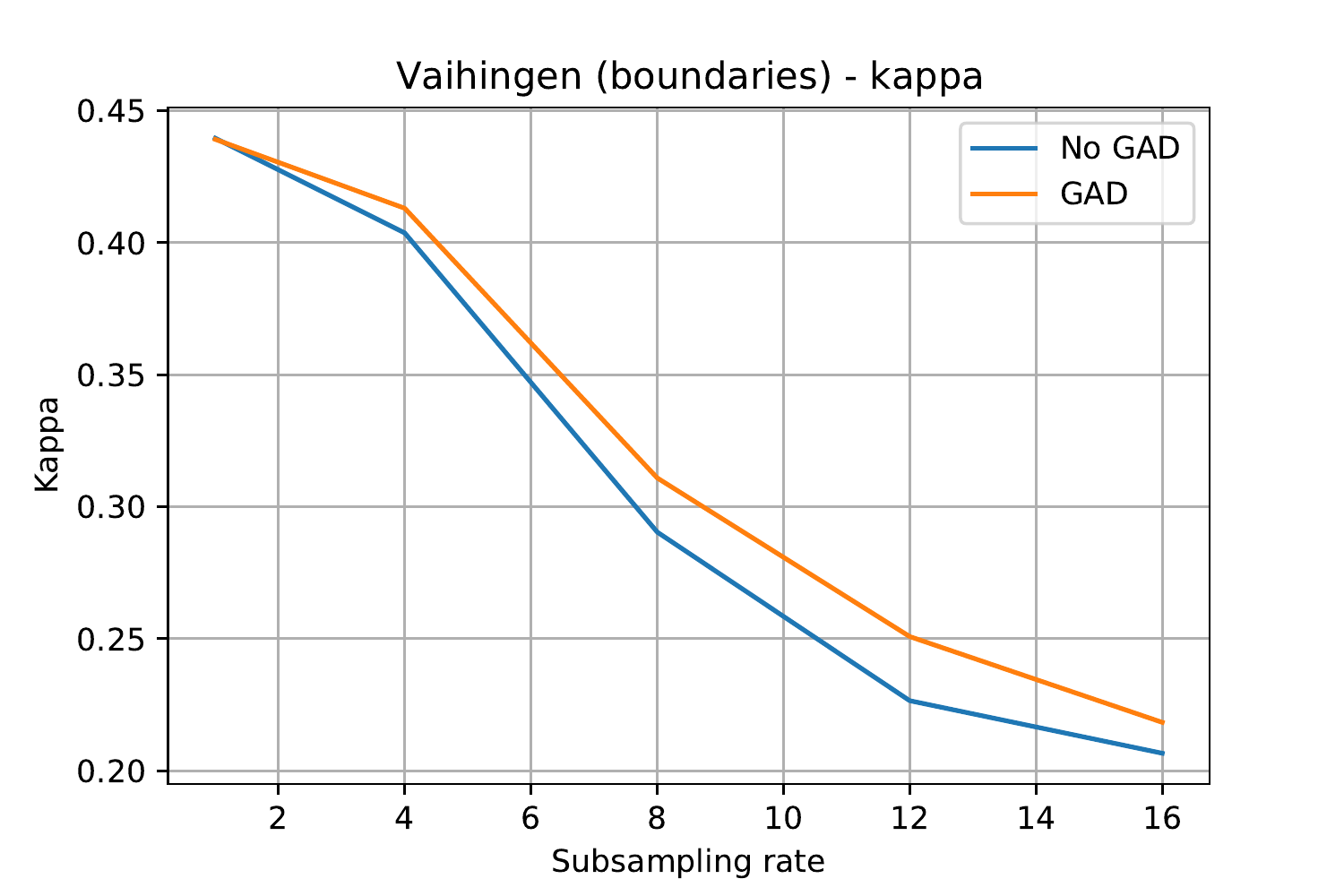}\\
    \end{minipage}
    \hfill
    \begin{minipage}{0.32\linewidth}
        \centering
        \includegraphics[width=\linewidth]{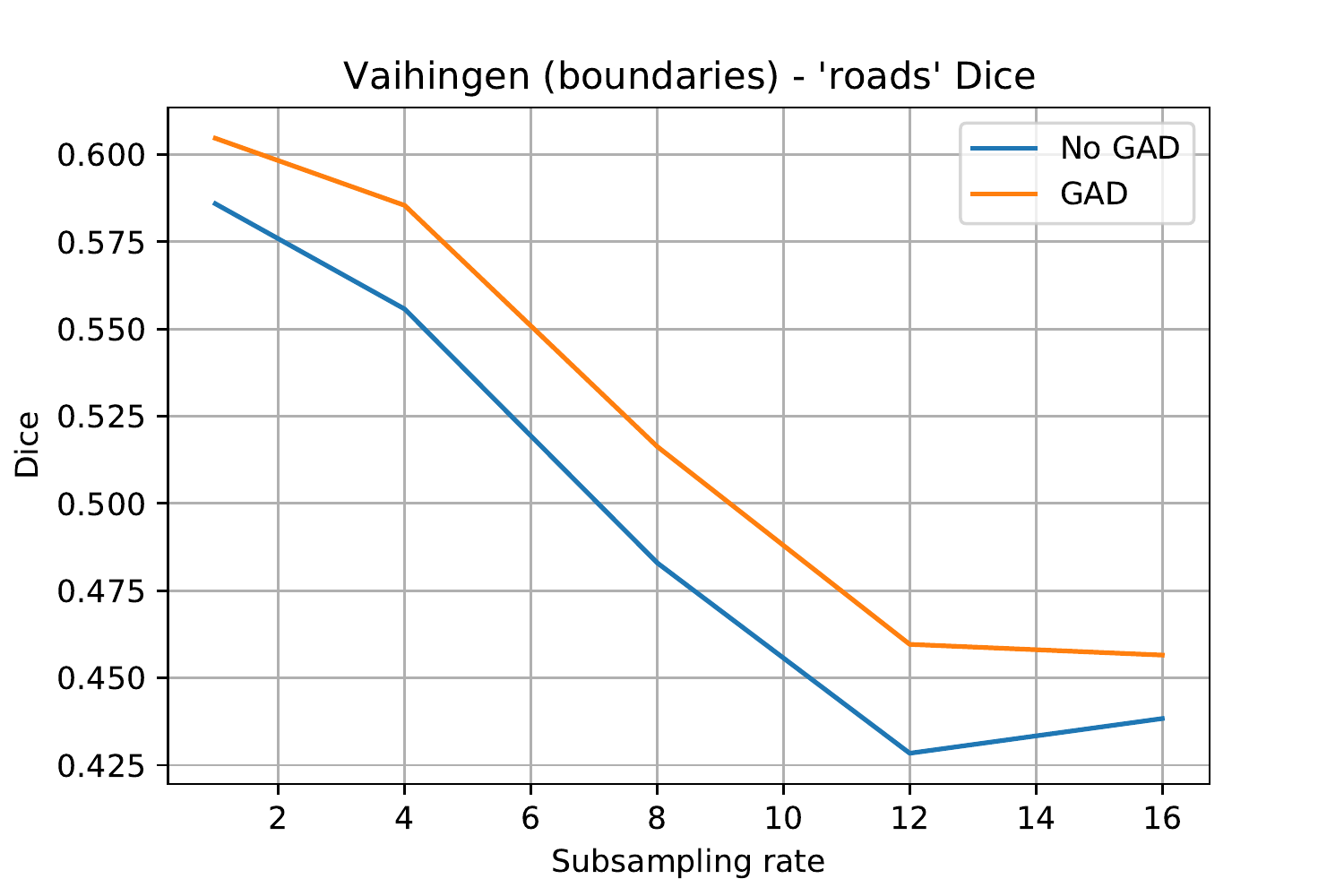}\\
    \end{minipage}
    \hfill
    \begin{minipage}{0.32\linewidth}
        \centering
        \includegraphics[width=\linewidth]{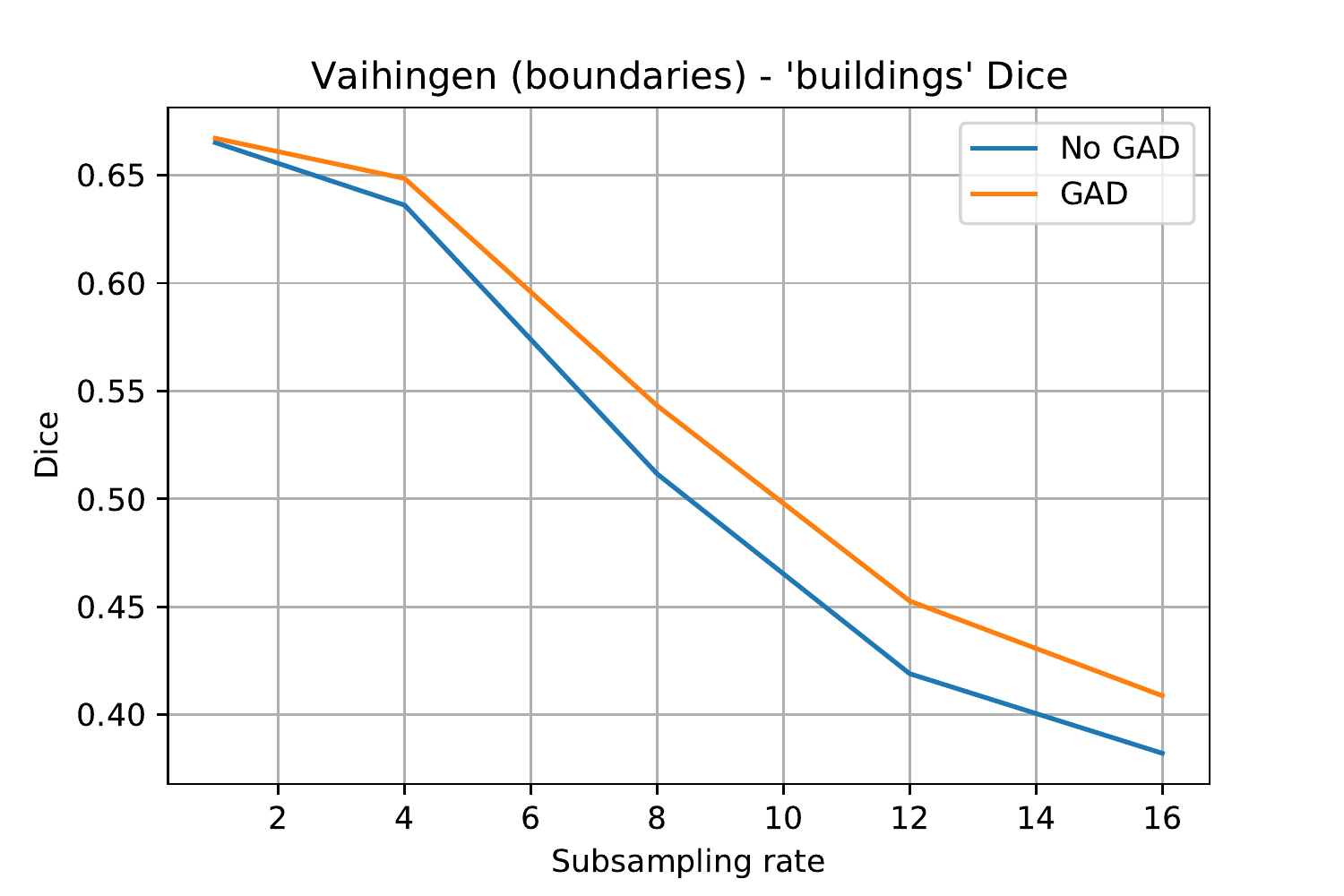}\\
    \end{minipage}\break

    \begin{minipage}{0.32\linewidth}
        \centering
        \includegraphics[width=\linewidth]{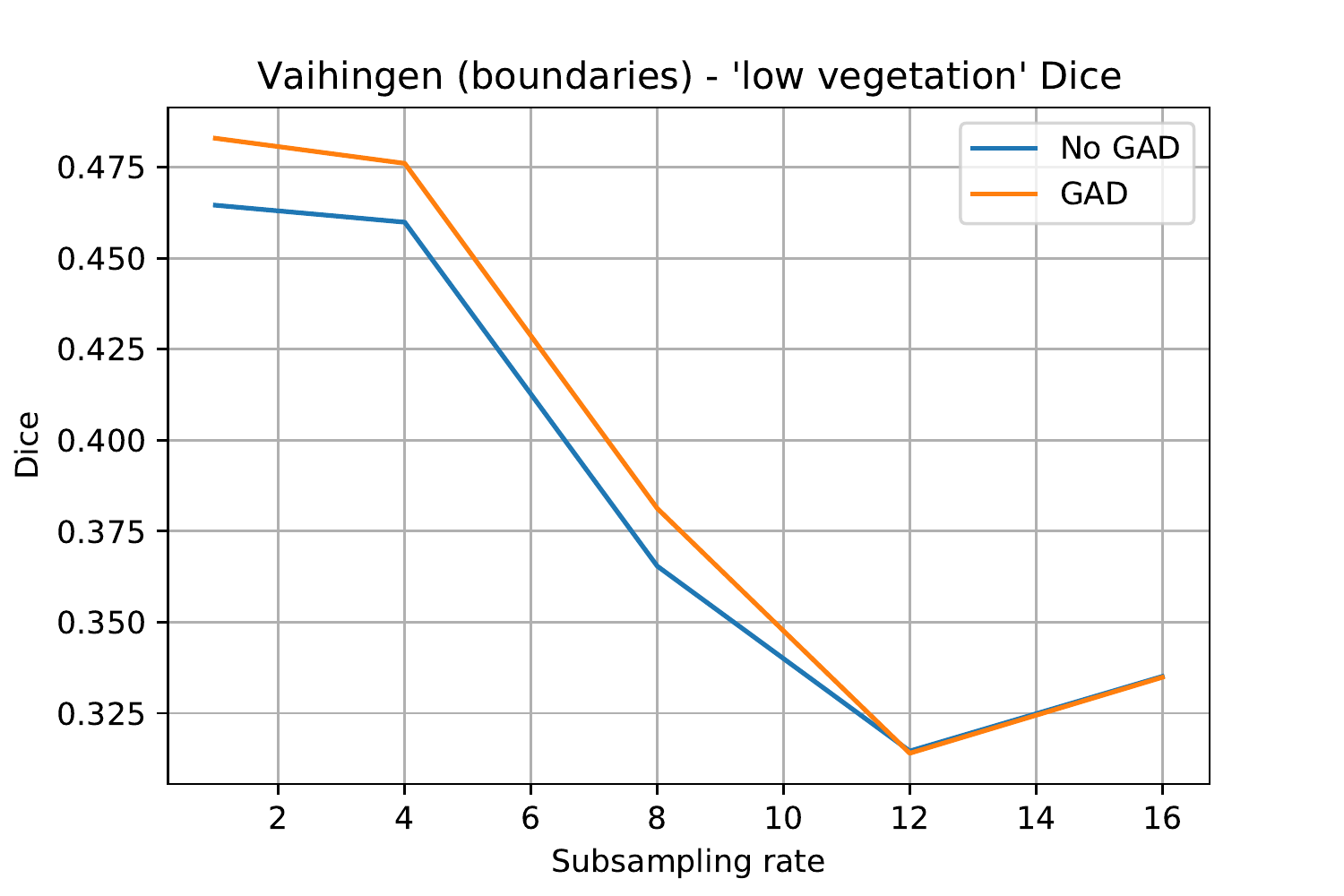}\\
    \end{minipage}
    \hfill
    \begin{minipage}{0.32\linewidth}
        \centering
        \includegraphics[width=\linewidth]{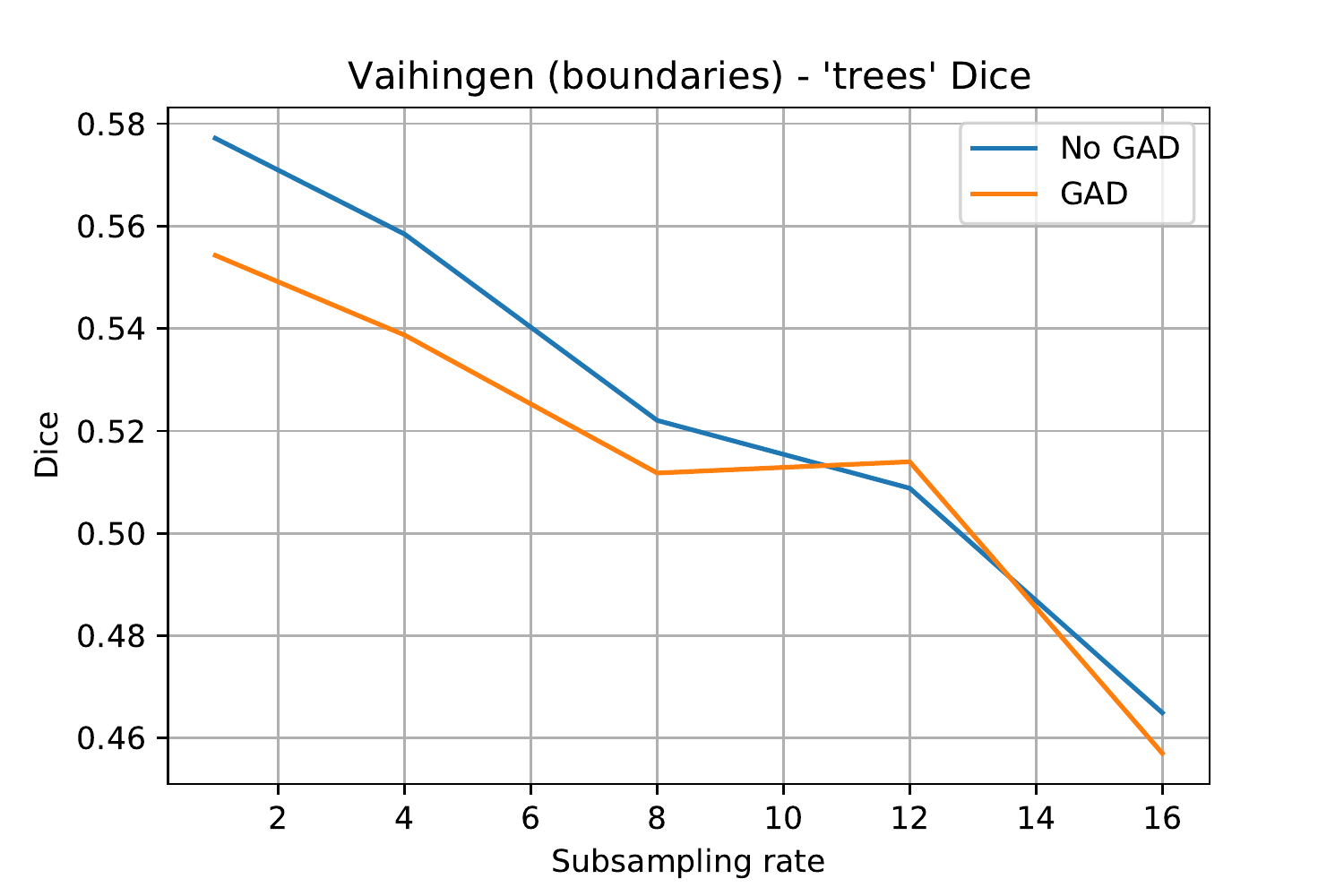}\\
    \end{minipage}
    \hfill
    \begin{minipage}{0.32\linewidth}
        \centering
        \includegraphics[width=\linewidth]{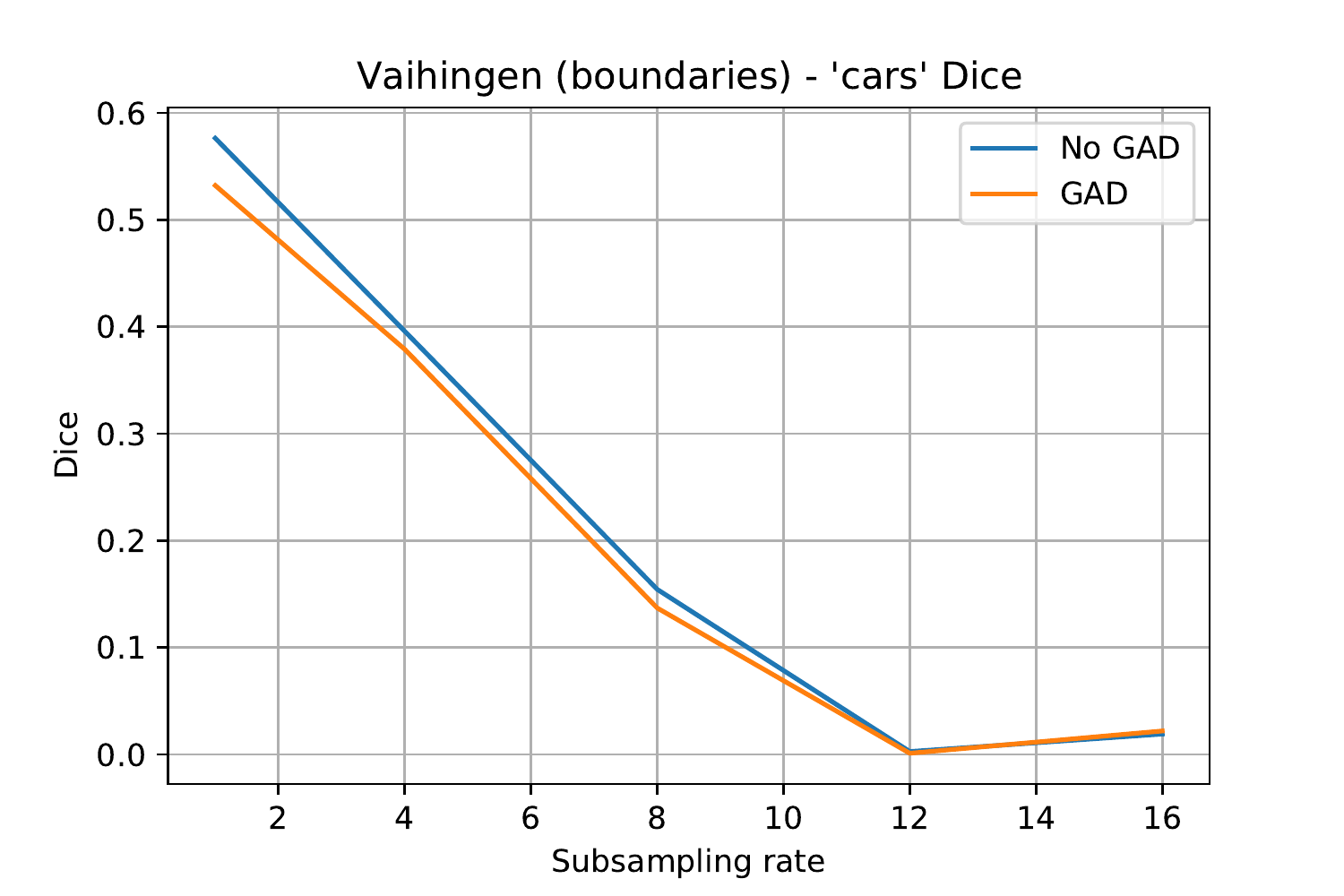}\\
    \end{minipage}\break

    \caption{Results for improving segmentation boundaries on the Vaihingen dataset considering only pixels around region boundaries. The effect of postprocessing results with GAD becomes clearer where one semantic region encounters another.}
    \label{fig:vaihingen_boundaries}
\end{figure}

Results on the Vaihingen dataset can be seen in Figs. \ref{fig:vaihingen_results}, \ref{fig:vaihingen}, and \ref{fig:vaihingen_boundaries}. Two main conclusions can be drawn from Fig.~\ref{fig:vaihingen_results}. First, objects with well defined boundaries in the guide image (e.g. buildings) benefit from GAD postprocessing. Second, objects with fuzzy borders or near objects with similar colors (e.g. trees next to grass) do not provide the necessary gradients that guide the anisotropic diffusion, and therefore do not benefit from GAD postprocessing.

Figure~\ref{fig:vaihingen_results} shows that the classes "buildings", "roads", and "low vegetation" profit from GAD postprocessing similarly to the results on the Inria dataset. Once again, the small gains can be explained by the fact that GAD acts only on region boundaries. The "trees" and "cars" classes did not benefit from GAD postprocessing, likely for different reasons. The "trees" regions often had colours very similar to the ones of to the neighbouring regions, which did not lead to sharp gradient maps. The "cars" class contained objects that are relatively small, and therefore could easily be eroded away by the anisotropic diffusion. At higher subsampling rates, cars would only occupy one or two pixels in the image, which explains why the network itself failed to detect them at these scales.

To highlight the impact of GAD postprocessing around region boundaries, the same metrics were calculated using only pixels around region boundaries. The locations of such pixels were calculated using the complement of the "gts\_eroded\_for\_participants" available with the Vaihingen dataset files. These results can be seen in Fig.~\ref{fig:vaihingen_boundaries}. These results show that the effects of GAD postprocessing is much stronger around region boundaries, which is coherent with what was expected. 

\section{Analysis}

The experiments presented in the previous section showed how GAD was successfully used in two different weakly supervised change detection settings. The results show an increase in performance in object-level segmentation from parcel-level labels through label cleaning, as well as the seldom explored task of weakly supervised image co-segmentation using classification labels.

The iterative training results made clear that it is of paramount importance to refer back to the ground truth data every time the training ground truth is being modified. Not doing so leads to a fast degradation in performance, since the network simply attempts to learn to copy itself and stops learning useful operations from the data. The results also showed that separating dubiously labelled pixels leads to a small increase in performance, likely due to the fact that we end up providing a cleaner and more trustworthy dataset at training time.

The guided anisotropic diffusion algorithm was compared against the Dense CRF algorithm for using information from the input images to improve semantic segmentation results. While both algorithms were successful when used in the proposed iterative training scheme, GAD outperformed Dense CRF at later hyperepochs for quantitative metrics. Both algorithms yielded visually pleasing results, each performing better in different test cases.

One possible criticism of the proposed iterative training method is that it would get rid of hard and important examples in the training dataset. It is true that the performance of this weakly supervised training scheme would likely never reach that of one supervised with perfectly clean data, but the results in Section~\ref{sec:experiments} show that using the proposed method we can consistently train networks that perform better than those naively trained with noisy data directly.

The proposed spatial attention operation was showed to be useful in improving the classification and weakly supervised segmentation results for datasets which are cropped using object locations as reference points. While this is a particular case, such datasets are often available or can be easily generated for remote sensing applications, where georeferenced data is widely available. The proposed ideas have been only tested in a two-class problem, but there is nothing that indicates that such methods would not work just as well in a multi-class context. Filtering the attention weights with the GAD algorithm further increased the classification performance of the network by increasing the coherence between the attention weights and the region where the building of interest is located in each image.

Finally, the usage of GAD as an edge enhancing postprocessing algorithm was tested in a semantic segmentation setting using two remote sensing datasets to compensate for networks trained with lower spatial resolution images. The results were mostly positive, and showed that GAD is effective at improving segmentation boundaries for classes with well defined edges and strong gradients, as long as the objects are not too small. This shows that GAD is a versatile tool for enhancing segmentation edges in a variety of settings.
\section{Conclusion}

In this paper we have proposed the guided anisotropic diffusion algorithm for improving semantic segmentation results by performing a cross-image edge preserving filtering. It was shown to improve semantic segmentation results on two standard aerial datasets, leading to better boundary accuracy for semantic segmentation results. We have then proposed two GAD-based weakly supervised change detection methods to demonstrate how it can help to recover from inaccurate segmentation labels or go beyond the available classification labels.

We first proposed an iterative training method for training networks with noisy data that alternates between training a fully convolutional network and leveraging its predictions to clean the training dataset from mislabelled examples.
We showed that the proposed method outperforms naive supervised training using the provided reference data for change detection. The GAD algorithm was used in conjunction with the iterative training method to obtain the best results in our tests. The GAD algorithm was compared against the Dense CRF algorithm, and was found to be superior in performance.
 
Finally, we proposed a spatial attention operation that can be easily incorporated into existing classification networks that significantly improve the classification and weakly supervised segmentation performances for datasets with object-aligned crops.

The proposed methods are useful when using data-based approaches in data-scarce domains, as is the case of change detection. We have observed improvement in all of our tests when approaching the problem from a weakly supervised perspective, as opposed to naive supervision. It would be interesting to test the efficacy of the proposed ideas outside the context of change detection. The proposed methods could be applied with minor adaptations to other applications to help mitigate the effects of data scarcity.

\bibliographystyle{spmpsci}      
\bibliography{refs.bib}   

\end{document}